\magnification=1200
\input epsf
\advance\hsize by 0.4truecm

\catcode`@=11 


\font\ninerm=cmr9
\font\eightrm=cmr8
\font\sixrm=cmr6

\font\ninei=cmmi9
\font\eighti=cmmi8
\font\sixi=cmmi6
\skewchar\ninei='177 \skewchar\eighti='177 \skewchar\sixi='177

\font\ninesy=cmsy9
\font\eightsy=cmsy8
\font\sixsy=cmsy6
\skewchar\ninesy='60 \skewchar\eightsy='60 \skewchar\sixsy='60

\font\ninebf=cmbx9
\font\eightbf=cmbx8
\font\sixbf=cmbx6

\font\ninett=cmtt9
\font\eighttt=cmtt8

\hyphenchar\tentt=-1 
\hyphenchar\ninett=-1
\hyphenchar\eighttt=-1

\font\ninebf=cmbx9
\font\eightbf=cmbx8
\font\sixbf=cmbx6

\font\ninett=cmtt9
\font\eighttt=cmtt8

\hyphenchar\tentt=-1 
\hyphenchar\ninett=-1
\hyphenchar\eighttt=-1

\font\ninesl=cmsl9
\font\eightsl=cmsl8

\font\nineit=cmti9
\font\eightit=cmti8



\newskip\ttglue
\def\tenpoint{\def\rm{\fam0\tenrm}%
  \textfont0=\tenrm \scriptfont0=\sevenrm \scriptscriptfont0=\fiverm
  \textfont1=\teni \scriptfont1=\seveni \scriptscriptfont1=\fivei
  \textfont2=\tensy \scriptfont2=\sevensy \scriptscriptfont2=\fivesy
  \textfont3=\tenex \scriptfont3=\tenex \scriptscriptfont3=\tenex
  \def\it{\fam\itfam\tenit}%
  \textfont\itfam=\tenit
  \def\sl{\fam\slfam\tensl}%
  \textfont\slfam=\tensl
  \def\bf{\fam\bffam\tenbf}%
  \textfont\bffam=\tenbf \scriptfont\bffam=\sevenbf
   \scriptscriptfont\bffam=\fivebf
  \def\tt{\fam\ttfam\tentt}%
  \textfont\ttfam=\tentt
  \tt \ttglue=.5em plus.25em minus.15em
  \normalbaselineskip=12pt
  \let\sc=\eightrm
  \let\big=\tenbig
  \setbox\strutbox=\hbox{\vrule height8.5pt depth3.5pt width\z@}%
  \normalbaselines\rm}

\def\ninepoint{\def\rm{\fam0\ninerm}%
  \textfont0=\ninerm \scriptfont0=\sixrm \scriptscriptfont0=\fiverm
  \textfont1=\ninei \scriptfont1=\sixi \scriptscriptfont1=\fivei
  \textfont2=\ninesy \scriptfont2=\sixsy \scriptscriptfont2=\fivesy
  \textfont3=\tenex \scriptfont3=\tenex \scriptscriptfont3=\tenex
  \def\it{\fam\itfam\nineit}%
  \textfont\itfam=\nineit
  \def\sl{\fam\slfam\ninesl}%
  \textfont\slfam=\ninesl
  \def\bf{\fam\bffam\ninebf}%
  \textfont\bffam=\ninebf \scriptfont\bffam=\sixbf
   \scriptscriptfont\bffam=\fivebf
  \def\tt{\fam\ttfam\ninett}%
  \textfont\ttfam=\ninett
  \tt \ttglue=.5em plus.25em minus.15em
  \normalbaselineskip=11pt
  \let\sc=\sevenrm
  \let\big=\ninebig
  \setbox\strutbox=\hbox{\vrule height8pt depth3pt width\z@}%
  \normalbaselines\rm}

\def\eightpoint{\def\rm{\fam0\eightrm}%
  \textfont0=\eightrm \scriptfont0=\sixrm \scriptscriptfont0=\fiverm
  \textfont1=\eighti \scriptfont1=\sixi \scriptscriptfont1=\fivei
  \textfont2=\eightsy \scriptfont2=\sixsy \scriptscriptfont2=\fivesy
  \textfont3=\tenex \scriptfont3=\tenex \scriptscriptfont3=\tenex
  \def\it{\fam\itfam\eightit}%
  \textfont\itfam=\eightit
  \def\sl{\fam\slfam\eightsl}%
  \textfont\slfam=\eightsl
  \def\bf{\fam\bffam\eightbf}%
  \textfont\bffam=\eightbf \scriptfont\bffam=\sixbf
   \scriptscriptfont\bffam=\fivebf
  \def\tt{\fam\ttfam\eighttt}%
  \textfont\ttfam=\eighttt
  \tt \ttglue=.5em plus.25em minus.15em
  \normalbaselineskip=9pt
  \let\sc=\sixrm
  \let\big=\eightbig
  \setbox\strutbox=\hbox{\vrule height7pt depth2pt width\z@}%
  \normalbaselines\rm}

\newinsert\quote

\newdimen\marginheight
\newinsert\margin
\font\marginfont=cmti7
\def\marginentry#1{
             \vadjust{\setbox\margin=\vbox{
              \parindent=0cm  
              \hsize=0.15\hsize 
              \marginfont #1}
             \marginheight=\ht\margin 
             \advance\marginheight by \dp\margin    
              \advance\marginheight by \lineskip
             \kern -\marginheight 
             \vbox to 
\marginheight{\rightline{\rlap{\quad\lower2\marginheight\box\margin}}
             \vss}
             }}

\newcount\pagenumber
\newcount\questionnumber
\newcount\sectionnumber
\newcount\appendixnumber
\newcount\equationnumber
\newcount\referencenumber
\newcount\subsecnumber
\newcount\footcount

\global\footcount=1
\def\foot#1{\interlinepenalty=-100 
\footnote{$^{\,\the\footcount}$}{\eightpoint #1 
\global\advance\footcount by 1}}

\def\ifundefined#1{\expandafter\ifx\csname#1\endcsname\relax}
\def\docref#1{\ifundefined{#1} {\bf ?.?}\message{#1 not yet defined,}
\else \csname#1\endcsname \fi}

\def\no{\eqno({\rm\sectionlabel} 
.\the\equationnumber){\global\advance\equationnumber by1}}

\def\beginref #1 {\par\vskip 2.4 pt\noindent\item{\bf\the\referencenumber .}
\noindent #1\par\vskip 2.4 pt\noindent{\global\advance\referencenumber by1}}
 
\def\ref #1{{\bf [#1]}} 

\def\numberpages{\pagenumber=1\output={\shipout\vbox{\centerline{\bf\the\pagenumber }
\par\vskip.5truecm\box255\vskip10truept\box\footins}
\global\advance\count0by1\global\advance\pagenumber by1}}                       

\def\sectionlabel{\ifnum\appendixnumber>0 \appendixlabel
\else\the\sectionnumber\fi}

\def\smalltitle #1 {\par\vskip12truept  minus 12truept
\noindent {\medit #1} \par\penalty 10000\vskip10truept  
minus 10truept\noindent 
}

\def\subsec #1 {\bf\par\vskip8truept  minus 8truept
\noindent \ifnum\appendixnumber=0 $\S\S\;$\else\fi
$\bf\sectionlabel.\the\subsecnumber$ #1
\global\advance\subsecnumber by1
\rm\par\penalty 10000\vskip6truept  minus 6truept\noindent}

\def\beginquote{\par\vskip0.8\baselineskip 
                 plus 0.8\baselineskip
                 minus 0.8\baselineskip
                \ninepoint\it\narrower\narrower}

\def\endquote{\par\vskip0.8\baselineskip 
                 plus 0.8\baselineskip
                 minus 0.8\baselineskip
                }

\def\beginsection #1
 {{\global\subsecnumber=1\global\advance\sectionnumber by1}\equationnumber=1
\par\vskip 1.0\baselineskip plus 1.0\baselineskip
 minus 1.0\baselineskip
\noindent$\S$ {\bfmed \the\sectionnumber . #1}
\par\penalty 10000\vskip 0.8\baselineskip plus 0.8\baselineskip 
minus 0.6\baselineskip \noindent}

\font\ssfont=cmss9 
\font\medit=cmti10 scaled\magstep1
\font\med=cmr10 scaled\magstep3
\font\bflarge=cmbx10 scaled\magstep3 
\font\bfmed=cmbx10 scaled\magstep2

\font\eufscript=eufm10
\def\scriptl#1{\hbox{\eufscript #1}}

\def\ham{{\cal H}}
\def\dbar{\overline\partial}
\def\det{det\,}

\def\dirac{\rlap/\partial_A}

\def\map{\hbox{\it Map}\,}


\def\longbar#1{\setbox1=\hbox{$#1$}
\setbox2=\vbox{\hrule width 0.8\wd1}
\raise0.5\ht1\hbox{${\lower\dp1\box2}\atop\box1$}}  
\def\mediumbar#1{\setbox1=\hbox{$#1$}
\setbox2=\vbox{\hrule width 0.6\wd1}
\raise0.5\ht1\hbox{${\lower\dp1\box2}\atop\box1$}}  

\def\instmodk{{\cal M}_k}
\def\instmodkbar{\longbar{{\cal M}}_k}

\def\theorem#1#2{
\par\vskip 0.55\baselineskip  minus 0.55\baselineskip\noindent 
{\bf Theorem} (#1){ \it #2}
\par\penalty 10000\vskip 0.4\baselineskip  minus 0.4\baselineskip\noindent}

\newdimen\tempsize
\def\skeingraphs{\global\tempsize=\hsize
\setbox1=\hbox{$\Biggr/$} 
\setbox3=\hbox{$\Biggl\backslash$}
\setbox2=\vbox{\hbox{$\big\backslash$}
\vskip3pt\hbox{\hskip5.75pt$\big\backslash$}}
\setbox4=\vbox{\hbox{\hskip5.75pt$\big/$}
\vskip3pt\hbox{$\big/$}}
\setbox5=\hbox to 4.1\wd2
{\global\hsize=35pt\raise8pt\vbox{\copy1}\vbox{\hskip-32pt\hbox{\copy2}}}
\setbox6=\hbox to 4.3\wd4{\global\hsize=35pt\raise8pt\vbox{\hbox{\copy3}}
\vbox{\hskip-34.4pt\hbox{\copy4}}}
\global\hsize=\tempsize
\matrix{\lower6pt\vbox{\copy5}&\lower6pt\vbox{\copy6}&\Biggl)\quad\Biggr(\cr
&&\cr
L_+\hfill&L_-\hfill&\;\,L_0\cr}}

\def\alexanderskeinrel{\global\tempsize=\hsize
\setbox1=\hbox{$\Biggr/$} 
\setbox3=\hbox{$\Biggl\backslash$}
\setbox2=\vbox{\hbox{$\big\backslash$}
\vskip3pt\hbox{\hskip5.75pt$\big\backslash$}}
\setbox4=\vbox{\hbox{\hskip5.75pt$\big/$}
\vskip3pt\hbox{$\big/$}}
\setbox5=\hbox to 4.1\wd2
{\global\hsize=35pt\raise8pt\vbox{\copy1}\vbox{\hskip-32pt\hbox{\copy2}}}
\setbox6=\hbox to 4.3\wd4{\global\hsize=35pt\raise8pt\vbox{\hbox{\copy3}}
\vbox{\hskip-34.4pt\hbox{\copy4}}}
\global\hsize=\tempsize
\lower6pt\vbox{\copy5}\hskip-1cm +\quad\lower6pt\vbox{\copy6}
\hskip-1cm +\quad(t^{1/2}-t^{-1/2})\;\,\Biggl)\quad\Biggr(\;\,=0}

\def\jonesskeinrel{\global\tempsize=\hsize
\setbox1=\hbox{$\Biggr/$} 
\setbox3=\hbox{$\Biggl\backslash$}
\setbox2=\vbox{\hbox{$\big\backslash$}
\vskip3pt\hbox{\hskip5.75pt$\big\backslash$}}
\setbox4=\vbox{\hbox{\hskip5.75pt$\big/$}
\vskip3pt\hbox{$\big/$}}
\setbox5=\hbox to 4.1\wd2
{\global\hsize=35pt\raise8pt\vbox{\copy1}\vbox{\hskip-32pt\hbox{\copy2}}}
\setbox6=\hbox to 4.3\wd4{\global\hsize=35pt\raise8pt\vbox{\hbox{\copy3}}
\vbox{\hskip-34.4pt\hbox{\copy4}}}
\global\hsize=\tempsize
t\;\lower6pt\vbox{\copy5}\hskip-1cm -\quad t^{-1}\;
\lower6pt\vbox{\copy6}
\hskip-1cm +\quad(t^{1/2}-t^{-1/2})\;\,\Biggl)\quad\Biggr(\;\,=0}


\newread\bib
\newwrite\reffs
\newcount\linecount
\newcount\citecount
\newcount\localauthorcount 
\newread\bib
\newwrite\references

\immediate\openout\references=temprefs

\newcount\linecount
\newcount\citecount
\newcount\articletype
\newcount\localauthorcount 
\newcount\numberbox
\newdimen\localindentsize

\def\specialarticlestyle{
\global\referencenumber=1
\def\eqlabel##1{\edef##1{\sectionlabel.\the\equationnumber}}
\def\seclabel##1{\edef##1{\sectionlabel}}                  
\def\feqlabel##1{\ifnum\passcount=1
\immediate\write\crossrefsout{\relax}  
\immediate\write\crossrefsout{\def\string##1{\sectionlabel.
\the\equationnumber}}\else \fi }
\def\fseclabel##1{\ifnum\passcount=1
\immediate\write\crossrefsout{\relax}   
\immediate\write\crossrefsout{\def\string##1{\sectionlabel}}\else\fi}
\def\docite##1 auth ##2 title ##3 jour ##4 vol ##5 pages ##6 year ##7{
\global\localindentsize=\parindent
\global\setbox\numberbox=\hbox{\the\localauthorcount. }
\global\advance\localindentsize by \wd\numberbox
\ifnum\localauthorcount=1 
\par\vskip 2.4 pt\noindent 
##2\par\penalty 10000\global\hangindent\localindentsize\global\hangafter=1
\the\localauthorcount.  ##3, {\it ##4}, {\bf ##5}, ##6, (##7).
\else \par\global\hangindent\localindentsize\global\hangafter=1
\the\localauthorcount.  ##3, {\it ##4}, {\bf ##5}, ##6, (##7). \fi}
\def\dobkcite##1 auth ##2 title ##3 publisher ##4 year ##5{
\global\localindentsize=\parindent
\global\setbox\numberbox=\hbox{\the\localauthorcount. }
\global\advance\localindentsize by \wd\numberbox
\ifnum\localauthorcount=1 
\par\vskip 2.4 pt\noindent ##2\par\penalty 10000\global\hangindent\localindentsize\global\hangafter=1
\the\localauthorcount. {\it ##3}, ##4, (##5).
\else \par\global\hangindent\localindentsize\global\hangafter=1
\the\localauthorcount. {\it ##3}, ##4, (##5). \fi}             
\def\doconfcite##1 auth ##2 title ##3 conftitle ##4 editor ##5 publisher ##6 
year ##7{
\global\localindentsize=\parindent
\global\setbox\numberbox=\hbox{\the\localauthorcount. }
\global\advance\localindentsize by \wd\numberbox
\ifnum\localauthorcount=1 
\par\vskip 2.4 pt\noindent ##2\par\penalty 10000\global\hangindent\localindentsize\global\hangafter=1
\the\localauthorcount. {\it ##3}, ##4, {edited by: ##5}, ##6, (##7).
\else \par\global\hangindent\localindentsize\global\hangafter=1
\the\localauthorcount. {\it ##3}, ##4, {edited by: ##5}, ##6, (##7).\fi} 
\def\cite[##1,##2]{\immediate\openin\bib=bib.tex\global\citecount=##1
\global\localauthorcount=##2
\global\linecount=0{\loop\ifnum\linecount<\citecount \read\bib
to\temp \global\advance\linecount by 1\repeat
\immediate\write\references{\temp}}\immediate\closein\bib}
}

\specialarticlestyle



\def\title{Topology and Physics---a historical essay}
\centerline{\bflarge \title}
\par\vskip25pt
\centerline{\med CHARLES NASH}
\par\vskip0.5\baselineskip
{\noindent
\ninepoint
Department of Mathematical Physics, National University of Ireland,  
Maynooth,  Ireland
}

\par\vskip\baselineskip

\beginsection{Introduction and early happenings} 
In this essay we wish to embark on the telling of a story which, almost certainly, stands only at its beginning.
We shall discuss the links between one very old subject, {\it physics}, and a much newer one, {\it topology}. Physics, being so much older,  has a considerably longer history than does topology. After all the bulk of 
topology didn't even exist before the beginning of the twentieth century. 
However, despite this disparity of antiquity between the two subjects, 
we shall still find it worthwhile to examine the situation before the 
twentieth century. We should also remind the reader that the term topology is not usually found in the literature before 1920 or so (cf. footnote 5 of this article), one finds instead the Latin terms
{\it geometria situs} and {\it analysis situs}.
\smalltitle{From graph theory to network theory}

Perhaps the  earliest occurrence of some topological noteworthiness was Euler's 
solution \ref{1} of the {\it bridges of K\"onigsberg problem} in 1736. This was 
nothing if not a physical problem though one with a rather amusing aspect. 
The problem was this: in the city of K\"onigsberg\foot{Since 1945,  when the 
Potsdam agreement passed the city to Russia,  K\"onigsberg has been called 
Kaliningrad. Kaliningrad  is a Baltic sea port on a separate piece of Russian 
territory that lies between Lithuania and Poland.} were seven bridges and the local 
populace were reputed to ask if one could start walking at any one of the 
seven so as to cross all of them precisely once and end at one's starting 
point. Euler had the idea of associating a graph with the problem (cf. fig. 1). 
He then observed that a positive  answer to this question requires the vertices in 
this graph to have an even  number of edges and so the answer  is 
{\it no}\foot{The point is that if one successfully traverses the 
graph in the manner required then the edges at each vertex divide 
naturally into pairs because one can label one as an entry edge and 
the other as an exit edge.}. 
\par\vskip1.5\baselineskip 
\epsfxsize=0.8\hsize
\centerline{\epsffile{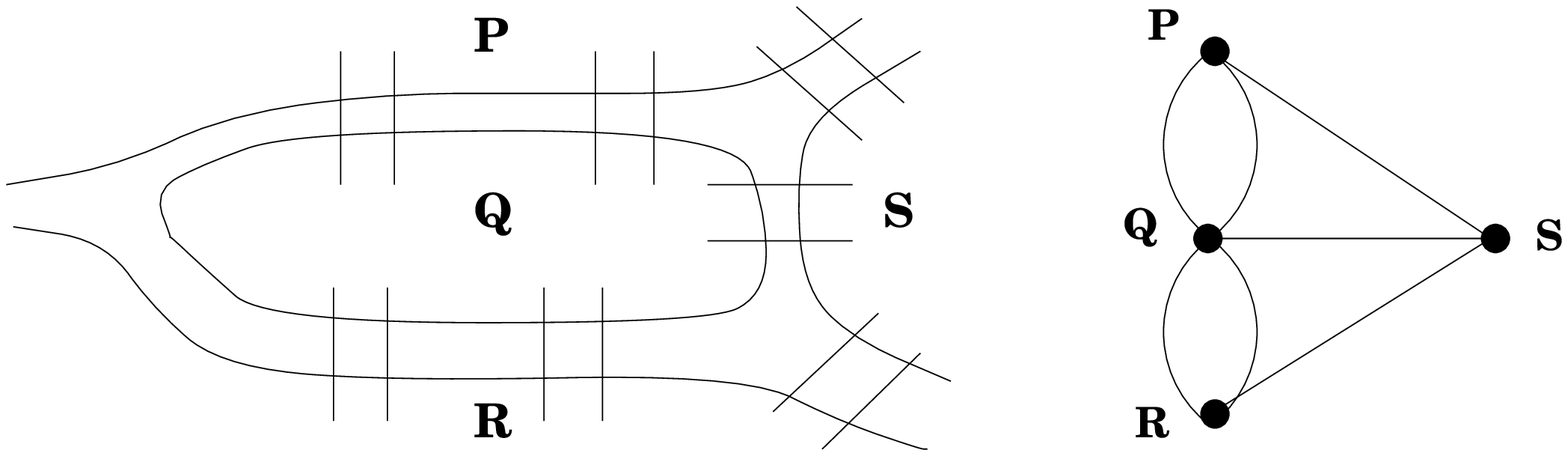}}
\par\penalty10000\vskip1.5\baselineskip
\centerline{\it Fig. 1. \bf The seven bridges of K\"onigsberg and its associated graph with seven edges}
\par\vskip1.5\baselineskip
\par
In  devising his answer Euler gave birth to what we now call graph theory 
and, in so doing, was considering one of the first problems 
in combinatorial topology;  and though 
graph theory could once be viewed as a combinatorial study of the 
topology of one dimensional complexes it is now an independent 
subject in its own right. 
\par
Among the parts of physics with close 
connections to graph theory is network theory. The earliest connection occurs in the work of Kirchhoff  \ref{1} in 1847 who,  as well as formulating his two famous  laws for  electric circuits, made use of 
a graph theoretic argument to solve the resulting equations for  a  general electrical network. 
\par
From these beginnings 
the links between  
graph theory and  physics have strengthened over the centuries.
\smalltitle{Electromagnetic theory and knots} 
In the nineteenth century we encounter a more substantial example 
of a physical phenomenon with a topological content. The 
physics concerned electromagnetic theory while the topology 
concerned the linking number of two curves. 
\par
In  electromagnetic theory the magnetic field ${\bf B}$ 
produced by a current $I$ passing through a wire obeys 
the pair of equations
$$\nabla\times{\bf B}=\mu_0\,{\bf J},\qquad
            \nabla. {\bf B}=0\no$$ 
where ${\bf J}$ is the current density. 
Now let us briefly summarise what is entailed in deriving the famous Amp\`ere law.  We integrate ${\bf B}$ round a closed curve $C$ which bounds a surface $S$, then we have
$$\eqalign{\int_C\bf B \cdot dl&=
\int_S\bf \nabla\times  B\cdot ds\cr
&=\mu_0\int_S \bf J\cdot ds\cr
&=\mu_0 I\cr}\no$$
where in the last step we use the fact that the integral 
$\int_S\bf J\cdot ds$ gives the total current passing through $S$ and this will be precisely $I$ if (as we assume) the wire cuts the surface $S$ just once. However the wire could cut the surface more than 
once---say $n$ times. When this is the case at each cut the the 
integral receives a contribution of $\mp\mu_0 I$ depending on whether 
${\bf J}$ is parallel or anti-parallel to ${\bf ds}$ at the cut. 
Thus, if $m\mu_0I$ denotes the algebraic sum of all these $n$ contributions then  the most general statement for Amp\`ere's law in this case  is
\eqlabel{\ampereone}
$$\int_C{\bf B\cdot dl}=\mu_0 m I,\quad m\in{\bf Z}\no$$
This integer $m$ is a familiar feature of textbook calculations of 
the magnetic field due to a solenoid.
\par
Now let us take an alternative route to calculating 
$\int_C \bf B\cdot dl$.
If we introduce the vector potential ${\bf A}$ where 
$$\bf B=\nabla\times \bf A\no$$
and impose the Coulomb gauge condition
$$\bf \nabla. A=0\no$$
then, to find ${\bf B}$,  we only have to solve, the equation
$${\bf\nabla^2 A}=-\mu_0{\bf J}\no$$
and this has the solution
$${\bf A(r)}={\mu_0\over 4\pi}\int_{{\bf R}^3} \bf d^3 r^\prime
{ J(r^\prime)\over \vert \bf r- r^\prime\vert}\no$$
But if the wire forms a closed curve $C^\prime$, say, with infinitesimal element of length $\bf dl^\prime$ then $\bf J$ has support only on the wire and so is related to the current $I$ by
$${\bf J(r^\prime)d^3 r^\prime}=I\,\bf dl^\prime\no$$
This allows us to express ${\bf B}$ as a line integral around $C^\prime$ giving us
$$\eqalign{\bf B( r)&={I\mu_0\over 4\pi}\,\bf\nabla\times\int_{C^\prime}
{ dl^\prime\over \vert r- r^\prime\vert} \cr
&=-{I\mu_0\over 4\pi}\int_{C^\prime}
  {{\bf (r-r^\prime)\times dl^\prime}\over
\vert {\bf r-r^\prime}\vert^3}\cr}\no$$ 
Now we introduce a second curve $C$ and integrate ${\bf B}$ round $C$ thereby obtaining
\eqlabel{\amperetwo}
$$\int_C{\bf B\cdot dl}
=-{I\mu_0\over 4\pi}\int_C \int_{C^\prime}
{({\bf r- r^\prime)\times dl^\prime\cdot dl}\over
\vert {\bf r- r^\prime}\vert^3}\no$$ 
Perusal of \docref{ampereone} and \docref{amperetwo} together shows us  immediately that

$$-{1\over 4\pi}\int_C \int_{C^\prime}
{({\bf r- r^\prime)\times dl^\prime\cdot dl}\over
\vert {\bf r-r^\prime}\vert^3}=m\no$$
For comparison with the work of Gauss below we wish to have this formula 
in a completely  explicit form with all its coordinate dependence 
manifest and so we write 
$${\bf r}=(x,y,z),\quad \hbox{and}\quad  {\bf r}^\prime=(x^\prime,y^\prime, z^\prime)\no$$
giving
\eqlabel{\linkno}
$$-{1\over 4\pi}\int\int{(x^\prime-x)(dy dz^\prime-dz dy^\prime)
+(y^\prime-y)(dz dx^\prime-dx dz^\prime)+
(z-z^\prime)(dx dy^\prime -dy dx^\prime)
\over 
\left[(x^\prime-x)^2+(y^\prime-y)^2+(z^\prime-z)^2\right]^{3\over2}} 
=m\no$$

That this is a topological statement is clear and the integer $m$
 is 
actually the {\it linking number} of the two curves $C$ and $C^\prime$ 
and so electromagnetic theory has provided with an explicit formula for 
the linking number and so an early result in knot theory.
\par
This result \docref{linkno} was known to Gauss in {\it exactly} the form that we have presented it above.
Gauss's work on this matter also came from work (in 1833) on electromagnetic theory
and is to be found in his {\it Nachlass} (Estate)  cf. Gauss \ref{1} where on p. 605 one finds the formula
(which I quote entirely unchanged from the printed version available in Gauss \ref{1} although one should remember that the original is handwritten rather than printed)
\par\vskip2\baselineskip
\setbox\quote=\vbox{
\ninepoint
\hsize=0.9\hsize 
\hfill ZUR ELECTRODYNAMIK. \hfill 605
\par\vskip\baselineskip
\centerline{[4.]}
{\obeylines
      \hfill Von der {\it Geometria Situs}, die L{\sevenrm EIBNITZ} ahnte und in die nur einem Paar
Geometern (E{\sevenrm ULER} und V{\sevenrm ANDERMONDE}) einen schwachen Blick zu thun verg\"onnt
war, wissen und haben wir nach anderthalbhundert Jahren noch nicht viel mehr 
wie nichts.
  \hfill     Eine Hauptaufgabe aus dem {\it Grenzgebeit} der {\it Geometria Situs} und der {\it Geo-
metria Magnitudinis} wird die sein, die Umschlingungen zweier geschlossener oder 
 unendlicher Linien zu z\"ahlen.
 \hfill Es seien die Coordinaten eines unbestimmten Punkts der ersten Linie $x$, $y$, $z$;  
der zweiten $x^\prime$, $y^\prime$, $z^\prime$ und
}
{\eightpoint 
$$\int\int{(x^\prime-x)(dy dz^\prime-dz dy^\prime)
+(y^\prime-y)(dz dx^\prime-dx dz^\prime+
(z-z^\prime)(dx dy^\prime -dy dx^\prime)
\over 
\left[(x^\prime-x)^2+(y^\prime-y)^2+(z^\prime-z)^2\right]^{3\over2}} 
=V$$
}
{\obeylines
dann ist dies Integral durch beide Linien ausgedehnt
$$=4 m \pi$$ 
und $m$ die Anzahl der Umschlingungen.
\hfill Der Werth ist gegenseitig, d. i. er bleibt derselbe, wenn beide Linien ge-
gen einander umgetauscht werden. 1833. Jan. 22.
}}
\box\quote
\noindent
This translates\foot{I am greatly indebted to Martin Mathieu for providing me with this translation.} to 
\par\vskip\baselineskip 
{\ninepoint
\hsize=0.9\hsize 
\hfill ON ELECTRODYNAMICS. \hfill 605
\par\vskip\baselineskip
\centerline{[4.]}
Concerning the {\it geometria situs}, foreseen by L{\sevenrm EIBNITZ}, 
and of which only a couple of geometers (E{\sevenrm ULER} 
and V{\sevenrm ANDERMONDE}) were allowed to catch a glimpse, we know 
and have obtained after a hundred and fifty years little more than 
nothing.
\par
A main task (that lies) on the border between {\it geometria situs} and 
{\it geometria magnitudinis} is to count the windings of two closed
or infinite lines.
\par
The coordinates of an arbitrary point on the first line shall be 
 $x$, $y$, $z$; (and) on the second $x^\prime$, $y^\prime$, $z^\prime$,  
and (let)
{\eightpoint 
$$\int\int{(x^\prime-x)(dy dz^\prime-dz dy^\prime)
+(y^\prime-y)(dz dx^\prime-dx dz^\prime+
(z-z^\prime)(dx dy^\prime -dy dx^\prime)
\over 
\left[(x^\prime-x)^2+(y^\prime-y)^2+(z^\prime-z)^2\right]^{3\over2}} 
=V$$
}
then this integral carried out over both lines equals
$$=4 m \pi$$
and $m$ is the number of windings.
\par
This value is shared, i.e., it remains the same if the lines are 
interchanged.
\par\hfill 1833. Jan. 22.
}
\par\vskip\baselineskip
We see on comparing Gauss's formula with \docref{linkno} that they are exactly the same modulo the fact that his integer $m$ is minus our $m$.
Gauss's remarks\foot{Additional references on the history of knot theory are Epple \ref{1--2} of which Epple \ref{2}  also discusses the work 
above of Gauss on knots and, in this connection, prints a fragment from one of Gauss's notebooks which show that Gauss spent some time  thinking about what the 
current literature now calls braids. I am indebted to Ioan James for sending me a preprint of Epple \ref{2}.} above also show that he understood the topological nature of his result which he quotes without reference to any electromagnetic quantities; in addition he bemoans the paucity of progress in topology in a 
manner which makes clear that he thinks that there is much to be discovered eventually.
\par
Maxwell was also aware of Gauss's result and mentions it in Maxwell \ref{1} when he discusses the conditions for the single valuedness of a function
defined by a line integral. It is clear that he realises the need for 
a topological restriction on  the domain of definition of the function.
On p. 17 of Maxwell \ref{1} he says
{\beginquote 
There are cases, however in which the conditions
$${dZ\over dy}-{dY\over dz}=0,\quad {dX\over dz}-{dZ\over dx}=0,
\quad{dY\over dx}-{dX\over dy}=0,\quad$$ 
which are those of $Xdz+Ydy+Zdz$ being a complete differential, are satisfied throughout a certain region of space, and yet the line-integral 
from $A$ to $P$ may be different for two lines, each of which lies wholly within that region. This may be the case if the region is in the form of 
the ring, and if the two lines from $A$ to $P$ pass through opposite 
segments of the ring .......... We are here led to considerations 
belonging to the Geometry of Position, a subject which, though its 
importance was pointed out by Leibnitz and illustrated by Gauss, has 
been little studied. The most complete treatment of this subject has 
been given by J. B. Listing.\foot{The work of Listing  referred to by 
Maxwell is  Listing 
\ref{1}. We 
note that Listing was the first to use the term  {\it Topology} (actually  ``Topologie'', since he wrote in German) in a 
letter  to a friend in 1836, cf. the detailed account of this on 
pp. 41--42 of Pont \ref{1}, cf. also Listing \ref{2}.}
\endquote}
Then on p. 43 of Maxwell \ref{2} Maxwell refers to Gauss's linking number 
formula and says
{\beginquote
It was the discovery by Gauss of this very integral, 
expressing the work done on a magnetic pole while describing a 
closed curve in presence of a closed electric current, and indicating 
the geometrical connexion between the two closed curves, that led him to
 lament the small progress made in the Geometry of Position since the 
time of Leibnitz, Euler and Vandermonde. We have now some progress to 
report, chiefly due to Riemann, Helmholtz, and Listing.
\endquote}
Maxwell also includes a figure for which  the linking number of  two oppositely oriented curves is zero.
\smalltitle{Knots, vortices and atomic theory}
The nineteenth century was to see another discussion of knots and 
physics before it came to an end. This was in the theory of vortex atoms 
proposed by Lord Kelvin (alias W. H. Thomson) in 1867;  cf. Thomson \ref{1} for his paper on atoms as vortices and Thomson \ref{2--4} for his work on vortices themselves including 
references to  knotted and linked vortices.
\par
Kelvin was influenced by an earlier fundamental paper by Helmholtz (1858) \ref{1} on vortices,  and a long and seminal paper of Riemann  (1857) \ref{1} 
on Abelian functions\foot{Actually, in Thomson \ref{2}, Kelvin specifically refers  to section 2 of this paper which is topological:  it discusses multiple 
connectedness for what we nowadays refer to as Riemann surfaces. This section
bears the title: {\it Lehrs\"atze aus der analysis situs f\"ur die Theorie der Integrale von zweigliedrigen vollst\"andigen Differentialien}
or {\it Theorems from analysis situs for the theory of integrals
over total differentials of functions of two variables. }}. 

\par
His idea really was that an atom was a kind of vortex.
He was sceptical about the chemists espousal of the Lucretius atom, (in 
Thomson \ref{1}) he says:
\par
{\beginquote 
Lucretius's atom 
does not explain any of the properties of matter without attributing 
them to the atom itself ..........  
The possibility of founding a theory of elastic solids and liquids on the dynamics of closely-packed vortex atoms may be reasonably anticipated.
\endquote}
and later in the same article
{\beginquote
A full mathematical investigation of the mutual action between
two vortex rings of any given magnitudes and velocities passing one
another in any two lines, so directed that they never come nearer
one another than a large multiple of the diameter of either, is a 
perfect mathematical problem; and the novelty of the circumstances 
contemplated
presents difficulties of an exciting character. Its solution will 
become the foundation of the proposed new kinetic theory of gases.
\endquote} 
\par
A significant part of the {\it ``difficulties of an exciting character''}
 referred  to by Kelvin above concerned the topological nature of 
vortices\foot{It seems clear that Kelvin thought of these vortex tubes as knotted tubes of the ether for the opening sentences of Thomson of \ref{2}, though they do not mention the word ether, describe an ideal fluid of that kind. One reads: {\it The mathematical work of the present paper
has been performed to illustrate the hypothesis, that space is continuously
occupied by an incompressible frictionless liquid acted on by no force, and that material phenomena of every kind depend solely on motions created 
in this liquid.} His belief in the ether lasted much further into the future cf.  Thomson \ref{5} which is dated 1900; this of course is to be expected 
as special relativity still lay five years ahead.}  
i.e. that they can be knotted and that several may be linked together  
(cf. fig. 2).
\par\vskip1.5\baselineskip 
\epsfxsize=0.4\hsize
\centerline{\epsffile{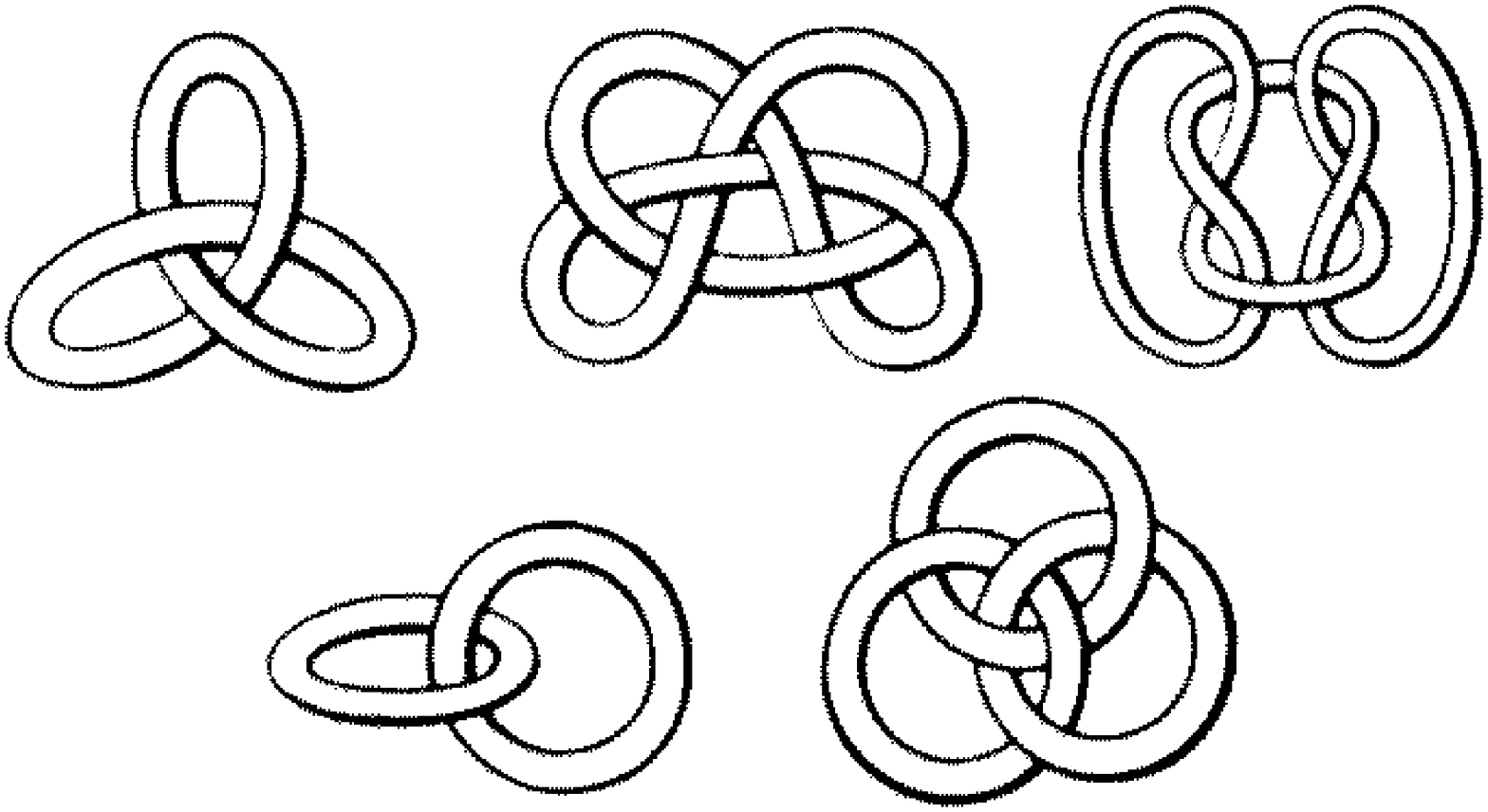}}
\par\penalty10000\vskip1.5\baselineskip
\centerline{\it Fig. 2. \bf Some of the vortex tubes considered 
by Kelvin in Thomson \ref{2}.}
\par\vskip1.5\baselineskip  
Atiyah \ref{1} has summarised very well the main points that Kelvin considered to be in favour of his theory.
Hence we provide a brief paraphrase of that summary here. The stability of atoms would be be accounted  by the stability under  deformation of the topological type of knots. The large variety of different knot types can 
accommodate all the different elements. Vibrational oscillations of knots could be the mechanism for atomic spectral lines.
\par
With regard to this latter point about spectral lines  Kelvin even 
gives (Thomson \ref{1}),  a rough upper bound for the rotation period of a sodium vortex based on the frequency of its celebrated yellow emission line. 
\par
Kelvin's contemporary Tait was thereby stimulated to do  extensive work on 
knot theory and to begin work on classifying knots cf. Tait \ref{1} where numerous complex knots are copiously illustrated as well as discussed. However, 
despite a lot of work,  many of his results  remained unproven  and were 
christened the {\it Tait conjectures}. It is clear now that these conjectures 
were out of range of the mathematical techniques of his day;  but 
many of them 
were finally disposed of in the 1980's by the work of Jones 
\ref{1} which we shall come to in section 6. 
This work caused a resurgence in 
knot theory in mathematics and also coincided with a renewal of the physicists
interest in knots. This is a rather special and interesting story 
cf. section 6. 
\smalltitle{The advent of Poincar\'e}
As the nineteenth century drew to a close topological matters were considerably enlivened by the work of Poincar\'e.  This work had some strong connections with physics as we shall now explain.
\par
Poincar\'e's  interest in physics endured throughout his all too short life but began
with Newtonian dynamics and, in particular, with the three body problem. 
\par
The Swedish mathematician Mittag-Leffler,  founder and chief editor of the journal {\it Acta Mathematica}, was the prime mover in the organisation of an international mathematical 
competition held to celebrate the $60^{th}$ birthday  of  King Oscar II of Sweden and Norway. The King was well disposed towards science and allowed
a competition to be announced in 1885. Four problems were proposed but entrants (who were, in theory, anonymous) were also at liberty to choose their own topic\foot{In brief the four problems (which were contributed by Hermite and Weierstrass)  were the $
n$-body problem, an analysis of Fuchs theory of 
differential equations, a problem in non-linear differential equations and a final (algebraic) problem on Fuchsian functions.}. 
The judges were 
Hermite, Mittag-Leffler and Weierstrass. Poincar\'e was declared the winner  in 1889---the King having officially approved the result on January $20^{th}$  the day before his $60^{th}$ birthday---there were eleven other entrants.
\par
Poincar\'e had chosen to work on the first problem and in particular the three body problem\foot{Poincar\'e actually studied the restricted three 
body problem. This meant that he took the first mass to be large, the 
second small and positive and the third negligible. He also took the 
first two masses to have a circular orbit about their common centre of 
mass while the third moved in the plane of the circles.}. 
His prize winning memoir\foot{The prize winning entry was to be 
published in {\it Acta Mathematica} and this was eventually done in Poincar\'e \ref{1}.
 However  the published version of Poincar\'e's work differs in some important respects from the entry that he submitted. This is because of an 
error discovered before publication by Phragm\'en. The mathematical and 
historical details of this particular  story are available in the 
excellent book  Barrow-Green \ref{1}.} 
bore the title {\it Sur le probl\`eme de trois corps et les \'equations de la dynamique}.  
\par
Before working on this problem Poincar\'e had worked on the theory of differential equations and had concentrated on obtaining {\it qualitative} results.
A key viewpoint he adopted and exploited was {\it geometric}---he thought of the solutions as defining geometric objects: e.g. {\it curves}. He 
then quickly obtained results of a topological nature. 
\par
For example he studied the singular points of these equations on surfaces of genus $p$ and introduced the notions of saddle points, nodes and foci to classify these singularities---in French he used the terms 
{\it cols}, {\it noeuds} and {\it foyers} respectively---then using $C$, 
$F$ or $N$ to denote the type of the singularity he proved (cf. Poincar\'e \ref{5} and \ref{2-6} (1880--1886) that
$$N+F-C=2-2p\no$$
which one recognises immediately as the index of a vector field being equated to the Euler-Poincar\'e characteristic of the Riemann surface.
\par
As evidence of the stimulus that physics gave to  investigations we quote from an analysis  Poincar\'e prepared in 1901 of his own scientific work. This was published after his death (Poincar\'e \ref{7}).  
{\beginquote
Pour \'etendre les r\'esultats pr\'eced\'ents aux 
\'equations d'ordre sup\'erieur au second, il faut renoncer \`a la repr\'esentation g\'eometrique qui nous a \'et\'e si commode, \`a moins 
d'employer le langage de l'hyperg\'eometrie \`a $n$ dimensions .......... Ce qu'il y a
de remarquable, c'est que le troisi\`eme et le quatri\`eme cas, 
c'est \`a dire ceux qui corresponde \`a la stabilit\'e, se 
rencontrent pr\'ecis\'ement dans les \'equations g\'en\'erales de la 
Dynamique .......... 
Pour aller plus loin, il me fallait cr\'eer un instrument 
destin\'e \`a remplacer l'instrument g\'eometrique qui me faisait 
d\'efaut quand je voulais p\'en\'etrer dans l'espace \`a plus de trois 
dimensions. C'est la principale raison qui m'a engag\'e \`a
aborder l'\'etude de l'Analysis situs; mes travaux \`a ce sujet seront expos\'es plus loin dans une paragraphe sp\'ecial. 
\endquote}
that is 
{\beginquote
To extend the preceding results to equations of higher than
second order, it is necessary to give up the geometric representation
which has been so useful to us unless one employs the language of 
hypergeometry of $n$ dimensions .......... What is remarkable is that in the third and fourth case, that is to say those that correspond to 
stability, are found precisely in the general equations of 
dynamics .......... To go further it was necessary to create a tool 
designed to replace the geometric tool which let me down when I 
wanted to penetrate spaces of more than three dimensions. This is the 
principal reason which led me  
 to  take up the study of Analysis situs; my work on this subject will be 
expounded further down in a special paragraph.
\endquote}
\par
As yet another insight 
into the way Poincar\'e was thinking about celestial mechanics we quote the following which is taken from the introduction 
to his paper {\it Analysis situs} of 1895 (Poincar\'e \ref{8}); this 
is the first of his epoch making papers on topology.
{\beginquote
D'autre part, dans une s\'erie de M\'emoires ins\'er\'es dans le {\ssfont Journal 
de Liouville}, et intitul\'es: {\ssfont Sur les courbes d\'efinies par les 
\'equations diff\'erentielles}, j'ai employ\'e l'Analysis situs ordinaire 
\`a trois dimensions \`a l'\'etude des \'equations diff\'erentielles. Les m\^emes recherches ont \'et\'e poursuivies par M. Walther Dyck. On 
voit ais\'ement que l'Analysis situs g\'eneralis\'ee permettrait de 
traiter de m\^eme les \'equations d'ordre sup\'erieur et, en 
particulier, celles de la M\'ecanique c\'eleste.
\endquote}
that is
{\beginquote
On the other hand, in a series of memoirs in  {\ssfont Liouville's Journal}\foot{Poincar\'e uses  the term Liouville's Journal to refer
 to the {\it Journal de Math\'ematiques} as found, for example, in 
Poincar\'e \ref{3}.}    under the title {\ssfont On curves defined by differential 
equations}, I have used ordinary analysis situs in three dimensions in 
the study of differential equations. The same research has been followed 
by Mr. Walther Dyck. One easily sees that generalised analysis situs 
would permit the treatment of equations of higher order and, in 
particular, those of celestial mechanics.
\endquote}
\par
Poincar\'e's work opened a new chapter in celestial mechanics;  the 
 strong topological content of his  papers on differential equations 
lead directly to his papers Poincar\'e \ref{8--18} on {\it analysis situs} which gave birth to the  subjects of algebraic and differential
topology. 
\smalltitle{Poincar\'e's geometric theorem}
Poincar\'e left unproved at the time of his death a (now) famous result 
usually referred to as {\it Poincar\'e's geometric theorem}. This theorem has both 
physical and topological content.  Shortly before he died,  
Poincar\'e wrote \ref{18} in order to describe the theorem and the 
reasons for his believing it to be true. 
\par
In the first paragraph  (of Poincar\'e  \ref{18}) he says
{\beginquote
Je n'ai jamais pr\'esent\'e au public un travail aussi inachev\'e;
 je crois donc n\'ecessaire d'expliquer en quelques mots les raisons qui m'ont
d\'etermin\'e \`a le publier, et d'abord celles qui m'avaient engag\'e \`a l'entreprendre. 
J'ai d\'emontr\'e il y a longtemps d\'ej\`a, l'existence des solutions p\'eriodiques du
probl\`eme des trois corps; le r\'esultat laissait cependant encore \`a d\'esirer; car
 si l'existence de chaque sorte de solution \'etait \'etablie pour les petites valeurs 
des masses, on ne voyait pas ce qui devait arriver pour des valuers plus grandes, quelles 
\'etaient celles de ces solutions qui subsistaient et dans quel ordre elles 
disparaissaient. En r\'efl\'echissant \`a cette question, je me suis assur\'e que la r\'eponse 
devait d\'ependre de l'exactitude ou la fausset\'e d'un certain 
th\'eor\`eme de g\'eometrie 
dont l'\'enonc\'e est tr\`es simple, du moins dans le cas du probl\`eme restreint et des 
probl\`emes de Dynamique o\`u il n'y a que deux degr\'es de libert\'e.
\endquote}
that is
{\beginquote 
I  have never presented to the public such an incomplete work; 
I believe it necessary therefore to explain in a few words the reasons which 
have decided me to publish it, and  first of all those (reasons) which had led me to undertake it. I  
proved, a long time ago now, the existence of periodic solutions to the three body problem; 
however the result left something to be desired; for if the existence of each sort of 
solution were established for small values of the masses, one couldn't see what 
might happen for larger values, (and) what
were those (values) of those solutions which persisted and in what order they disappeared. On reflecting on 
this question, I have ascertained  that the answer ought to depend on the truth or 
falsity of a certain geometrical theorem which is very simple to state, at least in the 
case of the restricted (three body) problem and in dynamical problems with only two 
degrees of freedom.
\endquote}
The geometric theorem is a {\it fixed point theorem}: it  states that a continuous, one to one, area preserving map $f$ from an annulus 
$0 < a \le  r \le b$ to itself has a pair of fixed points ($f$ is also required 
to have the property that it maps the two boundary circles in opposite senses). When applied to the restricted three body problem it proves the existence of infinitely many 
periodic solutions.
\par
Topology immediately enters because Poincar\'e's index theorem can easily be used to 
show---as Poincar\'e himself pointed out (Poincar\'e \ref{18})---that there must 
be an even number of fixed points; 
hence it is sufficient to prove that $f$ has at least one fixed point.
\par
In 1913, Birkhoff, who was to be a prime mover and founder  of the new 
subject of {\it dynamical systems}, proved the theorem cf. 
Birkhoff \ref{1}. 
\par
After Poincar\'e the  pursuit of problems with a joint 
dynamical and topological content was taken up by Birkhoff, 
Morse, Kolmogorov, Arnold and Moser and many, many others. 
  We shall have something to say about these 
matters later on in section 8.
\beginsection{A quiescent period}
\smalltitle{Dirac and Schwarzschild}
In the early twentieth century physics was undergoing the twin revolutions of quantum theory and special and general relativity. These revolutions imported much new mathematics into physics but topology, though itself 
growing at an explosive rate, did not figure prominently in the 
physics of this story. 
\par
Nevertheless two papers on physics of this period, do have a 
topological content, and  are worth noting.       In the first case this 
content is implicit and in the other it is explicit. These papers are, 
respectively,  that of Schwarzschild in 1916 (Schwarzschild \ref{1}) on 
solutions to Einstein's equations and that of Dirac in 1931 
(Dirac \ref{1}) on magnetic monopoles.
\par
We shall discuss Schwarzschild's paper in section 3 
but for the moment we want to deal with Dirac's paper because its 
topological content is  explicit from the outset and because it has 
proved to be so influential.
\smalltitle{Dirac's magnetic monopoles}
Dirac  begins his paper with some comments on  the r\^ole of 
mathematics in physics which are both philosophical and somewhat prophetic.
He says
{\beginquote
The steady progress of physics requires for its 
theoretical formulation a mathematics that gets continually more 
advanced. This is only natural and to be expected. What however was 
not expected by the scientific workers of the last century was the 
particular form that the line of advancement of the mathematics would 
take, namely, it was expected that the mathematics would get more 
and more complicated, but would rest on a permanent basis of axioms 
and definitions, while actually the modern physical developments have 
required a mathematics that continually shifts its foundations and gets 
more abstract. Non-euclidean geometry and non-commutative algebra, which
were at one time considered to be purely fictions of the mind and pastimes of logical thinkers, have now been found to be very necessary for 
the description of general facts of the physical world.  It seems likely that this process of increasing abstraction will continue in the future and that advance in physics is to be associated with a continual modification
and generalisation of the axioms at the base of the mathematics
rather than with a logical development of any one mathematical scheme
on a fixed foundation.
\endquote} 
The subject matter of Dirac's paper was both topological and electromagnetic: he found that there were magnetic monopole solutions to Maxwell's equations but that the magnetic charge $\mu_0$ of the monopole had to quantised; in addition there were two striking facts about this quantisation one 
mathematical and one physical. The mathematical novelty was that the 
quantisation was not due to the discreteness of the spectrum of an operator in Hilbert space but rather to topological considerations. The 
physical novelty was that the existence of even one of these monopoles 
would imply the quantisation of electric charge, something not hitherto  
achieved\foot{Electric charge, not being the eigenvalue of any basic operator,  is not quantised by the mechanism that quantises energy and angular momentum.}.    
\par
Dirac considered carefully the phase of a  wave function $\psi(x,y,z,t) $
of a particle in quantum mechanics. If $A$ and $\gamma$ are the amplitude and phase respectively then we have
$$\psi=A e^{i\gamma}\no$$
Once $\psi $ is normalised to unity, in the usual way, there remains a freedom to add a constant to the phase $\gamma$ without altering the physics of the particle. Dirac wanted to exploit this fact and argue that the 
absolute value of $\gamma$ has no physical meaning and only phase {\it differences} matter physically. In Dirac \ref{1} he wrote
{\beginquote
Thus the value of $\gamma$ at a particular point has no 
physical meaning  and only the difference between the values of 
$\gamma$ at two different points is of any importance.
\endquote}
He immediately introduces a generalisation, saying
{\beginquote
This immediately suggests a generalisation of the formalism. We may assume that $\gamma$ has no definite value at a particular point, 
but only a definite difference in value for any two points. We may go 
further and assume that this difference is not definite unless the two 
points are neighbouring. For two distant points there will  then be a definite phase difference only relative to some curve joining them and different curves will in general give different phase differences. The total
 change in phase when one goes round a closed curve need not vanish.
\endquote}
Dirac now does two more things: he finds that this change in phase round a closed curve will 
give rise to an ambiguity unless it takes the same value for all wave 
functions, and he goes on to equate this phase change to the flux of an electromagnetic field.  For a wave function in three {\it spatial} 
dimensions, to which Dirac specialises,  this  flux is just that of  a {\it magnetic field}. 
Yet closer scrutiny of the situation forces the consideration of the
possibility of $\psi$ vanishing (which it will do generically along a line in three dimensions) about which Dirac says 
{\beginquote
There is an exceptional case, however, occurring when the wave 
function vanishes, since then its phase does not have a meaning. As the 
wave function is complex, its vanishing will require two conditions, so 
that in general the points at which it vanishes will lie along a line. 
We call such a line a {\ssfont nodal line}.\foot{In the current literature
nodal lines are called Dirac strings} 
\endquote}   
Dirac now finds that to get something new he must relax the requirement 
that phase change round a closed curve be the same for all wave 
functions; he realises that it is possible to have it differ by integral 
multiples of $2\pi$ for different wave functions.
\par
The final stage of the argument is to compute the flux of the magnetic 
field ${\bf B}$ through a closed surface $S$ allowing also that 
nodal lines may lie totally within $S$ or may intersect with $S$.
\par
 Each nodal line is labelled by an integer $n$ which one 
detects by integrating round  a small curve enclosing the line. If $n_i$ are the integers for the nodal lines inside, or intersecting with, $S$ the $n_i$ are  
then related to the magnetic flux $\int_S{\bf B\cdot ds} $ by
\eqlabel{\phasechange}
$$\sum_i 2\pi  n_i+{2\pi e\over hc}\int_S{\bf B\cdot ds}=0\no$$ 
where $h$ , $c$ and $e$ are Planck's constant, the velocity of light 
and the charge on the electron respectively.\foot{Informally the reason 
that the RHS of \docref{phasechange} is zero is because it is the 
limit of the change in phase round a closed curve $C$  as $C$ shrinks to zero. $C$  bounds a surface $S$ so that when $C$ has finally shrunk to zero $S$ becomes closed.} 
Now if the nodal lines are closed they always cross $S$ an even number 
of times  and hence contribute zero to $\sum_i 2\pi n_i$ when one takes account of the sign of contributions associated with incoming and outgoing lines. Hence 
$\sum_i 2\pi n_i$ is only non-zero for those lines having end points within 
$S$. Finally Dirac then surrounds just {\it one} of these end points with a 
small surface $S$ so that \docref{phasechange} immediately implies 
that  magnetic flux emanates from this end point so that it is the location of a {\it magnetic monopole}. Such an end point will also be a singularity of the electromagnetic field.
\par
Now if a single  electric charge produces a field ${\bf E}$ then the size of its charge $q$ is given by
$$q={1\over 4\pi} \int_S{\bf E\cdot ds}\no$$ 
where $S$ is a small surface enclosing the charge. So, by analogy, the magnetic charge $\mu_0$ of a magnetic monopole is defined by writing
$$\mu_0={1\over 4\pi} \int_S{\bf B\cdot ds}\no$$
Applying this to our small surface $S$ and using \docref{phasechange} this 
gives at once the result that\foot{When reading Dirac \ref{1} one should be aware that the symbol $h$ denotes Planck's constant divided by $2\pi$; nowadays this quantity is usually denoted by $\hbar$.}
\eqlabel{\diraclabel}
$$\eqalign{2\pi n+{8\pi^2 e\over  h c}\mu_0&=0\cr
            \Rightarrow e\mu_0&=-{n h c\over 4\pi},\quad n\in{\bf Z}\cr}\no$$
\par
We see that the magnetic charge $\mu_0$ has a quantised strength with the fundamental quantum being 
$${h c\over 4\pi e}\no$$
which we note  is inversely proportional to the electric charge $e$. 
Dirac noted the key fact that the mere existence of such a monopole 
means that electric charge is quantised. In Dirac \ref{1} we find the passage
{\beginquote
The theory also requires a quantisation of electric charge, 
since any
charged particle moving in the field of a pole of strength $\mu_0$ 
must have for its charge some integral multiple (positive or negative) 
of $e$, in order that the wave functions describing the motion may exist.
\endquote}
\par
Dirac's remarkable paper  encouraged physicists  to consider   
particles which are simultaneously magnetically and electrically 
charged---such particles are called {\it dyons}. The  
 set of possible electric and magnetic charges form a skew 
lattice  covering an entire ${\bf R}^2$ 
cf. fig.  3. For convenience from now on we denote magnetic charge by 
  $g$ rather than Dirac's $\mu_0$ and in doing so we change 
to more conventional  units of magnetic and electric charge: 
in these units $\hbar=c=1$ and the 
charge $q$ on the electron is given by $q^2/(4\pi)=1/137$. 
Dirac's condition 
\docref{diraclabel} now reads 
$${eg\over 2\pi}=n\in{\bf Z}\no$$
\par\vskip1.5\baselineskip 
\epsfxsize=0.5\hsize
\centerline{\epsffile{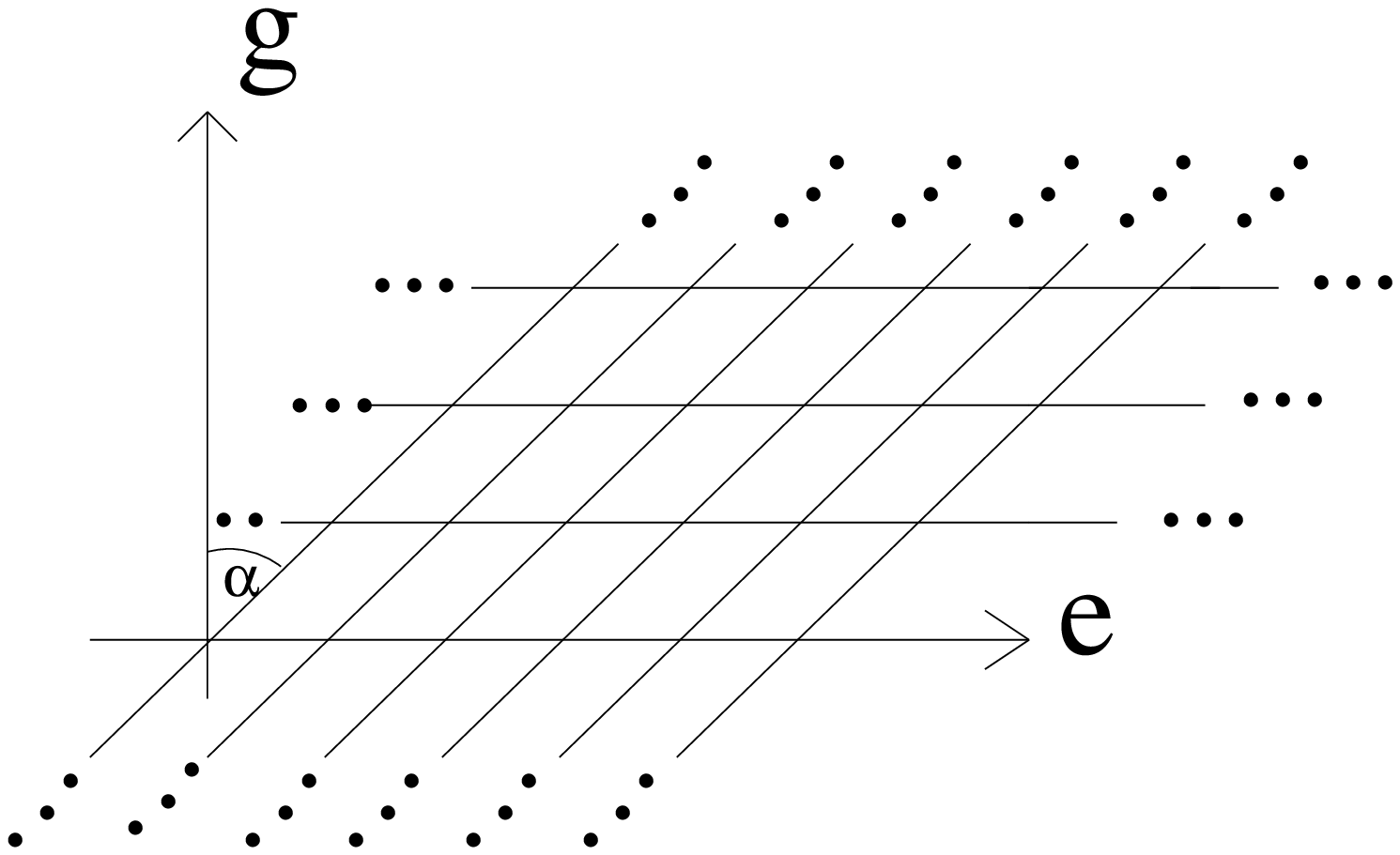}}
\par\penalty10000\vskip1.5\baselineskip
\centerline{\it Fig. 3. \bf The lattice of possible magnetic and electric charges} 
\par\vskip1.5\baselineskip 
The quantisation condition for dyons is a little more 
complicated than that for particles with only one type of charge. 
It has been studied independently  by Schwinger \ref{1} and 
Zwanziger \ref{1--2}. They found that if $(e_1,g_1)$ and $(e_2,g_2)$ 
represent the 
electric and magnetic charges of  a pair of dyons then 
$${(e_1g_2-e_2g_1)\over 2\pi}=n\in{\bf Z}$$
and these values we represent on the lattice of fig. 3. If the angle $\alpha$ of fig. 3 is precisely  zero then the lattice becomes rectangular 
rather than skew: this is practically forced if the theory is $CP$ invariant since 
$CP$ acts on $(e,g)$ to give $(-e,g)$ (actually $\alpha=\pi/4$ is also allowed). Hence the size of $\alpha$ is a 
measure of $CP$ breaking which is experimentally found to be small.  
\par
For monopoles coming from {\it non-Abelian} gauge theories serious consideration must be given to a specific topological mechanism 
for this $CP$ breaking. We shall expand on this remark in section 
4. 
\smalltitle{The topology of Dirac's monopoles}
It is very instructive to examine   Dirac's work from the  point of 
view of topology.
The mathematical setting is that of connections and  curvatures 
on fibre bundles together with the vital calculational tool of characteristic classes; this latter was only at the very beginning of its development when Dirac \ref{1} was published.
\par
The standard physical setting is as follows.
In electromagnetic theory the electric and magnetic fields form the 
components of the Maxwell field tensor $F_{\mu\nu}$ 
according to\foot{In our notation  $x^0$ denotes the time coordinate,  summation is implied for repeated indices and we use the 
convention that Greek and Latin indices run from $0$ to $3$ and 
$1$ to $3$ respectively.}
$$\eqalign{&E_i=F_{i0},\quad B_i={1\over 2}\epsilon_{ijk} F_{jk}\cr
           &F_{\mu\nu}=\partial_\mu A_{\nu}-\partial_\nu A_{\mu}\cr} 
 \no$$
\par
 However geometrically  
speaking the $ F_{\mu\nu}$ are also the components of a curvature tensor 
for the gauge potential $A_\mu$. This suggests immediately that one uses the curvature 2-form $F$ and the connection $1$-form $A$ about  which we 
know that
$$\eqalign{&F={1\over2} F_{\mu\nu} dx^{\mu}\wedge dx^{\nu},
\quad A=A_{\mu}dx^{\mu}\cr
&F=dA\cr}\no$$
Now because the electric field ${\bf E}$ is expressed as a gradient, 
while the magnetic field ${\bf B}$ is expressed as a  curl, then 
it is natural to associate ${\bf E}$ to a $1$-form and  
${\bf B}$  to a $2$-form. Hence we define the forms $E$ and $B$ by writing
 $$E=E_idx^i,\quad B={1\over2}F_{ij} dx^i\wedge dx^j\no$$
This means that the curvature 2-form $F$ is expressible as
\eqlabel{\maxtensor}
$$F=dx^0\wedge E +B\no$$ 
and Maxwell's equations simply assert the closure of $*F$ i.e.
$$d*F=0\no$$
where the  $*$ denotes the usual Hodge dual  with respect to the 
standard flat Minkowski metric on ${\bf R}^4$.
\par
But recall that Dirac specialises to a three dimensional situation 
by taking a time independent electromagnetic field.  We see at once from 
\docref{maxtensor} that, in the absence of time (or at a fixed time), 
$F$ pulls back to a $2$-form on ${\bf R}^3 $  which is just the magnetic field 
$2$-form $B$. Thus,  denoting for convenience this $2$-form on ${\bf R}^3 $ 
by $F_{\vert_{{\bf R}^3}} $, we have
$$F_{\vert_{{\bf R}^3}}=B\no$$
It is important to note that the magnetic field in the guise of $F_{\vert_{{\bf R}^3}}$
 is a curvature on ${\bf R}^3 $ but the same is not true of the electric field. So particularly for a {\it static} electromagnetic field the magnetic and electric fields are geometrically 
{\it very different}.  
\par
Dirac's monopole has to be singular  on ${\bf R}^3$  so its curvature  
$F_{\vert_{{\bf R}^3}}$  is a curvature defined on ${\bf R}^3-\{0\}$  with 
associated connection form $A_i dx^i$.  Now the gauge invariance of the monopole is the standard possibility of replacing $A$  by $A+df$ where $f$ is a function on ${\bf R}^3-\{0\}$, i.e. the gauge group is the Abelian group $U(1)$. 
\par
So, in bundle theoretic terms, we have a  connection on a $U(1)$ bundle $P$, say, over 
${\bf R}^3-\{0\}$; but ${\bf R}^3-\{0\}$ is $S^2\times {{\bf R}^+}$  (${{\bf R}^+}$ is the positive real axis) and so homotopy 
invariance   
means  we might as well consider $P$ to be a $U(1)$
bundle over $S^2$. Such a $P$ has an integral first Chern class $c_1(P)$
given in terms of its curvature form $F$ by the standard formula
$$c_1(P)=\int_{S^2}{F\over 2\pi},\quad c_1(P)\in {\bf Z}\no$$
\par
We can now illustrate everything by doing a concrete calculation: If we 
use spherical polar coordinates\foot{Were we to use Cartesian coordinates $(x,y,z)$ we would obtain 
$$A={C\over 2r}{(x dy-y dx)\over (z+r)}$$
Note that this expression  has a {\it genuine singularity} at $r=0$ and a 
{\it coordinate singularity} at $z+r=0$---i.e. the negative $z$-axis---this latter singularity corresponds to Dirac's nodal line or the Dirac string. This coordinate singularity can be shifted to the positive $z$-axis by the gauge transformation $f=\tan^{
-1}(y/x)$ yielding the connection 
$$A+df={C\over 2r}{(x dy-y dx)\over (z-r)}$$} 
$(r,\theta,\phi)$ and take 
$$A={C\over 2}(1-\cos(\theta))d\phi,\quad C\hbox{ a constant}\no$$
then the connection $A$ has curvature 
$$F={C\over 2}\sin(\theta) \, d\theta\wedge d\phi \equiv {C\over r^3}\epsilon_{ijk}x^i dx^j\wedge dx^k\no$$
Now $F$ of course is the same as the magnetic field $B$ or $F_{\vert_{{\bf R}^3}}$ introduced above but the integrality of the  Chern class of $P$ gives
$$\eqalign{\int_{S^2} {C\over 4\pi}\sin(\theta)
\; d\theta\wedge d\phi&=n\in{\bf Z}\cr
                \Rightarrow C&=n\cr}\no$$
so that the constant $C$ is quantised, and this ensures that $P$ is well defined. This then is Dirac's quantisation condition 
\docref{diraclabel} in  units 
 where magnetic charges take integral values (i.e. units 
of size $h c/ 4\pi e$).
\smalltitle{Aharonov and Bohm}
After 1931, and the appearance of Dirac's paper, neither physics nor topology stood remotely still but they largely went along separate ways. To pinpoint a significant instance of an interaction between topology and 
physics we pass forward nearly three decades to 1959 and the paper of
Aharonov and Bohm \ref{1}.
\par
The Aharonov--Bohm  effect \ref{1} is a  phenomenon  in which the non-triviality of a gauge field $A$ is measurable physically even though its curvature $F$ is zero.
Moreover this  non-triviality is  {\it topological} and can be expressed as a number $n$, say, which is a global topological invariant.
\par
To demonstrate this effect physically one arranges that  a {\it non simply connected} 
region $\Omega$ of space has zero  electromagnetic field $F$. We are using the same  notation  as in the discussion above of Dirac's magnetic monopole, i.e we have $$F=dA\no$$  
and
$$F={1\over2}F_{\mu\nu}dx^\mu\wedge dx^\nu,\;\; \hbox{and }\;A=A_\mu dx^\mu\no$$
where the $x_\mu$ are local, coordinates on $\Omega$.
\par
Given this $F$ and $\Omega$  
one can devise an experiment in which one measures a diffraction  pattern 
associated with the parallel transport  of the gauge field $A$ round a 
non-contractible loop $C$ in $\Omega$. 
\par
A successful experiment of precisely  this kind was  done 
by Brill and Werner  \ref{1} in 1960. The experimental 
setup---cf. fig. 4---was of the 
Young's slits type with electrons replacing photons and  with the 
addition of a very thin solenoid.  The electrons passed through the 
slits and on either side of the solenoid and an interference pattern 
was then detected. The interference pattern is first measured with the 
solenoid off. This pattern is then found to {\it change} when the 
solenoid is switched on even though the electrons always pass through a 
region where the field $F$ is zero. 
\epsfxsize=0.5\hsize
\par\penalty10000\vskip\baselineskip
\centerline{\epsffile{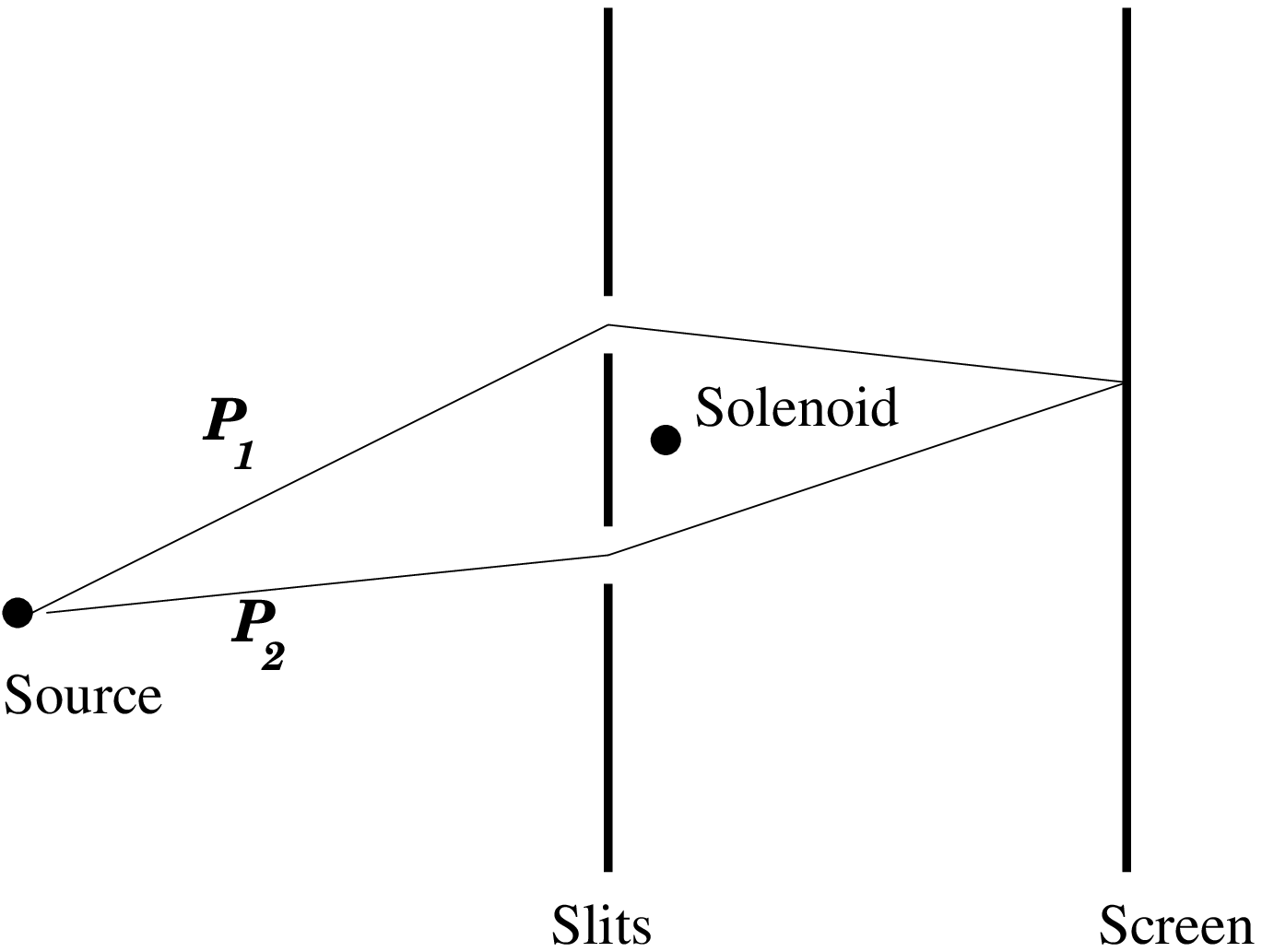}}
\centerline{\it Fig. 4. \bf A schematic Aharonov--Bohm experiment}
\par\vskip\baselineskip
This result is quite a dramatic demonstration of the fact that the connection, or gauge field, $A$ is a more fundamental object than the electromagnetic field, or curvature $F$; all the more so since the key point is a topological one\foot{It is possible 
that this result would not have been a complete surprise to Maxwell because, as is shown  in our first quotation from him above, he was well aware of the necessity for topological considerations, and even knew something of their nature,  when a function w
as defined by a line integral.}.
\par
A topological explanation  is easy to provide and begins with the notion of parallel transport round {\it closed} curves.
The action of parallel transport on multilinear objects, viewed  either as vectors, spinors,  tensors etc. or sections of the appropriate bundles, is via the operator $PT(C)$  where 
$$PT(C)=\exp\left[\int_C A\right]$$
Differential topology provides us immediately with the means to see that $PT(C)$ is non-trivial. The argument goes as follows: $F=dA$ so that the vanishing of $F$ gives us
$$\eqalign{dA&=0\cr
           \Rightarrow A&=df\quad \hbox{locally}\cr}  $$
Hence $A$ determines a de Rham cohomology class [A] and we have
$$[A]\in H^1(\Omega;{\bf R})$$
It is clear from Stokes' theorem that integral $\int_C A$ only depends on the homotopy class of the loop $C$; in addition  
the loop $C$  determines a homology class $[C]$ where 
$$[C]\in H_1(\Omega;{\bf R})$$
This means that the integral $\int_C A$ (which we can take to be the number $n$)  is just the dual pairing $(\bullet,\bullet)$ between cohomology and homology i.e.
$$([A],[C])=\int_C A$$ 
\par
Mathematically speaking we realise the solenoid by a cylinder $L$ so that
$$\eqalign{\Omega&= {\bf R^3}-L\cr
\Rightarrow H^1(\Omega;{\bf R})&=H^1({\bf R^3}-L;{\bf R})\cr
              &={\bf Z}\cr}$$ 
In the experiment referred to above the loop $C$ is the union of the electron paths $P_1$ and $P_2$, cf. fig. 4.
\par
The holonomy  group element $PT(C)$  is also of central interest
 elsewhere. It occurs in the study of the adiabatic periodic change of 
parameters of a quantum system described by Berry  
\ref{1} (1984) cf. also Simon \ref{1}. This phenomenon is known as 
Berry's phase. A relevant topological application here is an  
explanation of the quantum Hall effect, cf. Morandi \ref{1}.
\par
 The importance of  non-trivial flat  connections for physics 
extends to the non-Abelian or Yang--Mills case as we shall see in due course below.
\par
It is now time to discuss topological matters concerning general 
relativity.
\beginsection{Topology, general relativity and 
singularities---black holes and the big bang} 
\smalltitle{Chandrasekhar and gravitational collapse}
In 1965 Penrose \ref{1} made the biggest breakthrough since 
Chandrasekhar \ref{1} (1931)  in 
understanding the nature of gravitational collapse: He  proved the 
first theorem
which  showed that singularities of the 
gravitational field are a generic feature of gravitational 
collapse. Moreover Penrose's methods were topological.
\par
Before saying anything more about Penrose's paper we need 
a brief sketch of some of the salient features of  
 gravitational collapse. 
\par
Gravitational collapse is something that is worth investigating for 
very massive objects such as stars. This simple sounding idea is that,  
for a sufficiently massive body, the attractive force of gravity 
may be strong  enough to cause it  to start to {\it implode}. 
\par
To find something massive enough we have to choose a stellar object such as a star. 
Now, for a young active star, the burning of the nuclear fuel causes enough outward pressure to counteract all its gravitational inward pressure.
However, since the nuclear fuel will eventually be used up this line of 
thought suggests that one calculate what gravity can do once it is not 
opposed by the nuclear reactions. In 1931 Chandrasekhar  \ref{1} 
published his celebrated paper on this matter.  
\par
He took a star of mass $M$ to be a relativistic gas at temperature $T$ 
obeying a  relativistic equation of state. With the star's nuclear fuel 
all spent its cooling and contraction under gravity was opposed by the
degeneracy pressure of the electrons produced by Fermi-Dirac statistics. 
However Chandrasekhar found that this 
pressure could not resist gravity if $M$ was greater than about $1.4$ solar masses (in standard notation one writes this as $1.4 M_{\odot}$ where 
$M_{\odot}$ denotes the mass of the Sun). On the other hand for 
$M$ less than $1.4 M_{\odot}$  the star should cool and contract  to 
what is called a white dwarf.
Hence, for stars heavier than $1.4 M_{\odot}$, unless something special 
intervened---for example a mechanism causing matter to be ejected during cooling until the Chandrasekhar   limit is eluded---gravitational collapse  is predicted. 
\par
No one (and this included Einstein) was very comfortable with this 
result but it resisted all the  
attempts made to get round it or even to disprove it.
White dwarfs as ultimate fates of cooling stars were then supplemented
by neutron stars. 
\par
Neutrons stars are so dense that their protons and 
electrons have combined to form neutrons; these neutrons then have a 
degeneracy pressure which resists the gravitation of the cooling star 
just as the electrons do in a white dwarf.
The same ideas about the maximum mass of white dwarfs apply to neutron 
stars which then also  have a maximum mass, this varies from about 
$2M_{\odot}$  to $3M_{\odot}$, the precise value depending on one's 
knowledge of the nuclear force, or strong interactions,  at 
high densities.
\par
Stars which are heavy enough\foot{Apparently it may be possible for a 
large amount  of matter to be shed by stars as they collapse: a star may 
even need  a mass $M$ greater than $20M_{\odot}$ in 
order to be forced to gravitationally collapse. However there are stars 
known with mass $M$ ranging up to $100M_{\odot}$ so we do expect 
gravitationally collapsed stars to exist. For more details cf. chapter 
9 of Hawking and Ellis \ref{1}.}  are thought not to end up in the 
graveyard of white dwarfs or neutrons stars but instead continue their 
collapse and form black holes\foot{Black holes, in the sense of {\it dark stars}  
from which light cannot escape, were discussed in Newtonian physics by 
Michell \ref{1} (1784) and Laplace \ref{1} (1799) (cf. appendix A of 
Hawking and Ellis \ref{1} for a translation of Laplace \ref{1}). They took light to 
obey a corpuscular theory and computed the size of a star whose escape 
velocity was greater than that of light. For other details of interest  
cf. the article by Israel in Hawking and Israel \ref{1} and the 
excellent book for the layman by Thorne \ref{1}.}.
\par
The attitude to results about gravitational collapse as  such as 
Chandrasekhar's \ref{1,2} was that they were properties of the unrealistically high degree of symmetry of the solutions: collapse  was not expected in the real Universe where such symmetry would not be found.
This was also largely the attitude taken to the much later result of 
Oppenheimer and Snyder \ref {1} (1939). This was a a paper (in which 
spherical symmetry was assumed) that   
 produced the new result that  a star undergoing gravitational collapse
cut itself off from external observation as it contracted through a 
certain critical radius---the Schwarzschild radius. It contained, too, 
the facts about time asymptotically slowing to zero for an external 
observer but not for an observer moving with the star:  
{\beginquote
The total time of collapse for an observer co-moving with the star is finite .......... an external observer sees the star asymptotically shrinking to its gravitational radius.
\endquote}
\par
The Schwarzschild metric plays a central part in understanding 
gravitational collapse and we shall now sketch some of its main 
properties.
\smalltitle{The Schwarzschild metric }
Schwarzschild \ref{1} derived the form of  a (spatially) spherically 
symmetric 
metric. With  spherical polar coordinates $(r,\theta,\phi)$, and 
time $t$, it is determined by the line element
$$ds^2=-\left(1-{2m\over r}\right)dt^2+
{\left(1-{2m\over r}\right)}^{-1}dr^2+r^2(d\theta^2+\sin^2\theta\, d\phi^2)\no$$
This metric is meant to represent the empty space-time outside a 
spherically symmetric body of mass $m$\foot{One might think of the Schwarzschild 
metric as a solution to the one body problem in general relativity; 
unfortunately things are worse than in Newtonian gravity where the three 
body problem was so hard to solve analytically, in general relativity the two 
body problem has not (yet!) been solved analytically; this means that accurate 
approximation methods must be used to treat important problems such as 
binary stars. As regards  the Newtonian three body problem we 
should add that it was finally solved analytically in 1909 by 
Sundman \ref{1}, cf. also Sundman \ref{2,3} and Barrow-Green \ref{1}, 
p. 187. Unfortunately,  
though the Sundman solution is a great triumph,  his solution
is a convergent series whose convergence rate is incredibly slow: 
apparently some $10^p$ terms, with $p$ measuring in the millions, 
would be needed for practical work.}.
We  see at once a singularity at $r=0$ and one at $r=2m$. However the 
singularity at $r=0$ is {\it genuine}---for example the Riemann curvature 
tensor diverges there---but the singularity at $r=2m$ is only a 
{\it coordinate singularity} and disappears in an appropriately  
chosen  coordinate system. Incidentally this is precisely analogous to the two singularities we encountered for the Dirac monopole: one at $r=0$ and one at $z+r=0$; we found that $ r=0$ was a real singularity but that
$z+r=0$ was only a coordinate singularity.
\par
However the benign nature of the hypersurface  $r=2m$ was not realised 
for many years and it was usually misleadingly referred to as 
the ``Schwarzschild singularity''. This special value of $r$ is called 
the {\it Schwarzschild  radius} of the mass $m$. Schwarzschild \ref{2} 
himself\foot{Since this is a historical article I just add  that this 
presumably
was Schwarzschild's last article as he died of an illness while 
on the Russian front in 1916; he was born in 1876.}  went to 
the trouble of quoting the value for the Sun: it is very small, namely 
3km.   
\par
This smallness led to the conviction that $r=2m$ was irrelevant in
 practice because such a value of $r$ lay deep down in the interior of any realistic body.  Hence comfort was derived from the fact that the Schwarzschild metric was always used to 
describe the gravitational field in the empty space {\it outside} the 
star where $r$ was always much bigger than the Schwarzschild radius. 
This was fine if a star never started to collapse but not otherwise.
\par
The cosmologist Lema\^\i tre did notice in 1933 that $r=2m$ was not a 
real singularity but this seems to have gone unnoticed or not been 
appreciated for a long time. In Lema\^\i tre \ref{1} we find the words
{\beginquote
La singularit\'e du champ de Schwarzschild est donc une 
singularit\'e fictive 
\endquote} 
i.e.
{\beginquote
The singularity of the Schwarzschild field is therefore a 
fictitious singularity 
\endquote}
It is clear, though, that an $(r=const,t=const)$ surface does change its 
character precisely when $r$ passes through the value $2m$: For $r>2m$ 
such a surface is timelike while for $r<2m$ it is spacelike.
This does mean that there is {\it something} special about the 
Schwarzschild radius, the question is just what is this something? 
Penrose was able to provide the answer and use it to make a 
breakthrough in understanding 
gravitational collapse. The point is that, for $r<2m$, one can have what 
Penrose called a {\it trapped surface} and these we now consider.
\smalltitle{Penrose and trapped surfaces}  
Gravitational collapse still refused to go away and in the early 1960's 
with the discovery of gigantic energy sources dubbed quasars the subject 
again became topical. It was 
suggested 
(other more conventional explanations didn't seem to fit) that the
 energy source of a quasar  came from the gravitational collapse of an
immensely massive object of mass $10^6M_{\odot}$--$10^9M_{\odot}$. 
Presumably such a collapse would not be spherically symmetric, for example one would expect there to be non-zero angular momentum.
All this increased the need to study the possibility of 
gravitational collapse in general, i.e. without 
without any assumption of a special symmetry, spherical or otherwise.
\par
Fortunately this is precisely where Penrose's topologically obtained 
result comes to the rescue. Penrose deduced that gravitational collapse
to a space-time singularity was inevitable given certain reasonable 
conditions, and these conditions did {\it not} require any assumption about symmetry.
\par
In Penrose \ref{1} (1965) we find the statements
{\beginquote
It will be shown that, after a certain critical condition has 
been fulfilled, deviations from spherical symmetry cannot prevent 
space-time  singularities from arising .......... The argument will be 
to show that the existence of a trapped surface implies---irrespective 
of symmetry---that singularities develop.
\endquote}
Special attention has to be given to providing a definition of a 
singularity which is both mathematically and physically reasonable. 
In brief geodesic completeness is used as the basis for the definition 
of a singularity\foot{In  relativity, since the metric 
is Lorentzian, geodesic completeness exists in three varieties: null, 
timelike and spacelike. To be singularity free  both  null and 
timelike geodesic completeness are demanded of $\cal M$; spacelike 
completeness is not required because physical motion does not take place 
along spacelike curves. In addition to geodesic completeness one also 
requires a causality condition and a non-negative energy condition:  
The causality condition is usually stated as the absence in $\cal M$ of any closed timelike curves---causes always precede effects. The  non-negative energy condition is that, if 
$T_{\mu\nu}$ is the energy momentum tensor, then
$T_{\mu\nu}  t^\mu t^\nu\ge 0$ everywhere in $\cal M$ for all timelike vectors $t^\mu$---in the rest frame of such a $t^\mu$ this becomes the statement that the energy $T_{00}\ge0$ which is just another way of saying that
 gravity is always attractive.} 
of the space-time manifold $\cal M$; to see the significance of 
 such completeness just consider that if a particle travelled along an 
{\it incomplete} timelike geodesic then it could disappear suddenly from 
$\cal M$ in a finite time.
\par
It is impossible to give a detailed account here of the arguments so 
we shall only outline them; for a proper account cf. Hawking and Ellis 
\ref{1}.
\par
As the matter constituting the star contracts
it passes through its Schwarzschild radius $r=2m$ and after this has 
happened the matter lies totally within a spacelike sphere $S^2$ 
(cf. fig. 5). This $S^2$ is what is called a trapped surface; technically it is closed, compact, spacelike, two dimensional and has the property that null geodesics which intersect it
orthogonally converge in the future. 
\par
The space-time manifold $\cal M$ is the future time development of an 
initial {\it non-compact} Cauchy hypersurface. 
\par
Figure 5 shows a space-time diagram
of the collapse. In perusing the figure  the reader should bear in mind 
that {\it one spatial dimension is suppressed} and that the circular symmetry of the diagram is there only for aesthetic reasons; the whole 
point being that no symmetry is assumed. The initial Cauchy hypersurface is represented by the plane at the bottom of the diagram.
\par
The argument then computes the degree of an appropriate map which shows
that  null geodesic completeness implies that the 
future of the trapped surface is {\it compact}. However this is 
incompatible with the fact the initial Cauchy hypersurface is 
{\it non-compact}; this contradiction forces  $\cal M$ to have  
a singularity.
\epsfxsize=0.3\hsize
\par\vskip 0.5\baselineskip
\centerline{\epsffile{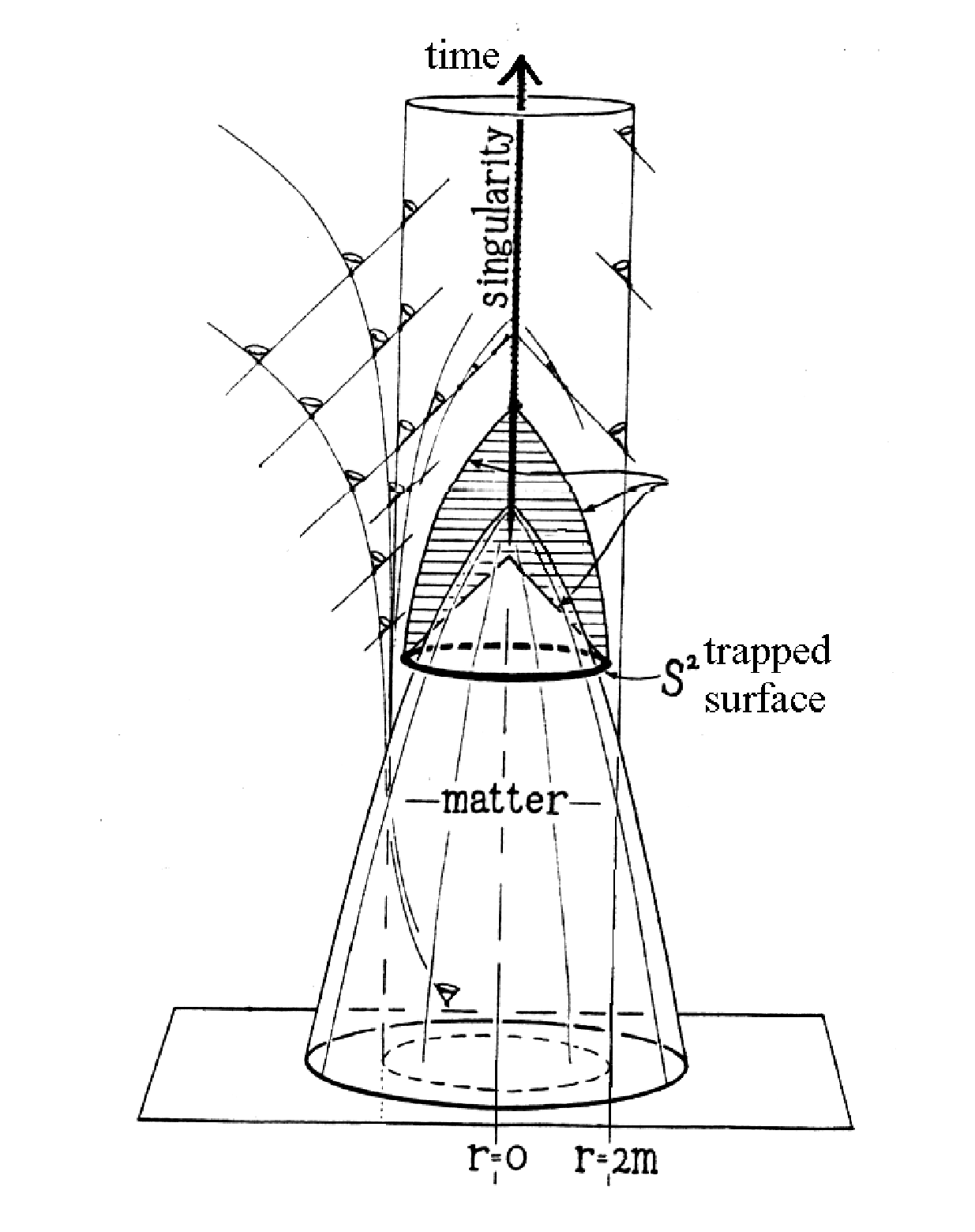}}
\par\penalty100000
\centerline{\it Fig. 5. \bf Gravitational collapse and a trapped surface $S^2$}
\par\vskip\baselineskip
Some insight into the  importance  of a trapped surface 
can be obtained from a physical discussion: Normally if light is 
emitted radially outwards from all points on the surface $S$, say, 
of a sphere then it creates an outward moving spherical wave with surface 
$S^\prime$; furthermore $S^\prime$ has a {\it bigger area} than $S$. 
However if $S$ is a trapped surface then one finds that  $S^\prime$ has 
a {\it smaller area} than $S$; this corresponds to the fact that gravity 
bends the light back and stops it escaping from the region inside $r=2m$.
Hence, as time progresses, $S$ evolves to a smaller and smaller surface
which eventually becomes a singularity, cf. again, fig. 5. The three surface $r=2m$ is called the (absolute) {\it event horizon} of the collapse.
\par
The use of the term black hole to describe such singularities is due to
Wheeler who coined it in 1968, cf. Thorne \ref{1}. The entire present 
day Universe may have originated in a past singularity known as the 
{\it big bang} a possibility for which there is considerable 
experimental evidence nowadays. This has resulted in the big bang being taken very seriously.  However without the use of topological methods to 
convince one that singularities are generic under certain reasonable 
conditions the 
big bang would have been  much more difficult to take seriously. 
\par
A further 
important paper on singularities was Hawking and Penrose \ref{1} (1970); 
the general situation is discussed at great length and in full detail in 
Hawking and Ellis \ref{1}.
\par
To round things off we point out that an important consequence 
 of this work on singularities is that the set of solutions to the 
general relativistic hyperbolic Cauchy problem, which are destined to 
evolve into singularities, form a set of positive measure in an 
appropriate topology.
\par
Just as topology was becoming a permanent bed fellow of relativity 
it also 
began to play a r\^ole in the Yang--Mills or non-Abelian gauge theories, these theories having  moved to centre stage in 
elementary particle theory. This was to be an even more important event for topology as it has led to a genuinely two sided interaction between theoretical physics and mathematics\foot{There was also some work in the 
1960's using algebraic topology to study singularities of 
Feynman integrals, cf. Hwa and 
Teplitz \ref{1} (1966), Froissart \ref{1} (1966) and Pham \ref{1} (1967) 
but this has not continued to any great extent.}
We begin this story in the next section.
\beginsection{Topology and Yang--Mills theory---the latter day explosion} 
\smalltitle{The rise of gauge invariance in the 1970's}
There is no doubt that a principal factor in the rise of topology in 
physics is due to the rise to supremacy\foot{This, of course, is another story and it is not our task to tell it here. However it is important for the reader to be aware  that the interest in non-Abelian gauge theories 
was rekindled almost overnight with the vital proof of the 
renormalisability of non-Abelian gauge theories by 't Hooft \ref{1} 
and 't Hooft and Veltman \ref{1} (1971--72). This led to the resurrection of earlier papers on the subject and to the so called 
{\it standard model} with gauge group $SU(2)\times U(1)$ of weak and 
electromagnetic interactions and to QCD or {\it quantum chromodynamics}, 
with gauge group $SU(3)$,  the favoured model for the confined quarks 
believed to be responsible for the strong interactions.}
of  gauge theories in 
physics.
Topology entered, in the main, via gauge theories:  physicists learned
that gauge theories had a formulation in terms of fibre bundles; they learned too that  much useful cohomological data  was possessed by 
these bundles.
\smalltitle{Nielsen, Olesen, Polyakov and 't Hooft}
An early important result of this period---which we may take to be post 
the papers of 't Hooft \ref{1} and 't Hooft and Veltman 
\ref{1} (1971--72)---is on magnetic monopoles in {\it non-Abelian} gauge theories. Two independent papers
't Hooft \ref{2} and Polyakov \ref{1} produced the first 
{\it non-Abelian monopole} (now referred to as the 't Hooft--Polyakov 
monopole). An earlier paper by Nielsen and Olesen \ref{1} on magnetic 
vortices in superconductors was an important influence: In 't Hooft \ref{2} 
the 
author opens with
{\beginquote
The present investigation is inspired by the work of Nielsen et 
al. \ref{1}, who found that quantized magnetic flux lines, in a 
superconductor, behave very much like the Nambu string \ref{2}.
\endquote}
\par 
The 't Hooft--Polyakov monopole, like the Dirac monopole, is a static object and lives in ${\bf R}^3$; however,  unlike the Dirac monopole, it has 
{\it no singularity} at the origin and is regular {\it everywhere} 
in ${\bf R}^3$. 
indeed as 't Hooft \ref{1} says
{\beginquote
Our way for formulating the theory of magnetic monopoles avoids the 
introduction of Dirac's string \ref{3}.
\endquote}
The  magnetic charge $g$ of the monopole is 
topologically quantised and is inversely proportional to  $q$ where 
$q$ is the electric charge of a heavy gauge boson in the theory.
The topology enters through the boundary condition at infinity in 
${\bf R}^3$. We shall now attempt to elucidate this by supplying some 
of the details of the mathematical setting.
\smalltitle{The topology of monopole boundary conditions} 
Monopoles are static, finite energy, objects which give the critical points of 
the energy of an appropriate system of fields defined on a three 
dimensional 
Riemannian manifold $M$.  In fact the usual choice for $M$ is the 
non-compact space ${\bf R}^3$. Analysis on a non-compact $M$ 
introduces some  
technical difficulties but these have not proved insurmountable. 
\par
The physical system studied consists of a Yang--Mills $G$-connection $A$, with curvature $F$, and a 
Higgs scalar field $\phi$ transforming according to the adjoint 
representation of $G$. If the Hodge dual with respect to the metric on 
$M$ is denoted by $*$, then the energy $E$ of the system is given by 
$$E={1\over2}\int_M\,\left\{-tr(F\wedge *F)-tr(d_A\phi\wedge 
*d_A\phi)+\lambda*(\vert\phi\vert^2-C^2)^2\right\}\no$$
where $d_A\phi$ is the covariant exterior derivative of the Higgs 
field, $tr$ denotes the trace in the Lie algebra $\scriptl {g}$ of $G$ and $\vert\phi\vert^2=-2tr(\phi^2)$. 
 The field equations for the critical points of this system are difficult 
to solve explicitly (indeed the 't Hooft--Polyakov monopole is 
constructed numerically) but many solutions are available in 
what is called the Prasad--Sommerfield limit (cf. Prasad and 
Sommerfield \ref{1}) where the scalar potential term vanishes. The 
energy is then
$$\eqalign{E\equiv E(A,\phi)&=-{1\over2}\int_M\,
\left\{tr(F\wedge *F)+tr(d_A\phi\wedge  *d_A\phi)\right\}\cr 
{}&={1\over2}\{\Vert F\Vert^2+\Vert d_A\phi\Vert^2\}\cr
            &={1\over2}\{
\Vert F\mp*d_A\phi \Vert^2  \pm2<F,*d_A\phi>\}\cr }\no$$
This shows that the absolute minima of $E$ are attained when the pair 
$(A,\phi)$ satisfy
$$F=\mp*d_A\phi\no$$
which is the celebrated  Bogomolny equation (Bogomolny \ref{1}).
 The expression\foot{Throughout this article the inner product $<\omega, \eta>$ between Lie algebra valued $p$-forms $\omega$ and $\eta$ on a 
Riemannian 
manifold $M$  has the standard definition: i.e.
$$<\omega,\eta>=-\int_M  tr(\omega\wedge * \eta>$$}  
\eqlabel{\magcharge}
$$<F,*d_A\phi>=-\int_M\,tr(F\wedge d_A\phi)\no$$
is the absolute minimum and looks like a topological charge. 
\par
Now suppose that  $M={\bf R}^3$ furnished with the 
Euclidean metric and also set $G=SU(2)$. For the energy $E$ to 
converge and to 
make the field equation problem well posed  we must specify 
boundary conditions at infinity. 
A standard boundary condition for 
$\phi$ is 
$$\lim_{r\rightarrow\infty}\vert\phi\vert\longrightarrow (C+O(r^{-2}))\no$$
where $r$ is the distance from the origin in ${\bf R}^3$. The integral \docref{magcharge} can now be 
non-zero and it does have a topological interpretation which forces it to take 
discrete values. More precisely, if $k$ is an integer, then
$${1\over4\pi C}<F,*d_A\phi>=k\no$$
This integer is the magnetic charge and can be thought of as the Chern class 
of a $U(1)$ bundle over a 
two sphere which is $S^2_{\infty}$, the two sphere at infinity; setting $C=1$ 
and using Stokes' theorem, we have
$$k={1\over4\pi}\int_{\hbox{\eightpoint \bf R}^3}\,tr(F\wedge d_A\phi)=
{1\over4\pi}\int_{S^2_{\infty}}\,tr(F\phi)\no$$
The condition $\vert\phi\vert=1$ on the boundary defines an $S^2$ inside 
the Lie algebra $su(2)$ and  $(F\phi)$, when evaluated at infinity, becomes the 
$U(1)$-curvature of a 
bundle over $S^2_{\infty}$ and $k$ is its Chern class. Alternatively, one can 
write $k$ as the  winding number, or degree, of a map  
$\hat \phi:S^2_{\infty}\rightarrow S^2_{su(2)}$, giving
$$\eqalign{\hat \phi:\,&S^2_{\infty}\longrightarrow S^2_{su(2)}\cr
                     &x\longmapsto \hat \phi(x)={\phi\over\vert\phi\vert}\cr
\hbox{and }\qquad\qquad 
k&=-{1\over2\pi}\int_{S^2_{\infty}}\,tr(\hat \phi d\hat \phi\wedge d\hat \phi)
\cr}\no$$
This boundary integer $k$ is the only topological invariant associated with 
the monopole system; the $SU(2)$ bundle over ${\bf R}^3$ is topologically trivial since ${\bf R}^3$ is contractible.
\smalltitle{Dyons and $CP$ breaking} 
In section 2 we promised to return to the subject of dyons  
and a {\it topological} mechanism for $CP$ breaking. This $CP$ 
breaking requires a non-Abelian monopole since it comes from the  
presence in the action of a multiple of the second Chern class 
$c_2(P)$ given by ($e$ and $\theta$ are real constants and for 
convenience we assume that the gauge group is $SU(N)$)
$${\theta e^2\over 16 \pi^2} tr(F\wedge F)={\theta e^2\over 2}c_2(P)\no$$ 
This has been discussed by Witten \ref{1} and it is immediate that such 
a term  is $CP$ non-invariant.  The electric charge  
of a dyon now involves the $\theta$ parameter: one finds that
for dyons with magnetic charge $g=2\pi n_0/e,\,n_0\in {\bf Z}$ their electric charge $q$ 
obeys the formula
$$q=\left(n-{\theta\over 2\pi}\right)e,\,n\in {\bf Z}\no$$
 This has the interesting feature  that  the {\it electric} 
charge of a dyon is not 
a {\it rational} multiple of a fundamental  electric charge unless 
the $CP$ violating Chern class coefficient $\theta$ is zero 
($\theta=\pi$ is also allowed but may be too large experimentally). In terms of
fig. 3 above it means that the angle $\alpha$  vanishes when $\theta$ is 
zero.
\par
 Still further insight into the r\^ ole played by monopoles in quantum field theory has been obtained by combining the electric and monopole charges
into a single {\it complex} parameter $e+ig$; we shall discuss this in section 8. 
\smalltitle{Gauge theories in four dimensions: Instantons}
Topology came even more to the fore in Yang--Mills theories with 
the publication in 1975 by Belavin et al. \ref{1} of topologically 
non-trivial solutions to the {\it Euclidean} Yang--Mills equations 
in four dimensions. Such solutions have come to be called 
{\it instantons}\foot{Quite a few early papers on the subject used the 
less attractive term pseudoparticle instead of instanton  
but luckily this usage was short-lived.}.
\par  
Of fundamental importance for these  solutions to the Euclidean 
Yang--Mills equations is that instantons are at the same time 
{\it non-perturbative} and {\it topological}.
\par
The Euclidean version of a quantum field theory is obtained from the 
Minkowskian version by replacing the Minkowski time $t$ by $it$. The 
relation between the two theories is supposed to be one of analytic 
continuation in the Lorentz invariant inner products $x_\mu y^\mu$; 
allowing these inner products to be  complex is the simple way to pass 
from one theory to the other. The existence of such a continuation makes 
tacit certain assumptions which require proof; significant progress in 
this technical matter was made in 1973--75, cf.
Osterwalder and Schrader \ref{1--3}, and Streater \ref{1}.
\par
The term instanton, though not quite precise,  is often  generalised to 
refer to a critical point of finite Euclidean action for any quantum 
field theory. Such solutions to the equations of motion---for this is 
what these critical points are---are closely related to quantum 
mechanical tunnelling phenomena. This property of instantons quickly 
attracted great interest because  tunnelling amplitudes are not calculable
 perturbatively but require a knowledge of the theory for large 
coupling as well as small. 
\par
This opening of the door into the room of 
non-perturbative techniques was a noteworthy event and we shall see below
that topology was a key ingredient to picking the lock on  this door.
\par
A good account of this is to be found in Coleman \ref{1} (1977) 
who showed his
pleasure at the progress made in his opening sentences:
{\beginquote
In the last two years there have been astonishing 
developments in quantum field theory. We have obtained control over
problems previously believed to be of insuperable difficulty and we 
have obtained deep (at least to me) insights into the structure of the 
leading candidate for the theory of strong interactions, quantum 
chromodynamics.
\endquote}
\par
In Coleman \ref{1} there is also an account of an important paper 
('t Hooft \ref{3}) which used instantons to solve an outstanding problem 
known as  ``the $U(1)$ problem'', thereby imbuing the fledgling 
instantons with considerable status. 
We shall now give a brief summary of some of the more salient features of
an instanton in the Yang-Mills case. 
\smalltitle{Profile of an instanton}
Our life can be made a little easier by choosing a  very specific 
setting: we have a non-Abelian 
gauge theory with $G$ a compact simple Lie group and action
\eqlabel{\ymaction}
$$S\equiv S(A)=\Vert F\Vert^2=-\int_M\,tr(F\wedge {}*F)\no$$ 
with $M$ a closed four dimensional orientable Riemannian manifold and 
$*$ the Hodge dual with respect to the Riemannian metric on $M$. 
Instantons are those $A$ which correspond to critical points of $S$; 
however we shall specialise the term here to mean only {\it minima} of 
$S$.
\par
First we should obtain the Euler--Lagrange equations of motion, i.e. the 
equation for the critical points. 
Let $A$ be an arbitrary connection through which passes the family  of 
connections 
$$A_t=A+ta\no$$
Expanding in the vicinity of $t=0$ gives
\eqlabel{\actionexpansion}
$$\eqalign{S(A_t)&=<F(A),F(A)>+t{d\over 
                     dt}\left.<F(A_t),F(A_t)>\right\vert_{t=0}+\cdots\cr
\hbox{and }\qquad F(A_t)&=F(A)+t(da+A\wedge a+a\wedge A)+t^2a\wedge a\cr
               {}&=F(A)+td_Aa+t^2a\wedge a\cr
\Rightarrow S(A_t)&=\Vert 
F(A)\Vert^2+t\{<d_Aa,F(A)>+<F(A),d_Aa>\}+\cdots\cr
{}&=S(A)+2t<F(A),d_Aa>+\cdots\cr}\no$$
$A$ is a critical point if 
$$\left.{dS(A_t)\over dt}\right\vert_{t=0}=0\no$$
That is, if
$$\eqalign{<F(A),d_Aa>&=0\cr
\Rightarrow <d^*_AF(A),a>&=0\cr
\Rightarrow d^*_AF(A)&=0,\quad\hbox{since $a$ is arbitrary}\cr}\no$$
However, $F(A)=dA+A\wedge A$ also satisfies the Bianchi identity 
$d_AF(A)=0$ and so we 
have the pair of equations 
$$d_AF(A)=0,\qquad\quad d^*_AF(A)=0\no$$
This is similar to the condition  for a form 
$\omega$ to be harmonic, which is
$$d\omega=0,\qquad\quad d^*\omega=0\no$$
It should be emphasised, though, that the Yang--Mills equations are not 
linear; thus they really express a kind of non-linear harmonic condition. 
\par
The most 
distinguished class of solutions to the Yang--Mills equations $d^*_AF(A)=0$ is that consisting of those connections whose curvature is self-dual or 
anti-self-dual. 
\par
To see how such solutions originate we point out that with respect to 
our inner product on $2$-forms $d^*_A$ has the property that
$$d^*_A=-*d_A*\no$$
so that the Yang--Mills equations become
$$d_A*F(A)=0\no$$
Thus if $F=\mp*F$ the Bianchi identities immediately imply that we have a 
solution to the Yang--Mills equations---we have managed to solve a 
non-linear second order equation by solving a non-linear first order 
equation. 
\par
It is also easy to see that these critical points are all {\it minima}  
of the action $S$; here are the details. 
First we (orthogonally) decompose $F$ into its self-dual and 
anti-self-dual parts $F^+$ and 
$F^-$, giving
$$\eqalign{F&={1\over2}(F+*F)+{1\over2}(F-*F)\cr
            &=F^++F^-\cr
\Rightarrow S&=\Vert (F^++F^-)\Vert^2\cr
             &=\Vert F^+\Vert^2+\Vert F^-\Vert^2\cr}\no$$
where the crossed terms in the norm contribute zero. 
\par
The {\it topological type} of the instanton is classified by the second 
Chern class $c_2(F)\in H^2(M;{\bf Z})$  of the bundle on which the 
connection $A$ is defined:  
Taking $G$ to be the group $SU(N)$
and evaluating $c_2(F)$ on $M$ we obtain the integer   
$$c_2(F)[M]={1\over8\pi^2}\int_M\,tr(F\wedge F)\in{\bf Z}\no$$
The {\it instanton number},  $k$, is defined to be minus this number
so  we find that 
$$\eqalign{k&=-{1\over8\pi^2}\int_M\,tr(F\wedge F) \cr
            &=-{1\over8\pi^2}\int_M\,tr\{(F^++F^-)\wedge(F^++F^-)\}\cr
            &={\Vert F^+\Vert^2-\Vert F^-\Vert^2\over8\pi^2}\cr}\no$$
The inequality $(a^2+b^2)\ge\vert a^2-b^2\vert$ shows that, for each 
$k$, the absolute minima of $S$ are attained when 
$$S=8\pi^2\vert k\vert\no$$
and this corresponds to $F^{\mp}=0$ or equivalently
$$F=\mp*F\no$$
and we have the celebrated self-dual and anti-self-dual conditions. 
Changing the 
orientation of $M$ has the effect of changing the sign of the $*$ 
operation 
and so interchanges $F^+$ with $F^-$. 
\par
Up to now, although we have not mentioned it,  for algebraic convenience 
we have set the coupling constant of the theory equal to unity.
But to understand anything non-perturbative the coupling must be present 
so we now temporarily cease this practice. Denoting the coupling constant by $g$
(the context should prevent any confusion with magnetic charge) 
the action $S$ is given by 
$$S\equiv S(A)={1\over g^2}\Vert F\Vert^2=-{1\over g^2}\int_M\,tr(F\wedge {}*F)\no$$
Hence if $A$ is an instanton then we immediately have
$$S(A)={8\pi^2 \vert k\vert\over g^2},\; k\in {\bf Z}\no$$
Finally the corresponding quantum mechanical amplitude is $\exp[-S]$ so that we have
$$\exp[-S(A)]=\exp\left[-{8\pi^2 \vert k\vert\over g^2}\right]\no$$
which we see at once is an {\it inverse} power series in $g^2$; moreover
topology is uppermost for  
we note that for this inverse power series to exist the instanton number 
$k$ must be non-zero.
\smalltitle{The mathematicians take a strong interest}
The pace of instanton research increased towards the end of the 1970's
due  in part to a keen interest being taken in the problems by some 
highly able and gifted mathematicians. As we shall see below this 
attack on the problems by two distinct groups was to 
prove highly beneficial to both physics {\it and} mathematics. 
In fact some particularly choice fruits of these labours fell into 
the garden of the mathematicians. 
\par
Thus far we have stressed the topological nature of the connections of
 Yang--Mills theory: the relevant mathematical structure is a fibre 
bundle and together with this comes cohomological characteristic class 
data giving discrete numerical invariants such as the instanton 
number $k$.
However, for instantons,  there remains a more prosaic object to study namely the 
non-linear partial differential equation for the instanton $A$ itself, 
i.e. the self-duality equation
$$F= *F\no$$
or, more explicitly,
$$
\partial_\mu A^a_\nu-
\partial_\nu A^a_\mu+ig f^{abc} A^b_\mu A^c_\nu=
{1\over2}\epsilon_{\mu\nu\alpha\beta}
(\partial^\alpha A^{a\beta}-
\partial^\beta A^{a\alpha}+ig f^{abc} A^{b\alpha} A^{c\beta})\no$$
A key change of viewpoint on the  self-duality equation changed the 
 focus away from differential equations; this was the breakthrough made 
by Ward \ref{1} (1977). 
\par
Ward showed that the solution of the self-duality partial differential 
equation was equivalent to the construction of an appropriate vector 
bundle. His paper (Ward \ref{1}) gives a brief summary at the beginning
{\beginquote
In this note we describe briefly how the information
of self-dual gauge fields may be ``coded'' into the structure of certain
complex vector bundles, and how the information may be extracted, 
yielding a procedure by which (at least in principle) all self-dual
solutions of the Yang--Mills equations may be generated. The construction 
arose as part of the programme of twistor theory \ref{3}; it is
the Yang--Mills analogue of Penrose's ``non-linear graviton'' construction
\ref{4}, which relates to self-dual solutions of Einstein's vacuum 
equations.
\endquote}
This discarding of the differential equation and its encoding into the 
transition functions of certain vector bundles  
immediately made the problem of more interest and accessibility to 
mathematicians. Atiyah and Ward \ref{1} showed how the problem was 
equivalent to one in algebraic geometry; there then followed a 
complete solution to the problem for $M=S^4$ by Atiyah, Drinfeld, 
Hitchin  and 
Manin \ref{1}, the situation for other four manifolds is treated in 
Atiyah, Hitchin and Singer \ref{1}. 
\par
Atiyah, who was to become a key figure in many subsequent developments 
of joint interest to mathematicians and physicists describes his 
introduction to Yang--Mills theories  as follows (taken from the preface 
to Atiyah \ref{2})
{\beginquote
My acquaintance with the geometry of Yang--Mills equations 
arose from lectures given in Oxford in Autumn 1976 by I. M. Singer, 
and I  am very grateful to him for arousing my interest in this 
aspect of theoretical physics.
\endquote}
It was not long before a large body of both mathematicians and physicists
were working on a large selection of problems related in some way to 
Yang--Mills theories. The next breakthrough was in mathematics rather 
than in physics and we turn to this in the  section that follows.
\beginsection{The Yang--Mills equations and four manifold theory}
\smalltitle{Donaldson's work}
In the 1980's interest in instantons continued strongly but there was a 
most striking result proved by Donaldson \ref{1} (1983) which used the 
Yang-Mills instantons to make a fundamental advance in the topology of 
four manifolds. 
\par
Donaldson's result concerned simply connected compact closed four 
manifolds $M$. We shall now give a short account of some of the result's 
main features so that the reader may be better able to appreciate its 
significance.
\par
In topology  one  distinguishes three types of manifold $M$: 
topological, piecewise-linear and differentiable (or smooth) 
which we can denote 
when necessary by
$M_{TOP}$, $M_{PL}$ and $M_{DIFF}$ respectively. There are topological 
obstacles to the existence of PL and DIFF structures on a given 
topological manifold $M$. The nature of these obstacles is quite well understood in dimension 5 and higher but, in dimension 4, the situation is quite different and much more difficult to comprehend. It is for 
this dimension that Yang--Mills theories and Donaldson's work 
have made such an important contribution. 
\par
On the subject of the importance of Yang--Mills theories for 
obtaining these  results Donaldson and Kronheimer \ref{1} (p. 27) 
have said the following in favour of Yang--Mills theory.
{\beginquote
These geometrical techniques will 
then be applied to obtain the differential--topological results
mentioned above. It is precisely this departure from standard techniques
which has led to the new results, and at present there is no way known 
to produce results such as these which does not rely on 
Yang--Mills theory.
\endquote}
\smalltitle{Donaldson and simply connected four manifolds}
We consider here compact closed four manifolds $M$.
For a simply connected four manifold $M$, $H_1(M;{\bf Z})$  and 
$H_3(M;{\bf Z})$  vanish and the non-trivial homological information is 
concentrated in the middle dimension in $H_2(M;{\bf Z}) $. A central 
object then is the {\it intersection form} defined by 
$$q(\alpha,\beta)=(\alpha\cup\beta)[M],\quad \alpha,\beta\in 
H_2(M;{\bf Z}) \no$$
with $\cup$ denoting cup product so that  
$(\alpha\cup\beta)[M]$ denotes the integer obtained by evaluating
$\alpha\cup\beta$  on the generating cycle $[M]$ of 
$H_4(M;{\bf Z})$ on $M$. Poincar\'e duality implies that the 
intersection  form  is always 
{\it non-degenerate} over ${\bf Z}$ and so
 has $\det q=\mp 1$---$q$ is then called unimodular. 
Also we refer to $q$, as {\it even}  if all its diagonal
entries are even, and as {\it odd}  otherwise.
A very powerful result of Freedman \ref{1} can now be called on---the 
intersection form $q$ very nearly determines the homeomorphism 
class of a simply connected $M$, and actually only fails to do so 
in the odd case where there are still just two possibilities. Further 
{\it every}  unimodular quadratic form occurs as the intersection 
form of some manifold.
\par
The relevant theorem is 
\theorem{Freedman \ref{1} (1982)} {A simply connected 4-manifold $M$ with even  intersection 
form $q$ belongs to a unique  homeomorphism class, while if 
 $q$ is odd there are precisely two 
non-homeomorphic $M$ with $q$ as their intersection form.}
An illustration of the impressive nature of Freedman's work is readily
available.  Recollect that the Poincar\'e conjecture in four dimensions 
is the statement that any homotopy 4-sphere, $S^4_h$ say, is actually 
{\it homeomorphic} \rm to the standard sphere $S^4$. Now  $S^4$ has 
trivial cohomology in two dimensions so its intersection form $q$ is 
the zero quadratic form which  we write  as  $q=\emptyset$.
But $S_h$, having the same homotopy type as 
$S^4$, has the same cohomology as $S^4$. So any  $S^4_h$
also has intersection form $q=\emptyset$. 
But Freedman's result says that for
a simply connected $M$ with even $q$ there is only {\it one 
homeomorphism class} for $M$, therefore $S^4_h$  {\it homeomorphic} to 
$S^4$ and we have 
established the conjecture. Incidentally this means that the Poincar\'e 
conjecture has now been proved for all $n$  except $n=3$---the case 
originally proposed by Poincar\'e.
\par
Now we come to Donaldson's work which concerns 
smoothability of four manifolds; one should also note that, when $q$ is a 
definite quadratic form, a choice of orientation can always render $q$ 
{\it positive} definite. Then we have the following theorem
\theorem{Donaldson \ref{1} (1983)} {A simply connected, smooth 
4-manifold, with positive definite intersection form $q$ has the 
property that $q$ is always 
diagonalisable over the integers to $q=diag(1,\ldots,1)$}
 Immediately one 
can go on to deduce that 
no simply connected, 4-manifold for which $q$ is even and positive definite 
can be smoothed! For example the Cartan matrix for the 
exceptional Lie algebra $e_8$ is given by 
$$E_8=\left(\matrix{2&-1&0&0&0&0&0&0\cr
                 -1&2&-1&0&0&0&0&0\cr
                  0&-1&2&-1&0&0&0&0\cr
                  0&0&-1&2&-1&0&0&0\cr
                  0&0&0&-1&2&-1&0&-1\cr
                  0&0&0&0&-1&2&-1&0\cr
                  0&0&0&0&0&-1&2&0\cr
                  0&0&0&0&-1&0&0&2\cr}\right)\no$$

Freedman's result guarantees that  there is a  manifold $M$ with 
intersection form $q=E_8\oplus E_8$. However Donaldson's theorem forbids 
such a manifold\foot{The reader may wonder why 
we did not discuss the four manifold with the simpler intersection form 
$q=E_8$. This manifold of course exists.  It  is not 
smoothable but this fact is due to a much older  result of Rohlin \ref{1} 
(1952) concerning smoothability and the signature of $q$.  Rohlin's 
theorem only provides a necessary condition for  smoothability, this is 
that the signature of an even $q$ must be  divisible by 16. The lack of sufficiency of this condition is shown by the example of $q=E_8\oplus E_8$ since one can verify that the signature of $q$ in this case is divisible by 16.} 
from existing smoothly.  
Before Donaldson's work surgery techniques had been extensively 
used to try to construct  smoothly the manifold with intersection form 
$E_8\oplus E_8$. We 
can now see that these techniques were destined to fail.
\par
In fact, in contrast to Freedman's theorem, which allows {\it all} 
unimodular quadratic forms to occur as the intersection form of some topological manifold, 
Donaldson's theorem says that in the positive definite, smooth, case only
{\it one }  quadratic form is allowed, namely the identity $I$. 
\par
One of the most striking aspects of Donaldson's work is that his proof uses
the Yang--Mills equations. 
We can only outline what is involved here, 
for more details cf. Donaldson and Kronheimer \ref{1},  Freed and 
Uhlenbeck \ref{1} and Nash \ref{1}. 
\par
In brief then let $A$ be a connection on a principal $SU(2)$-bundle 
over a simply connected 4-manifold $M$ with positive definite 
intersection form. If $S$
 is the usual  Euclidean Yang--Mills action $S$ of \docref{ymaction} 
one has
$$S=\Vert F\Vert^2=-\int_M\,tr(F\wedge *F)\no$$
Now given one instanton $A$ which minimises $S$ one can   
perturb about $A$ in an attempt to find more instantons. When this  is 
done  the space of all instantons can be fitted together to form a 
global {\it moduli space}  of finite dimension.   
For the instanton with $k=1$ which provides the absolute minimum of $S$, 
this moduli space ${\cal M}_1$, say, is a non-compact space of  
dimension  5, with singularities.
\par
We can now summarise the logic that is used to prove Donaldson's 
theorem: there
 are very strong 
relationships between $M$ and the moduli space ${\cal M}_1$; for 
example, let $q$ be regarded as an
$n\times n$ matrix with precisely $p$ unit eigenvalues (clearly $p\le n$ and 
Donaldson's theorem is just the statement that $p=n$), then
${\cal M}_1$ has precisely $p$ singularities which look like cones on the space 
${\bf C} P^2$. These combine to produce the result that the 4-manifold $M$ has the 
same topological signature $Sign\,(M)$ as $p$ copies of ${\bf C} P^2$; now $p$ 
copies of ${\bf C} P^2$ have signature $a-b$ where $a$ of the ${\bf C} P^2$'s are 
oriented in the usual fashion and $b$ are given the opposite orientation. Thus 
we have
$$Sign\,(M)=a-b\no$$
Now the definition of $Sign\,(M)$ is that it is the signature $\sigma(q)$ of the 
intersection form $q$ of $M$. But since, by assumption, $q$ is positive 
definite $n\times n$ then $\sigma(q)=n=Sign\,(M)$. So we can write
$$n=a-b\no$$
However, $a+b=p$ and $p\le n$ so we can  assemble this information in the 
form
$$n=a-b,\qquad p=a+b\le n\no$$
but one always has $a+b\ge a-b$ so now we have
$$n\le p\le n \Rightarrow p=n\no$$
and we have obtained Donaldson's theorem.
\beginsection{Physics and knots revisited---the Jones polynomial}
\smalltitle{Three manifolds and Floer, Jones and Witten}
In section one we discussed knots in our material  on the 
nineteenth century. It is now time to return to this subject. 
\par
In 1985 Jones \ref{1} made a great step forward in knot theory by 
introducing a new polynomial invariant of knots (and links), now known 
as the {\it Jones polynomial} and denoted by $V_L(t)$, where $L$ denotes 
the knot or link and $t$ is a real variable. Knot invariants of this 
kind had proved 
hard to find: the original one  was that of Alexander \ref{1} (1928), denoted by $\Delta_L(t)$.
\par
 The 
Jones polynomial originates in  certain finite dimensional von Neumann 
algebras which Jones denotes by $A_n$. A point of physical interest here 
is that, as Jones observed in his paper, D. Evans pointed out that 
some representations of these $A_n$'s had already been constructed 
in the physics literature in statistical mechanics, the relevant 
reference (which Jones gives) being Lieb and Temperley \ref{1}. 
The statistical mechanics concerns the Potts and ice-type models, 
cf. Baxter \ref{1}. 
This leads one to  speculate that the combinatorial structure of some 
models in statistical mechanics has a {\it topological origin}; this 
does seem to be borne out by subsequent work.
\par
The Jones polynomial proved powerful enough to decide many of the 
longstanding Tait conjectures on knots which we referred to in section 
one. The next event of joint topological and physical interest was a 
result by Witten \ref{2--3} (1988)
which gave a completely new formulation (and generalisation) 
of the Jones polynomial in terms of a certain kind of quantum field 
theory, nowadays known as a {\it topological quantum field theory}.
\par
A vital ingredient in this whole story is the work of Floer 
\ref{1--2} (1987--88) on a new homology invariant of three manifolds 
constructed from considerations of gauge theory and instantons.  
We shall meet this work again in section 8. We  
mention it now because of its influence on subsequent work. For the 
moment we just need to inform the reader that Floer considers the 
critical point behaviour of the function $f$ where $f$ depends on an 
$SU(2)$ connection $A$: 
$f$ is simply the Chern--Simons function obtained by integrating the 
Chern--Simons secondary characteristic class over a closed three 
manifold $M$. If ${\cal A}$ denotes the space of all connections 
$A$, we have 
\eqlabel{\csfunction} 
$$\eqalign{f:\,&{\cal A}\longrightarrow{\bf R}\cr
               &A\longmapsto f(A)\cr
\hbox{with}\qquad f(A)=&-
{1\over8\pi^2}\int_M\,tr(A\wedge dA+{2\over3}A\wedge A\wedge A)\cr}\no$$
Then from a very clever study of the Morse theory of this 
function $f$, whose domain is the {\it infinite dimensional} space 
$\cal A$, Floer 
obtains new homology groups $HF(M)$ known as {\it Floer } 
homology groups associated to the three manifold $M$.
\par
In 1987 Atiyah \ref{3} speculated that there was a relation between Floer's work 
and the Jones polynomial:  Towards the end of this paper we find  the 
following
{\beginquote
Finally let me list a few of the major problems that 
are still outstanding in the area. ..........   
More speculatively, I would like to end with
\par\noindent
      4) Find a connection with the link invariants of Vaughan Jones 
[11].
\par
 As circumstantial evidence that this is reasonable I will list some 
properties shared by Floer homology and the Jones polynomial.
\item{(i)} both are subtle 3-dimensional invariants,
\item{(ii)} they are sensitive to orientation of 3-space 
(unlike the Alexander polynomial),
\item{(iii)} they depend on Lie groups: $SU(2)$ in the first instance 
but capable of generalisation,
\item{(iv)} there are 2-dimensional schemes for computing these 3-dimensional invariants,
\item{(v)} whereas the variable in the Alexander polynomial corresponds to $\pi_1(S^1)$, the variable in the Jones polynomial appears 
to be related to $\pi_3(S^3)$, the origin of ``instanton numbers'',
\item{(vi)} both have deep connections with physics, specifically 
quantum field theory (and statistical mechanics).
\endquote}
In 1988 Witten \ref{2} rose to this challenge and found the relation 
that Atiyah had suspected existed. The content of Witten \ref{3} is 
described by its author with a certain amount of understatement. He 
says
{\beginquote
In a lecture at the Hermann Weyl Symposium last year \ref{1}, Michael 
Atiyah proposed two problems for quantum field theorists. The first 
problem  was to give a physical interpretation for Donaldson theory. The 
second problem was to find an intrinsically three dimensional definition 
of the Jones polynomial of knot theory. I would like to give a flavour 
of these two problems.
\endquote}
Our next task is to have a look at the methods that Witten used.
\smalltitle{Topological quantum field theories}
A topological quantum field theory (also called simply a {\it topological 
field theory}) is one which, at first sight, may seem
trivial physically: it has an action with no metric dependence. 
The absence of  a metric means that there are no distance 
measurements or forces and so no conventional dynamics. The Hamiltonian 
$\ham$ of the theory has only zero eigenstates and the Hilbert space of 
the theory is usually finite dimensional. The theory can, however, be 
non-trivial:  its non-triviality  is reflected in the existence of 
{\it tunnelling} between vacua.  
\par
The particular action chosen by Witten for obtaining the Jones polynomial 
was the well known Chern--Simons action $S$ given by
$$S={ik\over4\pi g^2}\int_M\,tr(A\wedge dA+{2\over3}A\wedge A\wedge 
A),\; k\in {\bf Z}\no$$
where  $A$ is an $SU(2)$ connection or gauge field and $M$ is 
a closed, compact three dimensional manifold. 
The partition function for this quantum field theory is $Z(M)$ where
$$Z(M)=\int{\cal DA}\,\exp\left[-{ik\over4\pi g^2}\int_M\,tr(A\wedge dA+{2\over3}A\wedge A\wedge A)
\right]\no$$
\par
This partition function itself is an invariant---the 
{\it Witten invariant}---of the three manifold $M$; however, at present, 
we want to study knots:
knots enter in the following way. Consider a closed curve $C$ embedded 
in $M$ so as to form a knot, $K$, say. Now ones takes the connection $A$, 
parallel transports it around  $C$ and constructs the holonomy 
operator $PT(C)$  as described above in the discussion of the 
Aharonov--Bohm effect in section 2. This time we have a 
non-Abelian connection and to obtain a gauge invariant operator 
we must  take 
the trace of $PT(C)$ giving what is called a {\it Wilson line}; we denote this by $W(R,C)$
where
$$W(R,C) =tr\,P\exp\left[\int_C\,A\right]\no$$ 
and $R$ denotes the particular representation carried by $A$.
There is a natural correlation 
function associated   with this knot namely
$$<W(R,C)>\,={1\over Z(M)}\int{\cal DA} \, W(R,C) \exp\left[-{ik\over4\pi g^2}\int_M\,tr(A\wedge dA+{2\over3}A\wedge A\wedge A)
\right]\no$$
and Witten \ref{2--3} shows that this determines the Jones polynomial 
$V_K(t)$ of the knot.
\par
Further if one has not {\it one } curve $C$ but a number of them, say
$C_1, \ldots C_p$ then one has a $p$ component link $L$, say, whose Jones polynomial $V_L(t)$ is determined by a multiple correlation function
of $p$ Wilson lines given by
$$\eqalign{&<W(R_1,C_1)\cdots W(R_p,C_p)>\,={1\over Z(M)}\int{\cal DA} \,\cr
&W(R_1,C_1) \cdots W(R_p,C_p)
\exp\left[-{ik\over4\pi g^2}\int_M\,tr(A\wedge dA+{2\over3}A\wedge A\wedge A)
\right]\cr}\no$$
\par
A nice thing that happens  if we step backwards slightly 
to the {\it Abelian} case is that one can recover the Gauss linking 
number: In the Abelian case $A$ is just a $U(1)$ connection and 
$S(A)$ becomes only quadratic in $A$ giving (we have set $g=1$)
$$S(A)={ik\over4\pi}\int_M\,tr(A\wedge dA)\no$$
so that
$$\eqalign{&<W(R_1,C_1)\cdots W(R_p,C_p)>\,=\cr
&\quad{1\over Z(M)}\int{\cal DA}\;W(R_1,C_1)\cdots W(R_p,C_p)
\exp\left[-{ik\over4\pi}\int_M\,tr(A\wedge dA)\right] \cr}\no$$
\par
The quadratic action, together with the exponential dependence on $A$ of a 
Wilson line, allows the entire integrand to be written as a Gaussian after 
completing the square. 
The calculation of the functional integral rests just 
on the calculation of a Green's function which, for $M=S^3$ (to which we now specialise), is an elementary 
computation. The result is that
\eqlabel{\abeliancase} 
$$\eqalign{&<W(n_1,C_1)\cdots W(n_p,C_p)>\,=\cr
&\quad\exp\left[{i\over 4k}\epsilon_{ijk}\sum_{l,m=1}^p 
n_ln_m\int_{C_l}\,dx^i\int_{C_m}\,dy^j\,
{(x-y)^k\over\vert x-y\vert^3}\right]\cr}\no$$
with $x^i$ and $y^j$ local coordinates on the knots $C_l$ and $C_m$. 
\par
We easily recognise the basic 
integral 
$${\epsilon_{ijk}\over4\pi}\int_{C_l}\,dx^i\int_{C_m}\,dy^j\,
{(x-y)^k\over\vert x-y\vert^3}\no$$ 
in \docref{abeliancase} as the linking number 
of Gauss that we met in section 1; and we note that we have met it again
in a physical context. Incidentally for another physical context in which the 
linking number appears  cf.  Wilczek and Zee \ref{1} (1983); 
in this paper a connection is made between the spin-statistics 
properties of particles and topology\foot{On this latter topic there is 
 more work. For some examples cf. Balachandran et al. \ref{1}, Berry and Robbins \ref{1}, Finkelstein and Rubinstein \ref{1}, 
Mickelsson \ref{1}, Tscheuschner 
\ref{1}, and references therein.}
\par
In the non-Abelian case 
we can also obtain a quadratic functional integral by 
studying the limit of small coupling $g$, or, completely equivalently, 
the limit of  large $k$. For any  topological field theory such a limit 
 has considerable significance.  When this limit is evaluated 
 for this theory one obtains another differential topological invariant:  
the Ray--Singer torsion of the connection $A$ on $M$, 
cf. Witten \ref{2--3}.
\par
Numerous topological field theories are now studied in the current 
literature, we shall meet another one in section 7;  indeed the whole 
notion of a topological quantum field theory has been axiomatised in 
Atiyah \ref{4}.
\par
Finally we point out some of the new and more general features of the
Witten approach to the Jones polynomial: Witten's definition is 
intrinsically three dimensional and not dependent on any two 
dimensional arguments for its validation. The group $SU(2)$ of the 
connection $A$ is not obligatory---it can be replaced by another Lie 
group $G$, say $G=SU(N)$; for the appropriate representation of $SU(N)$ 
this gives rise to a two variable generalisation of the Jones polynomial  
cf. Freyd, et al. \ref{1}.
There is an immediate 
generalisation to knots in {\it any} three manifold $M$ rather than the 
classical case of knots in $S^3$ (i.e. compactified ${\bf R}^3$). 
Invariants for three manifolds themselves immediately arise and so the 
theory is not really just one of knots (i.e. embeddings).
\beginsection{Yang--Mills and four manifolds once more}
\smalltitle{Donaldson again: polynomial invariants for four manifolds}
In the 1990's more progress was made in four dimensions with another 
result of Donaldson \ref{3} (1990); actually some of these results 
were announced considerably earlier in 1986 by Donaldson in his 
Field's medal address (cf. Donaldson \ref{2}).
\par
In section 5 we described Donaldson's use of the  moduli 
space ${\cal M}_1$  to derive smoothability results about $4$-manifolds. 
The space ${\cal M}_1$ only contains instantons with instanton number 
$k$ equal to one. In addition to this, by using all values of $k$ 
there exist  
moduli spaces $\instmodk$, $k=1,2,\ldots$ for instantons of any 
instanton number $k$. Donaldson's new invariants use all of the 
$\instmodk$ and in the process one obtains
powerful differential topological invariants of simply connected 
$4$-manifolds. Donaldson \ref{3} begins with 
{\beginquote
The traditional methods of geometric topology have not produced a clear
picture of the classification of smooth 4-manifolds. This gap has been 
partially bridged by methods using Yang--Mills theory or gauge theory. 
Riemannian manifolds carry with them an array of moduli spaces---finite 
dimensional spaces of connections cut out by the first order Yang--Mills 
equations. These equations depend on the Riemannian geometry of the 
4-manifold, but at the level of homology we find properties of the moduli 
spaces which do not change when the metric is changed continuously. Any 
two metrics can be joined by a path, so by default, these properties 
depend only on the differential topology of the 4-manifold, and furnish 
a mine of potential new differential topological invariants.
\endquote}
The ``mine of potential new differential topological invariants'', as 
Donaldson modestly puts it, is a reference to his new polynomial 
invariants. He goes on, in the same paper, to say
{\beginquote
Here we use 
infinite families of moduli spaces to define infinite numbers of 
invariants for simply connected manifolds with $b_2^+$ 
odd\foot{The number $b_2^+$ is 
defined to be the rank of the positive part of the intersection form.} 
and greater 
than 1. These invariants are distinguished elements in the ring:
$$S^*(H^2(X))$$
of polynomials on the cohomology of the underlying 4-manifold $X$.
\par
Equivalently, they can be viewed as symmetric multi-linear functions:
$$q:H_2(X;{\bf Z})\times\cdots\times H_2(X:{\bf Z})
\longrightarrow{\bf Z}$$
.......... Certainly one of the most striking facts is that we get 
infinitely many invariants for a single manifold. Discovering to what 
extent  these are independent (i.e. whether there are strong universal 
relations between them) appears to be an interesting target for future 
research.
\endquote}
\par
We just want to mention some results that have been obtained with the 
Donaldson invariants which serve to show that they are non-trivial and 
important. To be able to do this we must introduce some notation. 
\par
Let $M$ be a smooth, simply connected, orientable Riemannian 
four manifold without 
boundary and $A$ be an $SU(2)$ connection which is 
anti-self-dual\foot{We take anti-self-dual connections rather than 
self-dual connections so as to follow Donaldson's sign conventions.} 
so that 
$$F=-*F\no$$
Then  the dimension of the moduli space $\instmodk$ is 
the integer
$$\dim\instmodk=8k-3(1+b_2^+)\no$$ 
\par
A Donaldson invariant $q_d(M)$ is a symmetric 
integer polynomial of degree $d$ in the 2-homology $H_2(M;{\bf Z})$ of $M$
$$q_d(M):\underbrace{H_2(M)\times\cdots\times H_2(M)}_{d\rm\;factors}
\longrightarrow{\bf Z}\no$$
Given a certain map $m$ (cf. Donaldson \ref{3--4}  or 
Nash \ref{1})
$$m:H_2(M)\rightarrow H^2(\instmodk)\no$$ 
we use $m$ to define by $q_d(M)$  by using de Rham cohomology and differential forms.  Setting $d=\dim \instmodk/2$ we  
define $q_d(M)$ by 
$$\eqalign{q_d(M):&\,H_2(M)\times\cdots\times H_2(M)
\longrightarrow{\bf Z}\cr
           {}  &\, a_1\times\cdots \times a_d\longmapsto\int_{
\instmodkbar}\,m(a_1)\wedge\cdots\wedge m(a_d)\cr}\no$$
where $\instmodkbar$ denotes a compactification of the moduli space. We 
see that the $q_d(M)$ are symmetric integer valued polynomials of 
degree $d$ in 
$H^2(M)$, i.e. $q_d(M)\in Sym^d\,(H_2(M))\subset S^*(H(M))$; also, since 
$d=\dim\instmodk/2=(8k-3(1+b_2^+))/2$, we now understand why 
Donaldson required $b_2^+$ to be odd. 
\par
Now the Donaldson invariants are, {\it a priori}, not very easy to calculate 
since they require detailed knowledge of the instanton moduli space. However, 
if $M$ is a complex algebraic surface, a positivity argument shows that 
$$q_d(M)\not=0,\qquad\hbox{for }d\ge d_0\no$$ 
with $d_0$ some integer---in other words the $q_d(M)$ are all non-zero when 
$d$ is large enough. Conversely, if $M$ can be written as the connected 
sum
$$M=M_1\#M_2\no$$ 
where $M_1$ and $M_2$ both have $b_2^+>0$ then 
\eqlabel{\qvanishing} 
$$q_d(M)=0,\qquad\hbox{for all $d$}\no$$  
\par                                      
The $q_d(M)$ are {\it differential}  topological invariants rather than 
topological invariants; this means that they have the potential to 
distinguish homeomorphic manifolds which have distinct 
diffeomorphic structures. An example where the $q_d(M)$ are used to 
show that two homeomorphic manifolds are not diffeomorphic can be 
found in Ebeling \ref{1}. A possible  a physical context for 
this result can be found in Nash \ref{2}, 
 cf. too, Libgober and Wood \ref{1} for 
some earlier work related to Ebeling \ref{1} which was 
done before the $q_d(M)$ were defined. 
\par
The $q_d(M)$ can also be obtained from a topological quantum field theory as we shall now see below.
\smalltitle{Another topological field theory}
In 1988 Witten showed how to obtain the $q_d(M)$ as correlation functions in a 
BRST supersymmetric topological field theory.   We shall only 
give a brief statement of facts to give the reader some idea of 
what sort of action and physical fields are involved;   for 
a full account, cf. Witten \ref{4}. 
\par
The action $S$ for the theory is given by 
$$\eqalign{S=\int_M\,&d^4x
\sqrt{g}\,tr\left\{\right.{1\over4}F_{\mu\nu}F^{\mu\nu}+ 
{1\over4}F_{\mu\nu}^*F^{\mu\nu}+
{1\over2}\phi D_\mu D^\mu\lambda+iD_\mu\psi_\nu\chi^{\mu\nu}\cr
{}&-i\eta 
D_\mu\psi^\mu-{i\over8}\phi[\chi_{\mu\nu},\chi^{\mu\nu}]-
{i\over2}\lambda[\psi_\mu,\psi^\mu]-{i\over2}\phi[\eta,\eta]-
{1\over8}[\phi,\lambda]^2\left.\right\}\cr}\no$$
where $F_{\mu\nu}$ is the curvature of a connection $A_\mu$ and  
$(\phi,\lambda,\eta,\psi_\mu,\chi_{\mu\nu})$ are a collection of fields 
introduced in order to construct the right supersymmetric theory; $\phi$ and 
$\lambda$ are both spinless while the multiplet $(\psi_\mu,\chi_{\mu\nu})$ 
contains 
the components of a $0$-form, a $1$-form and a self-dual $2$-form 
respectively. The significance of this choice of multiplet is that the 
anti-instanton version of the instanton deformation complex 
used to calculate $\dim \instmodk$ contains precisely these fields. 
Even though $S$ contains a metric its 
correlation functions are independent of the metric $g$ so that $S$ can still 
be regarded as a topological field theory. This can be shown to follow from 
the fact that both $S$ and its associated energy momentum tensor 
$T\equiv(\delta S/\delta g)$ can be written as BRST commutators $S=\{Q,V\}$, 
$T=\{Q,V^\prime\}$ for suitable $V$ and $V^\prime$---cf. Witten \ref{4}.
\par
With this theory it is possible to show that the correlation functions are 
independent of the gauge coupling and hence we can evaluate them in a small 
coupling limit. In this limit the functional integrals are dominated by the 
classical minima of $S$, which for $A_\mu$ are just the instantons
$$F_{\mu\nu}=-F_{\mu\nu}^*\no$$
We also need $\phi$ and $\lambda$ to vanish for irreducible connections. 
If we expand all the fields around the minima up to quadratic terms and do the 
resulting Gaussian integrals, the correlation functions may be formally 
evaluated. Let us consider a correlation function 
$$<P>=\int\,{\cal DF}\,\exp[-S]\,P({\cal F})\no$$
where ${\cal F}$ denotes the collection of fields present in $S$ and $P({\cal 
F})$ is a polynomial in the fields. Now $S$ has been constructed so that the 
zero modes in the expansion about the minima are the tangents to the moduli 
space $\instmodk$; thus, if the ${\cal DF}$ integration is expressed as an 
integral over modes, all the 
non-zero modes may be integrated out first leaving a \it finite dimensional 
\rm integration over $\dim\instmodk$. The Gaussian integration over the 
non-zero modes is a Boson--Fermion ratio of determinants, a ratio which 
supersymmetry constrains to be $\mp1$ since Bosonic and Fermionic eigenvalues 
are equal in pairs. 
This amounts to expressing $<P>$ as 
$$<P>=\int_{\instmodk}\,P_n\no$$
where $P_n$ is an $n$-form over 
$\instmodk$ and $n=\dim\instmodk$. 
  If the 
original polynomial $P({\cal F})$ is chosen in the correct way then 
calculation of $<P>$ reproduces evaluation of the Donaldson polynomials.
\par
The next breakthrough in the topology of four manifolds came from physics
and was due to Seiberg and Witten (1994) (cf. Seiberg and Witten 
\ref{1--2} and Witten \ref{5}) and it is the next 
topic to which we turn.
\smalltitle{Physics again: Seiberg--Witten theory and four manifolds}
In Donaldson \ref{4} we find the most upbeat introduction to a review 
article on the Seiberg--Witten equations; it gives some idea of the 
excitement and the power of the methods associated with this 
latest breakthrough.
{\beginquote
Since 1982 the use of gauge theory, in the shape of the Yang--Mills 
instanton equations, has permeated research in 4-manifold topology. 
At first this use of differential geometry and differential equations 
had an unexpected and unorthodox flavour, but over the years the ideas 
have become more familiar; a body of techniques has built up through the 
efforts of many mathematicians, producing results which have uncovered 
some of the mysteries of 4-manifold theory, and leading to
substantial internal conundrums within the field itself. In the last 
three months of 1994  a remarkable thing happened: this research was turned on its head by the introduction of a new kind of 
differential--geometric equation by Seiberg and Witten: in the space of 
a few weeks long--standing problems were solved, new and unexpected results were found, along with simpler new proofs of existing ones, and new 
vistas for research opened up. This article is a report on some of these 
developments, which are due to various mathematicians, notably 
Kronheimer, Mrowka, Morgan, Stern and Taubes, building on the 
seminal work of Seiberg [S] and Seiberg and Witten [SW].
\endquote}
We shall say a little about both the physics and the mathematics 
relating to the Seiberg--Witten equations; however we shall make the 
remarks about the mathematics here but leave the remarks 
about the physics until section 9 where they fit in more 
naturally.
\par
Seiberg and Witten's work allows one to produce another physical theory, 
in addition to Witten \ref{4}, with which to compute Donaldson invariants.
In Witten \ref{4}, as just described above, Donaldson theory is obtained 
from a twisted $N=2$ supersymmetric Yang--Mills theory. Seiberg and 
Witten  produce a duality which amounts to an equivalence between the 
{\it strong coupling} limit of this $N=2$ theory and the 
{\it weak coupling} limit of a theory of {\it Abelian} monopoles. 
This latter theory is much easier to compute with leading (on the 
mathematical side) to the advances described in Donaldson \ref{4} and 
Witten \ref{5}.
\par
If we choose an oriented, compact, closed, Riemannian manifold $M$ then 
the data we need for the Seiberg--Witten equations are a connection $A$ 
on a line bundle $L$ over $M$ and a  ``local spinor'' field $\psi$. The 
Seiberg--Witten equations are then
\eqlabel{\seiwitteqs}
$$\dirac\psi=0,\qquad F^+_A=-{1\over 2}\longbar{\psi} \Gamma\psi\no$$
where $\dirac$ is the Dirac operator and $\Gamma$ is made from the 
gamma matrices $\Gamma_i$ according to 
$\Gamma={1\over2}[\Gamma_i,\Gamma_j]dx^i\wedge dx^j$. We call $\psi$
a local spinor because global spinors may not exist on $M$; however 
orientability guarantees that a $spin_{\bf c}$  structure does exist 
and $\psi$ is the appropriate section for this $spin_{\bf c}$ structure.
We note that $A$ is just a $U(1)$ Abelian connection and so $F=dA$, 
with $F^+$ just being the self-dual part of $F$.
\par
We shall now have a brief look at one example of a new result using the 
Seiberg-Witten equations. 
The equations clearly provide the absolute minima for the action
$$S=\int_M\left\{\vert \dirac \psi\vert^2+
{1\over2}\vert F^++{1\over 2}\longbar{\psi}\Gamma\psi\vert^2\right\}\no$$
If we use a Weitzenb\"ock formula to relate the Laplacian 
$\nabla_A^*\nabla_A$ to $\dirac^*\dirac$ plus curvature terms we find 
that $S$  satisfies
$$\eqalign{\int_M\left\{\vert \dirac \psi\vert^2+
{1\over2}\vert F^++{1\over 2}\longbar{\psi}\Gamma\psi\vert^2\right\}&=
\int_M\left\{\vert\nabla_A\psi\vert^2+{1\over 2}\vert F^+\vert ^2+
{1\over 8}\vert \psi\vert^4 +{1\over 4}R\vert \psi\vert^2\right\}\cr
&=\int_M\left\{\vert\nabla_A\psi\vert^2+{1\over 4}\vert F\vert ^2+
{1\over 8}\vert \psi\vert^4 +{1\over 4}R\vert 
\psi\vert^2\right\}+\pi^2 c_1^2(L)\cr}\no$$
where  $R$ is the scalar curvature of $M$.  The action 
now looks like one  for monopoles---indeed in Witten \ref{5}, Witten 
refers to \docref{seiwitteqs} as ``the monopole equations''.
But now suppose that $R$ is 
{\it positive} and that the pair $(A,\psi)$ is a solution to the 
Seiberg--Witten equations: then the LHS is zero and all the integrands 
on the RHS are positive so the solution  must 
obey $\psi=0$ and $F^+=0$. It turns out that if $M$ has $b_2^+>1$  then 
a perturbation of the metric can  preserve the positivity of $R$ but change $F^+=0$  to be plain $F=0$ rendering  the connection $A$  flat. 
Hence, in these circumstances, the solution $(A,\psi)$ is the trivial 
one. This means that we have a new kind of vanishing theorem in 
four dimensions.
\theorem{Witten \ref{3} (1994)}{No four manifold with $b_2^+>1$ and 
non-trivial Seiberg--Witten invariants admits  a metric of 
positive scalar curvature.}
We referred just now to the Seiberg--Witten invariants
and unfortunately we cannot define them here. However we do want to 
say that they are rational numbers $a_i$ and there are formulae relating 
the Donaldson polynomial invariants $q_d$ to the $a_i$. 
\par
Many more new results have been found involving, for example, symplectic 
and K\"ahler manifolds, cf. Donaldson \ref{4}; the story, however, is 
clearly not at all finished. 
\beginsection{Dynamics and topology since Poincar\'e}
\smalltitle{Dynamical systems and Morse theory}
In this section we want to return  to Poincar\'e and consider 
 that part of his topological legacy which sprang from his 
work on dynamics. We shall only be able to look at two areas and these 
are the theory of {\it dynamical systems} and {\it Morse theory}. This 
is, of necessity, somewhat selective, nevertheless these two subjects do 
represent mainstream developments which descend directly from 
Poincar\'e's work on dynamics and topology\foot{Some more detailed historical material, of relevance here, is that of Dahan-Dalm\'edico 
\ref{1--2}.}. It should also be 
borne in mind that there is a large overlap between the two subjects.
\par
It is still true that the $n$-body problem attracts much attention from 
mathematicians, including those using topological techniques.  
A few references of interest here are Smale \ref{2--3} (1970) and 
Saari and Xia \ref{1} (1996). 
\smalltitle{Dynamical systems}
 Poincar\'e's pioneering work on celestial mechanics prepared the way for
 the present day subject of dynamical systems with Birkhoff as the actual 
founder. In this subject one 
studies an immense diversity of sophisticated mathematical problems 
usually no longer connected with celestial or Newtonian mechanics. 
\par
A very rough idea of what is
involved goes as follows: Recall first that the celestial mechanics of $n$
bodies has a motion that is described by a set of differential equations
together with their initial data. One then varies the initial data 
and asks how the motion changes. 
\par
Now the modern mathematical setting is to view the
orbits of the $n$ bodies as integral curves for their associated
differential equations. Then one regards the {\it qualitative study} of 
the orbits as being a study of the {\it global geometry} of the space of
integral curves as their  initial conditions vary smoothly. Integral
curves $\gamma(t)$ are associated with vector fields $V(t)$ via the
differential equation
$${d\gamma(t)\over dt}=V(\gamma(t))\no$$
\par
Hence one is now studying the vastly more general subject of the global
geometry of the space of flows  of a vector field $V$ on a manifold $M$.
\par
Two notions play a distinguished part in the theory of
dynamical systems: closed integral curves and singular points.
It is natural to regard two flows on $M$ as {\it equivalent} if there is a
homeomorphism of $M$ which takes one flow into the other; one can also
 insist that this homeomorphism is smooth, i.e. a diffeomorphism. 
Finally an equivalence class of flows in the homeomorphic sense is a
{\it topological dynamical system},  and one in the diffeomorphic sense is
 {\it a smooth, or differentiable, dynamical system}. 
\par
As we explained in section 1 Birkhoff proved Poincar\'e's geometric theorem in 1913; a subsequent piece of work of great importance and influence 
was Birkhoff's proof of what is called his {\it ergodic theorem} in 1931, cf. Birkhoff \ref{2}.
\par
The subsequent blossoming of ergodic theory can be dated from this time.
 Ergodic
theory originates largely in nineteenth century studies in the kinetic
theory of gases. However it has now been axiomatised, expanded, 
refined and reformulated so that it has links with many parts of 
mathematics as well as retaining some with physics. 
\par
Some dynamical systems exhibit ergodic
behaviour, a notable class of examples being provided by 
{\it geodesic flow} on surfaces of constant negative curvature. This 
involves too the study of the flows by a discrete encoding known 
as symbolic dynamics, use of   one dimensional interval maps 
cf. Bedford et al. \ref{1}. Classical and quantum chaos, 
and the distinction between the two, are also studied in this context.
\par
A vast body of the theory of dynamical systems concerns {\it Hamiltonian
systems}. These of course have their origin in ordinary dynamics but
exist now in a much wider context. To have a Hamiltonian system  $M$ 
must be even dimensional, possess a 
Hamiltonian function
$$H:M\longrightarrow{\bf R}\no$$
 and have 
a closed non-degenerate  symplectic form $2$-form  $\omega$ 
appropriately related to $H$. 
The perturbation theory of Hamiltonian systems underwent an enormous
development in the 1950's and 1960's with the work particularly of 
Kolmogorov, Arnold and Moser and the creation of what is known as 
KAM theory (cf. Broer et al.  \ref{1}).
\par
Gradient dynamical systems were used by Thom \ref{1--3} (1969--71) 
in his work on what is now called {\it Catastrophe theory}. Thom 
took the system
$${d\gamma(t)\over dt}=grad\,V(\gamma(t))\no$$
where $V$ is a potential  function. Thom classified the possible critical
points of $V$ into seven types known as the seven elementary 
catastrophes; he then proposed to use these dynamical systems as 
models for the behaviour of a large class of 
physical, chemical and biological systems. In many cases the 
models are not at all adequate,  
nevertheless, there are some
successes. However the seminal
nature of Thom's work is clear though as it is the beginning of the
classification theory for singularities. In this connection there are 
the two results of Arnold \ref{1--2} (1973, 1978) which closely 
relate the classification of  singularities to the Weyl groups of the 
various compact simple Lie groups.
\smalltitle{Morse theory: the topology of critical points}
The aim in Morse theory is to study the relation between critical points and 
topology. More specifically one extracts topological information from a study 
of the critical points of a smooth real valued function 
\eqlabel{\smoothfunction}
$$f:M\longrightarrow{\bf R}\no$$
where $M$ is a compact manifold usually without boundary. For a suitably 
behaved class of functions $f$ there exists quite a tight relationship between 
the number and type of 
critical points of $f$ and topological invariants of $M$ such as the 
Euler--Poincar\'e characteristic, the Betti numbers and other cohomological data.
This relationship can then be used in two ways: one can take certain special 
functions whose critical points are easy to find and use this information to 
derive results about the topology of $M$; on the other hand, if the topology 
of $M$ is well understood, one can use this topology to infer the existence 
of critical points of $f$ in cases where $f$ is too complex, or too 
abstractly defined, to allow a direct calculation.
\par
Taking a function $f$ the equation for its critical points is
$$df=0\no$$
We {\it assume} that all the critical points $p$ of $f$ are non-degenerate; 
this means that the Hessian matrix $Hf$ of second derivatives is invertible at 
$p$, or 
$$\det Hf(p)\not=0\quad\hbox{ where }\quad Hf(p)=\left[\left.
{\partial^2 f/ \partial x^i\partial 
x^j}\right\vert_{p}\right]_{n\times n}\no$$
Each critical point $p$ has an index $\lambda_p$ which is defined to be the 
number of {\it negative}  eigenvalues of $Hf(p)$.  We can then 
associate to the function $f$ and 
its critical points $p$ the Morse 
series $M_t(f)$ defined by
$$M_t(f)=\sum_{\hbox{all}\; p} t^{\lambda_p}=\sum_i m_i t^i \no$$ The topology of $M$ now enters via $P_t(M)$: the 
Poincar\'e series of $M$.  We have
$$P_t(M)=\sum_{i=0}^n\dim H^i(M;{\bf R})t^i=\sum_{i=0}^n b_i t^i \no$$
The  fundamental result of Morse theory is the statement that 
$$M_t(f) \ge P_t(M)\no$$
from which so many things follow, to mention just one simple example
$$m_i\ge b_i\no$$
showing that the number of critical points of index $i$ is bounded below 
by the Betti number $b_i$.
\par
Successful applications of Morse theory in mathematics are impressive and 
widespread; a few notable examples are the proof by Morse \ref{1}  (1934)
that there exist infinitely many geodesics joining a pair of points on a  
sphere $S^n$ endowed with any Riemannian metric, Bott's  \ref{1--2} 
(1956, 1958)
proof of his celebrated periodicity theorems on the homotopy of Lie groups, 
Milnor's construction \ref{1} (1956) of the first exotic spheres, 
and the proof by Smale \ref{1} (1961) of the Poincar\'e conjecture for 
$\dim M\ge 5$. 
\smalltitle{Supersymmetric quantum mechanics and Morse theory}
In 1982  Witten \ref{6}  constructed a quantum mechanical point of 
view on Morse theory which has proved very influential. It also 
provides a  
point of departure for the Floer theory discussed below. 
In summary Witten gives a quantum mechanical proof of the Morse 
inequalities on $M$; however an important extra feature is that 
the cohomology of $M$ is also explicitly constructed. 
\par
Witten takes as Hamiltonian ${\cal H}$: the Hodge Laplacian on forms, 
i.e. one has
$$\ham={\textstyle \bigoplus\limits_{p\ge0}}\, 
\Delta_p\equiv 
{\textstyle \bigoplus\limits_{p\ge0}}\,(d d^*+d^*d)_p\no$$ 
The  Bosons and Fermions of the supersymmetry are the spaces 
 $H^B$ and $H^F$ formed by even  and odd forms while the 
supersymmetry algebra is generated by two operators $Q_1$ and $Q_2$ 
which are constructed from $d$ and $d^*$. The appropriate definitions are
$$\eqalign{Q_1&=(d+d^*)\qquad Q_2=i(d-d^*)\cr
           H^B&={\textstyle \bigoplus\limits_{p\ge0}}\,
                 \Omega^{2p}(M)  \qquad
     H^F={\textstyle \bigoplus\limits_{p\ge0}}\,
                 \Omega^{2p+1}(M)\cr}\no$$
\par
A Morse function $f$ is now incorporated into the model without 
changing the supersymmetry algebra by replacing $d$ by $d_t$ where
$$d_t=e^{-ft}d e^{ft}\no$$
It is a routine matter to verify that this conjugation of $d$ by $e^{ft}$ 
leaves the algebra unchanged.  The proof of the Morse inequalities rests on an 
analysis of the spectrum of the associated Hamiltonian, which is now
$$\ham_t=d_td^*_t+d^*_td_t={\textstyle \bigoplus\limits_{p\ge0}}\,
\Delta_p(t)\no$$
\par
One needs additional physics to carry out the rest of the work. 
It turns out that this all comes from the consideration of  quantum mechanical tunnelling between critical points of $f$. One  
considers $grad\;f$ as a vector field on $M$ and then studies 
the integral curves of this vector  field,  
that is the  solutions $\gamma(t)$ of the differential equation
$${d\gamma(s)\over ds}=-grad\,f(\gamma(s))\no$$
We can give no more details here but, as has been emphasised by Witten \ref{6}, 
these ideas are applicable in quantum field theory as well as in  
quantum mechanics. In that case one has to deal with functions in 
infinite dimensions and it was not long before a significant result 
along these lines emerged; this was the work of Floer \ref{1--2} 
(1988) which we now examine.
\smalltitle{Floer homology and Morse theory}
In section 6, cf. \docref{csfunction}, we referred to Floer's work and 
his Morse theoretic study
 of the function 
$$\eqalign{f:\,&{\cal A}\longrightarrow{\bf R}\cr
               &A\longmapsto f(A)\cr
\hbox{with}\qquad f(A)=&-
{1\over8\pi^2}\int_M\,tr(A\wedge dA+{2\over3}A\wedge A\wedge A)\cr}\no$$
We review now some of the details. 
\par
The critical points of $f$ are given by   
$$df(A)=0\no$$
where the exterior derivative is now taken to be acting in the 
space
${\cal A}$ of $SU(2)$ connections on $M$. If $A$ is such a critical 
point then we can write $A_t=A+t a$ and 
obtain
$$f(A_t)=f(A)-{t\over4\pi^2}\int_M\,tr(F(A)\wedge a)+\cdots\no$$
Hence we can conclude  that 
$$df(A)=-{F(A)\over4\pi^2}\no$$
and so the critical points of the Chern--Simons function are the 
{\it flat connections} on $M$. 
\par
As long as $\pi_1(M)\not=0$ then flat connections on $M$ are 
not trivial, since they can have non-zero holonomy round a non-trivial 
loop on $M$. The holonomy of each flat connection is an $SU(2)$ element parametrised 
by a loop on $M$; in this way it defines a representation of 
$\pi_1(M)$ in $SU(2)$ and the space of {\it inequivalent} such 
representations is  the quotient 
$$Hom\,(\pi_1(M),SU(2))/Ad\,SU(2)\no$$
\par
Having found a critical point Morse theory requires us to calculate 
its index and so we must also calculate the Hessian of $f$: the snag is 
that this gives an operator which is unbounded from below rendering the 
index  formally infinite. This is not entirely unexpected since 
we are working in infinite dimensions.
\par
Floer gets round this very cleverly by realising that he only needs a
{\it relative index} which he can compute via spectral flow and the 
Atiyah--Singer index theorem. He takes two critical points $A_P$ and 
$A_Q$ in ${\cal A}$ and joins then with a steepest descent path $A(t)$; i.e. a path   
which obeys the equation
$${dA(t)\over dt}=-grad\,f(A(t))\no$$
with $grad$ denoting the gradient operator on the space ${\cal A}$.  
The consequence of all this for the Morse theory 
construction is that he is able to construct a homology complex and 
associated homology groups $HF_p(M)$. However the topology of the 
situation 
dictates  that the relative Morse index of $f$ is 
only well defined $mod\,8$. This means that $HF_p(M)$ are graded 
$mod\,8$ and one only obtains eight 
homology groups:  $HF_p(M)$, $p=0,\ldots 7$; for more details cf. Nash 
\ref{1}.
\par
Morse theory has also been successfully applied to other problems in 
Yang--Mills theory. Some important papers are Atiyah and Bott \ref{1} 
(1982) on 
Yang--Mills theories on Riemann surfaces where an equivariant Morse 
theory was required, and Taubes \ref{1--2} (1988, 1985) on pure 
Yang--Mills theory 
and Yang--Mills theory for monopoles. In all these examples one has to 
grapple with the infinite dimensionality of the quantum field theory.
\smalltitle{Knots again}
Vassil'ev \ref{1--2} (1990) has developed an approach to knot theory 
using singularity theory. 
Vasil'ev constructs a huge new class of knot invariants and we shall now
give a sketch of what is involved.
\par
A knot is a smooth embedding of a circle into ${\bf R}^3$. So a knot gives a
map
$$f:S^1\longrightarrow{\bf R}^3\no$$
so that $f$ belongs to the space ${\cal F}$ where ${\cal
F}=\map(S^1,{\bf R}^3)$.
 Not all elements of ${\cal F}$ give knots since a  knot map $f$ is
not allowed to self-intersect or be singular. Let $\Sigma$ be the subspace of
${\cal F}$ which contains either self-intersecting or singular maps,
then the subspace of knots is the {\it complement}
$${\cal F} -\Sigma\no$$
Now any element of $\Sigma$ can be made smooth by a simple one parameter
deformation, hence $\Sigma$ is a {\it hypersurface} in $\cal F$ and is
known as the {\it discriminant}. As the discriminant $\Sigma$ wanders
through $\cal F$ it skirts along the edge of the complement ${\cal F} -\Sigma$ and
divides it into many different connected components. Clearly knots in
the same connected component can be deformed into each other and so are
equivalent (or isotopic). 
\par
Now any knot {\it invariant} is, by the previous sentence, a function
which is {\it constant} on each connected component of ${\cal F} -\Sigma$.
Hence the task of constructing all (numerical) knot invariants is the same as
finding all functions on ${\cal F} -\Sigma$  which are constant on each
connected component.  
But topology tells us at once that this is just 
the $0$-cohomology of ${\cal F} -\Sigma$. In other words
$$H^0({\cal F} -\Sigma)=\hbox{The space of knot invariants}\no$$ 
Vasil'ev  provides a method for computing most, and possibly all, of 
$H^0({\cal F} -\Sigma)$. There are connections, too, to physics    
cf. Bar-Natan, \ref{1}.
\beginsection{Strings, mirrors and duals} 
\smalltitle{String theory, supersymmetry and unification of interactions}
String theory has, by now, a fairly long history very little of 
which we can mention here. We shall, in the main, limit ourselves to 
remarks which relate in some way to topology. 
\par
Topology enters string theory at the outset because a moving string 
sweeps out a two  dimensional surface and, in quantum theory, all such 
surfaces must be summed over whatever their topology.
This leads to the Polyakov expression for  the partition function $Z$ 
of the Bosonic string which contains a sum over the genera $p$ of Riemann 
surfaces $\Sigma$. One has
$$Z=\sum_{genera}\int{\cal D}g{\cal D}\phi\, 
\exp\left[-{1\over2}\int_\Sigma<\partial\phi,\partial\phi>_g\right]
=\sum_{p=0}^{\infty}Z_p\no$$
where the functional integral is over all metrics $g$ on $\Sigma$ 
and the string's  position  $\phi$. We have
 $\phi\equiv \phi^\mu(x^1,x|^2),\,\mu=1,\ldots d$  so the $\phi^\mu$ 
can be thought of as specifying an embedding of $\Sigma$ in a 
$d$-dimensional space-time $M$. It is well known that the theory is only conformally invariant when $d=26$: the critical dimension; however if one 
includes Fermions then this critical dimension changes to $d=10$.
\par
 A string theory is also a two dimensional conformal field 
theory and this latter subject is very important for mathematics
as well as physics. It has been axiomatised in a very fruitful and influential way 
by Segal \ref{1} (1988). It involves important representation theory 
of infinite dimensional groups such as $LG=\map(S^1, G)$ and $Diff(S^1)$. 
We regret that we have been unable to trace its 
history in this article because of lack of space; its importance 
is immediately apparent when one reads the literature on string theory 
as well as that of many  statistical mechanical models, it is also 
a key notion used in calculation and conceptual work in string theory. 
Finally, closely connected to conformal field theories, are the subjects 
of Kac--Moody algebras, vertex operator algebras and quantum groups; 
these all have close connections with physics but, although they have 
topological aspects, their algebraic properties are more prominent 
and this, as well as considerations of space, is another reason why 
we have had to omit them from this essay. 
Conformal field theory is also a vital ingredient in the surgery 
argument used in Witten \ref{2--3} to compute the Jones polynomial 
and, viewed from this standpoint,  conformal field theories can be 
seen  to provide a   link between the infinite dimensional 
representational theory just mentioned and the topology of 
two and three manifolds.
\par
String theory came into its own with the incorporation of supersymmetry 
in the early 1980's  cf. Green, Schwarz and Witten \ref{1--2}.
\par
Quantum field theories with chiral Fermions sometimes exhibit a 
pathological behaviour  when coupled to gauge fields, gauge invariance may break down, this is referred to as an anomaly. Such anomalies, too,  
have played a major part in shaping present  state of 
string theory;  a key paper here  is that of Green and Schwarz \ref{1} 
(1984) who discovered a remarkable anomaly cancellation mechanism which 
thereby singles out  five distinguished supersymmetric $d=10$ string 
theories.  Anomalies also have an important topological aspect
involving the Atiyah--Singer index theorem for families of Dirac 
operators. We have not had space to discuss this  here,  cf. 
Nash \ref{1} for  more  details. 
\par
The low energy limit of a string theory is meant to be 
a conventional quantum field theory which should be a 
theory that describes all known 
interactions including gravity. Supersymmetric string theories
do succeed in including gravity  and so offer the best chance so far for 
a quantum theory of gravity as well as, perhaps,  for an 
eventual  unified theory of 
all interactions.
\par
The five string theories singled out by the work of Green and Schwarz 
\ref{1} are all supersymmetric and contain Yang--Mills fields. These 
five 10 dimensional theories are denoted by type I, type IIA, type IIB, 
$E_8\times E_8$ heterotic and $SO(32)$ heterotic.
\par
Only four of the ten dimensions of space-time $M$ are directly 
observable; so the remaining six are meant to form a small (i.e. 
small compared with the string scale) compact six 
dimensional space $M_6$, say. The favoured physical choice for $M_6$ is 
that it be a three complex dimensional {\it Calabi--Yau} manifold---this 
means that it is complex manifold of a special kind: it is K\"ahler with 
holonomy group contained in $SU(3)$. The favoured status of Calabi--Yau 
manifolds has to do with what is called {\it mirror symmetry} which we now briefly review as it is of considerable interest to both 
mathematicians (for example algebraic geometers) and physicists.
\smalltitle{Mirror symmetry}
Mirror symmetry refers to the property that Calabi--Yau manifolds come in 
dual pairs which were conjectured to give equivalent string theories.
These dual pairs are then called {\it mirror manifolds}.
Mirror symmetry can also be profitably thought of as  providing a 
{\it transform}. In other words a difficult problem on one 
Calabi--Yau manifold may be much easier, but equivalent to, 
one on its dual. We give an example of a result obtainable this way 
below; it concerns the number of curves of prescribed degree and genus on a 
Calabi--Yau manifold.
\par
The term  {\it mirror manifolds} was coined in  Greene and 
Plesser \ref{1} (1990). This paper also contains  the first, 
and as yet only, known construction of such 
dual pairs. 
\par
If $M$ and $N$ are Calabi--Yau manifolds of complex dimension $n$ ($n=3$ 
in the string theory cited above), 
and if $h^{(p,q)}(M)$ denote the 
Hodge numbers\foot{Hodge numbers are the dimensions of the various 
Dolbeault cohomology groups of $M$, i.e. 
$h^{(p,q)}(M)=\dim H^{(p,q)}_{\dbar}(M)=\dim H^q(M;\Omega^p)$ where 
$\Omega^p$ is the sheaf of holomorphic $p$-forms on $M$.} of a 
complex manifold $M$,  then mirror symmetric pairs satisfy 
$$h^{(p,q)}(M)=h^{(n-p,q)}(N)\no$$
The term mirror symmetry originates in the fact that this represents a 
reflection symmetry about the diagonal in the Hodge diamond formed by the
$h^{(p,q)}$'s. 
This reflection property of the Hodge numbers is not sufficient to ensure that
$M$ and $N$ are mirror manifolds, one must prove that the associated conformal field theories are also identical, cf. Greene and Plesser \ref{1-2} for more information. 
\par
The introduction (by Greene, Vafa and Warner \ref{1}) of manifolds which are complete intersections in  {\it weighted} projective spaces is an important part of this story, cf. too, Candelas, Lynker and Schimmrigk \ref{1}; previously complete intersections had been studied in 
ordinary projective space, cf., for example, Green, H\"ubsch and L\"utken \ref{1} and references therein. 
\par
A bold prediction of mirror symmetry (following from the equality of the three point functions on a Calabi--Yau $M$ and its mirror) concerned the numbers $n$ of 
 curves  of genus $g$ and degree $d$ on a Calabi---Yau manifold; 
these numbers could be read off from a rather sophisticated instanton 
calculation, cf. Candelas, Ossa et al. \ref{1} (1991) and the articles 
in Yau \ref{1} for more details. The startling  nature of these 
predictions can be partly appreciated by browsing the following table 
(this is half of a table appearing in  Bershadsky et al. \ref{1}).  The 
particular Calabi--Yau manifold is three dimensional and is 
a certain quotient by 
${\bf Z}_5 \times{\bf Z}_5\times{\bf Z}_5$ of a quintic hypersurface in 
${\bf CP}^4$ whose equation in homogeneous coordinates is
$$z_1^5+z_2^5+\cdots+z_5^5=0\no$$
{\beginquote
\centerline{\vbox{\offinterlineskip
\hrule
\halign{&\vrule#&
   \strut\quad\hfil#\quad\cr
height2pt\cr
&Degree\hfil&&$g=0$&&$g=1$&\cr
height2pt&\omit&\cr
\noalign{\hrule}
height2pt&\omit&\cr
&n=0&&5&&50/12&\cr
&n=1&&2875&&0&\cr
&n=2&&609250&&0&\cr
&n=3&&317206375&&609250&\cr
&n=4&&242467530000&&3721431625&\cr
&n=5&&229305888887625&&12129909700200&\cr
&n=6&&248249742118022000&&31147299732677250&\cr
&n=7&&295091050570845659250&&71578406022880761750&\cr
&n=8&&375632160937476603550000&&154990541752957846986500&\cr
&n=9&&503840510416985243645106250&&324064464310279585656399500&\cr
&\vdots&&\vdots&&\vdots&\cr
& large n && $a_0n^{-3}(\log n)^{-2} e^{2 \pi n \alpha}$
&& $a_1n^{-1} e^{2 \pi n\alpha}$ &\cr
height2pt&\omit&\cr
\noalign{\hrule}
height2pt&\omit&\cr
}}}\par\penalty100000
\centerline{\bf Table showing numbers of curves of genus $g$ on a 
quintic hypersurface}
\centerline{\bf as  predicted by mirror symmetry}
\endquote}
 Mathematical verifications of these spectacular, physically obtained,  
numbers were at first only available 
for small values of the degree. However there has now been a beautiful 
confirmation of all of them by Givental \ref{1} (1996); this work also 
extends the ideas to non-Calabi--Yau manifolds; a key paper, whose results
are used in Givental \ref{1}, is that 
of Kontsevich \ref{1} (1995). 
\par
The solutions methods employed for these problems involving curve counting, 
or {\it enumerative geometry}, 
 are closely connected with another development of joint physical and mathematical interest: this is the subject of {\it quantum cohomology}. Quantum cohomology originates in quantum field theory. One 
considers a quantum cohomology ring $H^*_q(M)$  which is a natural 
deformation of the standard cohomology ring $H^*(M)$  of a manifold 
$M$.  The $q$ in $H^*_q(M)$ is a real parameter which can be taken to 
zero and, when 
this is done, one recovers the standard cohomology ring $H^*(M)$; 
thus $q\rightarrow 0$ represents the {\it classical limit}, one also  
recognises similarities with the deformations of Lie algebras 
 known as quantum groups.  
 For some physical and mathematical background material 
on this highly interesting new area 
cf. Morrison and Plesser \ref{1} and Kontsevich and Manin \ref{1} 
and references therein. 
\smalltitle{Dyons again and the tyranny of dualities}
Several new kinds of duality emerged from about 1994 onwards. Their 
origin can be traced back to the subject of dyons and in particular to a 
paper of Montonen and Olive \ref{1} (1977); on the other hand the 
present work in the subject is due in great part to the papers of 
Seiberg and Witten \ref{1--2} and the insights they offer into   
strong coupling problems  such as the celebrated conundrum  of 
the mechanism for quark confinement.
\par
Montonen and Olive proposed a duality between a theory of magnetic 
monopoles and one containing gauge fields. They were motivated in part by a 
semiclassical analysis of dyons and gauge fields. Let $g$ denote the coupling of the gauge field\foot{Previously we used $g$ to denote magnetic 
charge, since this conflicts with our present choice we shall go back to 
Dirac's notation and  use $\mu$ for magnetic charge.} then a  dyon with 
(magnetic, electric) quantum numbers $(n,m)$ has mass $M$ given by
$$M^2=V^2\left(n^2+{16\pi^2\over g^4}m^2\right)\no$$
where $V$ is the Higgs vacuum expectation value. If we interchange $g$ 
with $1/g$, $n$ with $m$, and $V$ and $4\pi V/g^2$ then $M$ is invariant; 
there is also an exchange of the gauge group $G$ with $\widehat G$: a 
group with weight lattice dual to that of $G$.
\par
Montonen and Olive astutely observed this invariance and boldly 
conjectured, with accompanying reasons, that it led to a duality 
possessed by a full quantum theory of dyons. They proposed that the two 
quantum field theories passed between by the interchanges above were 
really dual: i.e. it was only necessary to calculate one of them to 
obtain full knowledge of the other. This is a very dramatic and 
attractive conjecture because the interchange of $g$ with $1/g$ is 
an exchange of strong and weak coupling: i.e. the (intractable) 
strong coupling limit of one theory ought to be calculable as the 
(tractable) weak coupling limit of a theory with appropriately 
altered spectrum and quantum numbers.
Osborn \ref{1} (1979) found that an $N=4$ supersymmetric 
$SU(2)$ gauge theory was the best candidate for which the conjecture 
might hold; nevertheless Seiberg and Witten's results are for an $N=2$ theory.
\par
It transpires that the electric and magnetic charges $e$, $\mu$ 
and the $CP$ breaking angle $\theta$ live more  
naturally together as the single complex variable
$$e+i\mu=e_0(n+\tau),\quad\hbox{ where } \tau={\theta\over 2\pi}+
{4\pi i \over g^2} \no$$
A general point on the dyon lattice of Fig. 3. is now given by
$$e_0(m\tau +n),\qquad\hbox{ where }
m,n\in{\bf Z}\no$$
This addition of the angle $\theta$ allows the interchange (or ${\bf Z}_2$) symmetry of Montonen 
and Olive to be  promoted to a full $SL(2,{\bf Z})$  symmetry under which
$$\tau\mapsto {a\tau +b\over c\tau+d},\quad \hbox{ with }
\cases{a,b,c,d\in{\bf Z}&\cr 
       ad-bc=1&\cr}\no$$
\par
Now the conventional string viewpoint of the physics is that this
$SL(2,{\bf Z})$ symmetry, and the mirror symmetry, are only a low 
energy manifestation of a richer symmetry of the full string theory. 
There is now considerable successful work in this direction 
which involves string dualities known by the symbols $S$, $T$ and $U$. 
Two theories which are $S$ dual have the property that their weak and strong coupling limits are equivalent, two theories which are $T$ dual have 
the property that one compactified on a large volume is equivalent 
to the other compactified on a small volume, finally $U$ duality 
corresponds to a theory compactified on a large, or small, volume being 
equivalent to another at strong, or weak, coupling respectively.
\par
Certain pairs of the five basic superstring 
theories are thought to be dual to one another in this framework. It 
is conjectured that these five theories are just different 
manifestations of a single eleven dimensional theory known as M theory.
\smalltitle{Black hole postscript}
It has also recently become of interest to study what happens when 
strings have ends that move on $p$-dimensional membranes 
(called Dirichlet branes or $D$-branes because of the boundary 
condition imposed at their ends). This has made possible a study of 
black holes in this string setting. Included in this study is the ability 
to calculate some of their quantum properties and their entropy 
(Strominger and Vafa \ref{1} (1996)). The topological nature of 
space-time here makes contact with the non-commutative geometry of 
Connes \ref{1}.

\beginsection{Concluding remarks}
In the twentieth century,  some time after the early successes of the new 
quantum mechanics and relativity, 
 quantum field theory encountered difficult mathematical problems. 
Efforts to solve these problems led to the birth of the  
subject of axiomatic quantum field theory. In  
this approach the main idea was to tackle the formidable 
problems of quantum field theory head on using the most 
powerful mathematical tools available; the bulk of these 
tools being drawn from analysis. 
\par
It is now evident that  the way forward in these problems is 
considerably illuminated if, in addition to analysis, one uses 
differential topology. We have also seen that this inclusion of 
topology has produced profound results in mathematics as well as physics.
\par
Not since  Poincar\'e, Hilbert and Weyl 
took an interest in physics has such lavish attention been visited on the 
physicists by the mathematicians.
The story this time is rather different: In the early part of the 
twentieth century the physicists imported Riemannian geometry for 
relativity, thereby of course accelerating its rise to  be  an essential
 pillar of the body of mathematics; Hilbert spaces were duly digested 
for quantum mechanics as was the notion of symmetry in its widest possible sense  leading to a systematic use of  the theory of group representations.
\par
The Riemannian geometry used by physicists in relativity 
was first used implicitly in a local manner; but it was inevitable 
that global issues would arise eventually. This, of course, 
entails topology and so more mathematics
has to be learned by the physicist but the singularity theorems of 
general relativity more than justify the intellectual investment 
required.
\par
Thus far, then,  the gifts,  were mainly from the mathematicians to 
the physicists.
For the last quarter of the twentieth century things are rather 
different,  the physicists have been able to give as well as to receive.
 The  Yang--Mills equations have  been the source of many new results in 
three and four dimensional differential geometry and topology. 
\par
In sum, the growing interaction between topology and physics has 
been a very healthy thing for both subjects. Their joint futures 
look very bright.
It seems fitting that we should leave the  last word to Dirac \ref{1}
{\beginquote
It seems likely that this process of increasing abstraction will continue in the future and that advance in physics is to be associated with a continual modification
and generalisation of the axioms at the base of the mathematics
rather than with a logical development of any one mathematical scheme
on a fixed foundation.
\endquote}

{\obeylines















 }
\par\vskip\baselineskip
\centerline{\bflarge References}
\par\penalty100000
\par\vskip\baselineskip   
\ninepoint
{ \global \localindentsize =\parindent \global \setbox \numberbox =\hbox {1. } \global \advance \localindentsize by \wd \numberbox \par \vskip 2.4 pt\noindent Aharonov Y. and Bohm D.\par \penalty 10000\global \hangindent \localindentsize \global \hangafter =1 1. Significance of electromagnetic potentials in quantum theory, {\fam \itfam \nineit Phys. Rev.}, {\fam \bffam \ninebf 115}, 485--491, (1959). } 
{ \global \localindentsize =\parindent \global \setbox \numberbox =\hbox {1. } \global \advance \localindentsize by \wd \numberbox \par \vskip 2.4 pt\noindent Alexander J. W.\par \penalty 10000\global \hangindent \localindentsize \global \hangafter =1 1. Topological invariants of knots and links, {\fam \itfam \nineit Trans. Amer. Math. Soc.}, {\fam \bffam \ninebf 30}, 275--306, (1928). } 
{ \global \localindentsize =\parindent \global \setbox \numberbox =\hbox {1. } \global \advance \localindentsize by \wd \numberbox \par \vskip 2.4 pt\noindent Arnold V. I.\par \penalty 10000\global \hangindent \localindentsize \global \hangafter =1 1. Normal forms of function of functions close to degenerate critical points, The Weyl groups $A_k$, $D_k$, $E_k$ and Lagrangian singularities, {\fam \itfam \nineit Funct. Anal. Appl.}, {\fam \bffam \ninebf 6}, 254--272, (1973). } 
{ \global \localindentsize =\parindent \global \setbox \numberbox =\hbox {2. } \global \advance \localindentsize by \wd \numberbox \par \global \hangindent \localindentsize \global \hangafter =1 2. Critical points of functions on manifolds with boundary, the simple Lie groups $B_k$, $C_k$ and $F_4$., {\fam \itfam \nineit Russ. Math. Surv.}, {\fam \bffam \ninebf 33}, 99--116, (1978). } 
{ \global \localindentsize =\parindent \global \setbox \numberbox =\hbox {1. } \global \advance \localindentsize by \wd \numberbox \par \vskip 2.4 pt\noindent Atiyah M. F.\par \penalty 10000\global \hangindent \localindentsize \global \hangafter =1 1. {\fam \itfam \nineit The geometry and physics of knots}, Cambridge University Press, (1990). } 
{ \global \localindentsize =\parindent \global \setbox \numberbox =\hbox {2. } \global \advance \localindentsize by \wd \numberbox \par \global \hangindent \localindentsize \global \hangafter =1 2. {\fam \itfam \nineit Geometry of Yang-Mills Fields}, Accademia Nazionale dei Lincei, (1979). } 
{ \global \localindentsize =\parindent \global \setbox \numberbox =\hbox {3. } \global \advance \localindentsize by \wd \numberbox \par \global \hangindent \localindentsize \global \hangafter =1 3. {\fam \itfam \nineit New invariants of 3 and 4 dimensional manifolds}, Symposium on the mathematical heritage of Hermann Weyl, May 1987, {edited by: Wells R. O. }, Amer. Math. Soc., (1988).} 
{ \global \localindentsize =\parindent \global \setbox \numberbox =\hbox {4. } \global \advance \localindentsize by \wd \numberbox \par \global \hangindent \localindentsize \global \hangafter =1 4. Topological quantum field theories, {\fam \itfam \nineit Inst. Hautes \accent 19 Etudes Sci. Publ. Math. }, {\fam \bffam \ninebf 68}, 175--186, (1989). } 
{ \global \localindentsize =\parindent \global \setbox \numberbox =\hbox {1. } \global \advance \localindentsize by \wd \numberbox \par \vskip 2.4 pt\noindent Atiyah M. F. and Bott R.\par \penalty 10000\global \hangindent \localindentsize \global \hangafter =1 1. The Yang-Mills equations over Riemann surfaces, {\fam \itfam \nineit Phil. Trans. Roy. Soc. Lond. A}, {\fam \bffam \ninebf 308}, 523--615, (1982). } 
{ \global \localindentsize =\parindent \global \setbox \numberbox =\hbox {1. } \global \advance \localindentsize by \wd \numberbox \par \vskip 2.4 pt\noindent Atiyah M. F., Hitchin N. J., Drinfeld V. G. and Manin Y. I.\par \penalty 10000\global \hangindent \localindentsize \global \hangafter =1 1. Construction of Instantons, {\fam \itfam \nineit Phys. Lett.}, {\fam \bffam \ninebf 65A}, 185--187, (1978). } 
{ \global \localindentsize =\parindent \global \setbox \numberbox =\hbox {1. } \global \advance \localindentsize by \wd \numberbox \par \vskip 2.4 pt\noindent Atiyah M. F., Hitchin N. J. and Singer I. M.\par \penalty 10000\global \hangindent \localindentsize \global \hangafter =1 1. Self-duality in four dimensional Riemannian geometry, {\fam \itfam \nineit Proc. Roy. Soc. Lond. A.}, {\fam \bffam \ninebf 362}, 425--461, (1978). } 
{ \global \localindentsize =\parindent \global \setbox \numberbox =\hbox {1. } \global \advance \localindentsize by \wd \numberbox \par \vskip 2.4 pt\noindent Atiyah M. F. and Ward R. S.\par \penalty 10000\global \hangindent \localindentsize \global \hangafter =1 1. Instantons and Algebraic Geometry, {\fam \itfam \nineit Commun. Math. Phys.}, {\fam \bffam \ninebf 55}, 117--124, (1977). } 
{ \global \localindentsize =\parindent \global \setbox \numberbox =\hbox {1. } \global \advance \localindentsize by \wd \numberbox \par \vskip 2.4 pt\noindent Balachandran A. P., Daughton A., Gu Z.-C., Sorkin R. D., Marmo G., and Srivastava A. M.\par \penalty 10000\global \hangindent \localindentsize \global \hangafter =1 1. Spin-statistics theorems without relativity or field theory, {\fam \itfam \nineit Int. J. Mod. Phys.}, {\fam \bffam \ninebf A8}, 2993--3044, (1993). } 
{ \global \localindentsize =\parindent \global \setbox \numberbox =\hbox {1. } \global \advance \localindentsize by \wd \numberbox \par \vskip 2.4 pt\noindent Bar-Natan D.\par \penalty 10000\global \hangindent \localindentsize \global \hangafter =1 1. Vassiliev and quantum invariants of braids. The interface of knots and physics, {\fam \itfam \nineit Proc. Sympos. Appl. Math.}, {\fam \bffam \ninebf 51}, 129--144, (1996). } 
{ \global \localindentsize =\parindent \global \setbox \numberbox =\hbox {1. } \global \advance \localindentsize by \wd \numberbox \par \vskip 2.4 pt\noindent Barrow-Green J.\par \penalty 10000\global \hangindent \localindentsize \global \hangafter =1 1. {\fam \itfam \nineit Poincar\accent 19 e and the three body problem}, American Mathematical Society, (1997). } 
{ \global \localindentsize =\parindent \global \setbox \numberbox =\hbox {1. } \global \advance \localindentsize by \wd \numberbox \par \vskip 2.4 pt\noindent Baxter R. J.\par \penalty 10000\global \hangindent \localindentsize \global \hangafter =1 1. {\fam \itfam \nineit Exactly solved models in statistical mechanics}, Academic Press, (1982). } 
{ \global \localindentsize =\parindent \global \setbox \numberbox =\hbox {1. } \global \advance \localindentsize by \wd \numberbox \par \vskip 2.4 pt\noindent Bedford T., Keane M. and Series C. (eds.)\par \penalty 10000\global \hangindent \localindentsize \global \hangafter =1 1. {\fam \itfam \nineit Ergodic theory, symbolic dynamics and hyperbolic spaces}, Oxford University Press, (1991). } 
{ \global \localindentsize =\parindent \global \setbox \numberbox =\hbox {1. } \global \advance \localindentsize by \wd \numberbox \par \vskip 2.4 pt\noindent Belavin A. A., Polyakov A. M., Schwarz A. S. and Tyupkin Y. S.\par \penalty 10000\global \hangindent \localindentsize \global \hangafter =1 1. Pseudoparticle solutions of the Yang-Mills equations, {\fam \itfam \nineit Phys. Lett.}, {\fam \bffam \ninebf 59B}, 85--87, (1975). } 
{ \global \localindentsize =\parindent \global \setbox \numberbox =\hbox {1. } \global \advance \localindentsize by \wd \numberbox \par \vskip 2.4 pt\noindent Berry M. V.\par \penalty 10000\global \hangindent \localindentsize \global \hangafter =1 1. Quantal phase factors accompanying adiabatic changes, {\fam \itfam \nineit Proc. Roy. Soc. London A}, {\fam \bffam \ninebf 392}, 45--57, (1984). } 
{ \global \localindentsize =\parindent \global \setbox \numberbox =\hbox {1. } \global \advance \localindentsize by \wd \numberbox \par \vskip 2.4 pt\noindent Berry M. V. and Robbins J. M.\par \penalty 10000\global \hangindent \localindentsize \global \hangafter =1 1. Indistinguishability for quantum particles: spin, statistics and the geometric phase, {\fam \itfam \nineit Proc. Roy. Soc. Lond. A}, {\fam \bffam \ninebf 453}, 1771--1790, (1997). } 
{ \global \localindentsize =\parindent \global \setbox \numberbox =\hbox {1. } \global \advance \localindentsize by \wd \numberbox \par \vskip 2.4 pt\noindent Bershadsky M., Cecotti S., Ooguri H. and Vafa C.\par \penalty 10000\global \hangindent \localindentsize \global \hangafter =1 1. Kodaira-Spencer theory of gravity and exact results for quantum string amplitudes, {\fam \itfam \nineit Commun. Math. Phys.}, {\fam \bffam \ninebf 165}, 311--428, (1994). } 
{ \global \localindentsize =\parindent \global \setbox \numberbox =\hbox {1. } \global \advance \localindentsize by \wd \numberbox \par \vskip 2.4 pt\noindent Birkhoff G. D.\par \penalty 10000\global \hangindent \localindentsize \global \hangafter =1 1. Proof of Poincar\accent 19 e's geometric theorem, {\fam \itfam \nineit Trans. Amer. Math. Soc.}, {\fam \bffam \ninebf 14}, 14--22, (1913). } 
{ \global \localindentsize =\parindent \global \setbox \numberbox =\hbox {2. } \global \advance \localindentsize by \wd \numberbox \par \global \hangindent \localindentsize \global \hangafter =1 2. Proof of the ergodic theorem, {\fam \itfam \nineit Proc. Nat. Acad. Sci. U. S. A.}, {\fam \bffam \ninebf 17}, 656--660, (1931). } 
{ \global \localindentsize =\parindent \global \setbox \numberbox =\hbox {1. } \global \advance \localindentsize by \wd \numberbox \par \vskip 2.4 pt\noindent Bogomolny E. B.\par \penalty 10000\global \hangindent \localindentsize \global \hangafter =1 1. Stability of Classical solutions, {\fam \itfam \nineit Sov. J. Nucl. Phys.}, {\fam \bffam \ninebf 24}, 861--870, (1976). } 
{ \global \localindentsize =\parindent \global \setbox \numberbox =\hbox {1. } \global \advance \localindentsize by \wd \numberbox \par \vskip 2.4 pt\noindent Bott R.\par \penalty 10000\global \hangindent \localindentsize \global \hangafter =1 1. An application of Morse theory to the topology of Lie groups, {\fam \itfam \nineit Bull. Soc. Math. France}, {\fam \bffam \ninebf 84}, 251--281, (1956). } 
{ \global \localindentsize =\parindent \global \setbox \numberbox =\hbox {2. } \global \advance \localindentsize by \wd \numberbox \par \global \hangindent \localindentsize \global \hangafter =1 2. The stable homotopy of the classical groups, {\fam \itfam \nineit Ann. Math.}, {\fam \bffam \ninebf 70}, 313--337, (1959). } 
{ \global \localindentsize =\parindent \global \setbox \numberbox =\hbox {1. } \global \advance \localindentsize by \wd \numberbox \par \vskip 2.4 pt\noindent Brill D. R. and Werner F. G.\par \penalty 10000\global \hangindent \localindentsize \global \hangafter =1 1. Significance of electromagnetic potentials in the quantum theory in the interpretation of electron fringe interferometer observations, {\fam \itfam \nineit Phys. Rev. Lett.}, {\fam \bffam \ninebf 4}, 344--347, (1960). } 
{ \global \localindentsize =\parindent \global \setbox \numberbox =\hbox {1. } \global \advance \localindentsize by \wd \numberbox \par \vskip 2.4 pt\noindent Broer H. W., Hoveijn, Takens F. and van Gils S. A.\par \penalty 10000\global \hangindent \localindentsize \global \hangafter =1 1. {\fam \itfam \nineit Nonlinear dynamical systems and chaos}, Birkhauser, (1995). } 
{ \global \localindentsize =\parindent \global \setbox \numberbox =\hbox {1. } \global \advance \localindentsize by \wd \numberbox \par \vskip 2.4 pt\noindent Candelas P., Lynker M. and Schimmrigk R.\par \penalty 10000\global \hangindent \localindentsize \global \hangafter =1 1. Calabi--Yau manifolds in weighted $P_4$, {\fam \itfam \nineit Nucl. Phys.}, {\fam \bffam \ninebf B341}, 383--402, (1990). } 
{ \global \localindentsize =\parindent \global \setbox \numberbox =\hbox {1. } \global \advance \localindentsize by \wd \numberbox \par \vskip 2.4 pt\noindent Candelas P., de la Ossa X. C., Green P. S. and Parkes L.\par \penalty 10000\global \hangindent \localindentsize \global \hangafter =1 1. A pair of Calabi--Yau manifolds as an exactly soluble superconformal theory, {\fam \itfam \nineit Nucl. Phys.}, {\fam \bffam \ninebf B359}, 21--74, (1991). } 
{ \global \localindentsize =\parindent \global \setbox \numberbox =\hbox {1. } \global \advance \localindentsize by \wd \numberbox \par \vskip 2.4 pt\noindent Chandrasekhar S.\par \penalty 10000\global \hangindent \localindentsize \global \hangafter =1 1. The maximum mass of ideal white dwarfs, {\fam \itfam \nineit Astrophys. J.}, {\fam \bffam \ninebf 74}, 81--82, (1931). } 
{ \global \localindentsize =\parindent \global \setbox \numberbox =\hbox {2. } \global \advance \localindentsize by \wd \numberbox \par \global \hangindent \localindentsize \global \hangafter =1 2. The highly collapsed configurations of a stellar mass, {\fam \itfam \nineit Mon. Not. R. Astron. Soc.}, {\fam \bffam \ninebf 95}, 207--225, (1935). } 
{ \global \localindentsize =\parindent \global \setbox \numberbox =\hbox {1. } \global \advance \localindentsize by \wd \numberbox \par \vskip 2.4 pt\noindent Coleman S.\par \penalty 10000\global \hangindent \localindentsize \global \hangafter =1 1. {\fam \itfam \nineit The uses of instantons}, The whys of subnuclear physics: Erice summer school 1977, {edited by: Zichichi A.}, Plenum Press, (1979). } 
{ \global \localindentsize =\parindent \global \setbox \numberbox =\hbox {1. } \global \advance \localindentsize by \wd \numberbox \par \vskip 2.4 pt\noindent Connes A.\par \penalty 10000\global \hangindent \localindentsize \global \hangafter =1 1. {\fam \itfam \nineit Noncommutative geometry}, Academic Press, (1994). } 
{ \global \localindentsize =\parindent \global \setbox \numberbox =\hbox {1. } \global \advance \localindentsize by \wd \numberbox \par \vskip 2.4 pt\noindent Dahan-Dalm\accent 19 edico A.\par \penalty 10000\global \hangindent \localindentsize \global \hangafter =1 1. La renaissance des syst\accent 18 emes dynamiques aux Etats-Unis apr\accent 18 es la deuxi\accent 18 eme guerre mondiale: l'action de Solomon Lefschetz, {\fam \itfam \nineit Rend. Circ. Mat. Palermo (2) Suppl. No 34}, {\fam \bffam \ninebf }, 133--166, (1994). } 
{ \global \localindentsize =\parindent \global \setbox \numberbox =\hbox {2. } \global \advance \localindentsize by \wd \numberbox \par \global \hangindent \localindentsize \global \hangafter =1 2. Le difficile h\accent 19 eritage de Henri Poincar\accent 19 e en syst\accent 18 emes dynamiques; Henri Poincar\accent 19 e: science et philosophie (Nancy, 1994), {\fam \itfam \nineit Publ. Henri-Poincar\accent 19 e-Arch.}, {\fam \bffam \ninebf }, 13--33, (Akademie Verlag, Berlin, 1996). } 
{ \global \localindentsize =\parindent \global \setbox \numberbox =\hbox {1. } \global \advance \localindentsize by \wd \numberbox \par \vskip 2.4 pt\noindent Dirac P. A. M.\par \penalty 10000\global \hangindent \localindentsize \global \hangafter =1 1. Quantised singularities in the electromagnetic field, {\fam \itfam \nineit Proc. Roy. Soc. Lond.}, {\fam \bffam \ninebf A133}, 60--72, (1931). } 
{ \global \localindentsize =\parindent \global \setbox \numberbox =\hbox {1. } \global \advance \localindentsize by \wd \numberbox \par \vskip 2.4 pt\noindent Donaldson S. K.\par \penalty 10000\global \hangindent \localindentsize \global \hangafter =1 1. An Application of Gauge Theory to Four Dimensional Topology, {\fam \itfam \nineit J. Diff. Geom.}, {\fam \bffam \ninebf 18}, 279--315, (1983). } 
{ \global \localindentsize =\parindent \global \setbox \numberbox =\hbox {2. } \global \advance \localindentsize by \wd \numberbox \par \global \hangindent \localindentsize \global \hangafter =1 2. {\fam \itfam \nineit The Geometry of 4-Manifolds}, Proc. of the International Congress of Mathematicians, Berkeley 1986, {edited by: Gleason A. M.}, Amer. Math. Soc., (1987).} 
{ \global \localindentsize =\parindent \global \setbox \numberbox =\hbox {3. } \global \advance \localindentsize by \wd \numberbox \par \global \hangindent \localindentsize \global \hangafter =1 3. Polynomial invariants for smooth four manifolds, {\fam \itfam \nineit Topology}, {\fam \bffam \ninebf 29}, 257--315, (1990). } 
{ \global \localindentsize =\parindent \global \setbox \numberbox =\hbox {4. } \global \advance \localindentsize by \wd \numberbox \par \global \hangindent \localindentsize \global \hangafter =1 4. The Seiberg--Witten equations and 4-manifold topology, {\fam \itfam \nineit Bull. Amer. Math. Soc.}, {\fam \bffam \ninebf 33}, 45--70, (1996). } 
{ \global \localindentsize =\parindent \global \setbox \numberbox =\hbox {1. } \global \advance \localindentsize by \wd \numberbox \par \vskip 2.4 pt\noindent Donaldson S. K. and Kronheimer P. B.\par \penalty 10000\global \hangindent \localindentsize \global \hangafter =1 1. {\fam \itfam \nineit The geometry of four manifolds}, Oxford University Press, (1990). } 
{ \global \localindentsize =\parindent \global \setbox \numberbox =\hbox {1. } \global \advance \localindentsize by \wd \numberbox \par \vskip 2.4 pt\noindent Ebeling W.\par \penalty 10000\global \hangindent \localindentsize \global \hangafter =1 1. An example of two homeomorphic, nondiffeomorphic complete intersection surfaces, {\fam \itfam \nineit Inventiones Math.}, {\fam \bffam \ninebf 99}, 651--654, (1990). } 
{ \global \localindentsize =\parindent \global \setbox \numberbox =\hbox {1. } \global \advance \localindentsize by \wd \numberbox \par \vskip 2.4 pt\noindent Epple M.\par \penalty 10000\global \hangindent \localindentsize \global \hangafter =1 1. Branch points of algebraic functions and the beginnings of modern knot theory , {\fam \itfam \nineit Historia Math.}, {\fam \bffam \ninebf 22}, 371--401, (1995). } 
{ \global \localindentsize =\parindent \global \setbox \numberbox =\hbox {2. } \global \advance \localindentsize by \wd \numberbox \par \global \hangindent \localindentsize \global \hangafter =1 2. Orbits of asteroids, a braid and the first link invariant, {\fam \itfam \nineit Math. Intell.}, {\fam \bffam \ninebf }, , (to appear, 1998). } 
{ \global \localindentsize =\parindent \global \setbox \numberbox =\hbox {1. } \global \advance \localindentsize by \wd \numberbox \par \vskip 2.4 pt\noindent Euler L.\par \penalty 10000\global \hangindent \localindentsize \global \hangafter =1 1. Solutio problematis ad geometriam situs pertinentis, {\fam \itfam \nineit Commentarii Academiae Scientiarum Imperialis Petropolitanae}, {\fam \bffam \ninebf 8}, 128--140, (1736). } 
{ \global \localindentsize =\parindent \global \setbox \numberbox =\hbox {1. } \global \advance \localindentsize by \wd \numberbox \par \vskip 2.4 pt\noindent Finklestein D. and Rubinstein J. \par \penalty 10000\global \hangindent \localindentsize \global \hangafter =1 1. Connection between spin, statistics and kinks, {\fam \itfam \nineit J. Math. Physics}, {\fam \bffam \ninebf 9}, 1762--1779, (1968). } 
{ \global \localindentsize =\parindent \global \setbox \numberbox =\hbox {1. } \global \advance \localindentsize by \wd \numberbox \par \vskip 2.4 pt\noindent Floer A.\par \penalty 10000\global \hangindent \localindentsize \global \hangafter =1 1. A Relative Morse Index for the Symplectic Action, {\fam \itfam \nineit Comm. Pure Appl. Math.}, {\fam \bffam \ninebf 41}, 393--407, (1988). } 
{ \global \localindentsize =\parindent \global \setbox \numberbox =\hbox {2. } \global \advance \localindentsize by \wd \numberbox \par \global \hangindent \localindentsize \global \hangafter =1 2. An Instanton Invariant for 3-Manifolds, {\fam \itfam \nineit Commun. Math. Phys.}, {\fam \bffam \ninebf 118}, 215--240, (1988). } 
{ \global \localindentsize =\parindent \global \setbox \numberbox =\hbox {1. } \global \advance \localindentsize by \wd \numberbox \par \vskip 2.4 pt\noindent Freedman M. H.\par \penalty 10000\global \hangindent \localindentsize \global \hangafter =1 1. The topology of 4-dimensional manifolds, {\fam \itfam \nineit Jour. Diff. Geom.}, {\fam \bffam \ninebf 17}, 357--453, (1982). } 
{ \global \localindentsize =\parindent \global \setbox \numberbox =\hbox {1. } \global \advance \localindentsize by \wd \numberbox \par \vskip 2.4 pt\noindent Freed D. S. and Uhlenbeck K. K.\par \penalty 10000\global \hangindent \localindentsize \global \hangafter =1 1. {\fam \itfam \nineit Instantons and Four-Manifolds}, Springer-Verlag, (1984). } 
{ \global \localindentsize =\parindent \global \setbox \numberbox =\hbox {1. } \global \advance \localindentsize by \wd \numberbox \par \vskip 2.4 pt\noindent Freyd P., Yetter D.; Hoste J.; Lickorish W. B. R., Millet K.; and Ocneanu A.\par \penalty 10000\global \hangindent \localindentsize \global \hangafter =1 1. A new polynomial invariant of knots and links, {\fam \itfam \nineit Bull. Amer. Math. Soc.}, {\fam \bffam \ninebf 12}, 239--246, (1985). } 
{ \global \localindentsize =\parindent \global \setbox \numberbox =\hbox {1. } \global \advance \localindentsize by \wd \numberbox \par \vskip 2.4 pt\noindent Froissart M.\par \penalty 10000\global \hangindent \localindentsize \global \hangafter =1 1. {\fam \itfam \nineit Applications of algebraic topology to physics}, Mathematical theory of elementary particles, {edited by: Goodman R. and Segal I.}, M. I. T. Press, (1966). } 
{ \global \localindentsize =\parindent \global \setbox \numberbox =\hbox {1. } \global \advance \localindentsize by \wd \numberbox \par \vskip 2.4 pt\noindent Gauss C. F.\par \penalty 10000\global \hangindent \localindentsize \global \hangafter =1 1. Zur mathematischen theorie der electrodynamischen Wirkungen (1833), {\fam \itfam \nineit Werke. K\accent 127 on\-ig\-lichen Gesell\-schaft der Wissen\-schaften zu G\accent 127 ott\-ingen}, {\fam \bffam \ninebf 5}, 605, (1877). } 
{ \global \localindentsize =\parindent \global \setbox \numberbox =\hbox {1. } \global \advance \localindentsize by \wd \numberbox \par \vskip 2.4 pt\noindent Givental A. B.\par \penalty 10000\global \hangindent \localindentsize \global \hangafter =1 1. Equivariant Gromov--Witten invariants, {\fam \itfam \nineit Internat. Math. Res. Notices}, {\fam \bffam \ninebf 13}, 613--663, (1996). } 
{ \global \localindentsize =\parindent \global \setbox \numberbox =\hbox {1. } \global \advance \localindentsize by \wd \numberbox \par \vskip 2.4 pt\noindent Green M. B. and Schwarz J. H.\par \penalty 10000\global \hangindent \localindentsize \global \hangafter =1 1. Anomaly cancellations in supersymmetric $D=10$ gauge theory and superstring theory, {\fam \itfam \nineit Phys. Lett.}, {\fam \bffam \ninebf 149B}, 117--122, (1984). } 
{ \global \localindentsize =\parindent \global \setbox \numberbox =\hbox {1. } \global \advance \localindentsize by \wd \numberbox \par \vskip 2.4 pt\noindent Green P., H\accent 127 ubsch T. and L\accent 127 utken C. A.\par \penalty 10000\global \hangindent \localindentsize \global \hangafter =1 1. All the Hodge numbers for all Calabi--Yau complete intersections, {\fam \itfam \nineit Class. Quant. Grav.}, {\fam \bffam \ninebf 6}, 105--124, (1989). } 
{ \global \localindentsize =\parindent \global \setbox \numberbox =\hbox {1. } \global \advance \localindentsize by \wd \numberbox \par \vskip 2.4 pt\noindent Green M. B., Schwarz J. H. and Witten E.\par \penalty 10000\global \hangindent \localindentsize \global \hangafter =1 1. {\fam \itfam \nineit Superstring Theory vol. 1}, Cambridge University Press, (1987). } 
{ \global \localindentsize =\parindent \global \setbox \numberbox =\hbox {2. } \global \advance \localindentsize by \wd \numberbox \par \global \hangindent \localindentsize \global \hangafter =1 2. {\fam \itfam \nineit Superstring Theory vol. 2}, Cambridge University Press, (1987). } 
{ \global \localindentsize =\parindent \global \setbox \numberbox =\hbox {1. } \global \advance \localindentsize by \wd \numberbox \par \vskip 2.4 pt\noindent Greene B. R. and Plesser M. R.\par \penalty 10000\global \hangindent \localindentsize \global \hangafter =1 1. Duality in Calabi-Yau moduli space, {\fam \itfam \nineit Nucl. Phys.}, {\fam \bffam \ninebf B338}, 15--37, (1990). } 
{ \global \localindentsize =\parindent \global \setbox \numberbox =\hbox {2. } \global \advance \localindentsize by \wd \numberbox \par \global \hangindent \localindentsize \global \hangafter =1 2. {\fam \itfam \nineit An introduction to mirror manifolds}, Essays on Mirror Manifolds, {edited by: Yau S.-T.}, International Press, Hong Kong, (1992).} 
{ \global \localindentsize =\parindent \global \setbox \numberbox =\hbox {1. } \global \advance \localindentsize by \wd \numberbox \par \vskip 2.4 pt\noindent Greene B. R., Vafa C. and Warner N. P.\par \penalty 10000\global \hangindent \localindentsize \global \hangafter =1 1. Calabi--Yau manifolds and renormalization group flows, {\fam \itfam \nineit Nucl. Phys.}, {\fam \bffam \ninebf B324}, 371--390, (1989). } 
{ \global \localindentsize =\parindent \global \setbox \numberbox =\hbox {1. } \global \advance \localindentsize by \wd \numberbox \par \vskip 2.4 pt\noindent Hawking S. W. and Ellis G. F. R.\par \penalty 10000\global \hangindent \localindentsize \global \hangafter =1 1. {\fam \itfam \nineit The large scale structure of space-time}, Cambridge University Press, (1973). } 
{ \global \localindentsize =\parindent \global \setbox \numberbox =\hbox {1. } \global \advance \localindentsize by \wd \numberbox \par \vskip 2.4 pt\noindent Hawking S. W. and Israel W. (eds.)\par \penalty 10000\global \hangindent \localindentsize \global \hangafter =1 1. {\fam \itfam \nineit Three hundred years of gravitation}, Cambridge University Press, (1987). } 
{ \global \localindentsize =\parindent \global \setbox \numberbox =\hbox {1. } \global \advance \localindentsize by \wd \numberbox \par \vskip 2.4 pt\noindent Hawking S. W. and Penrose R.\par \penalty 10000\global \hangindent \localindentsize \global \hangafter =1 1. The singularities of gravitational collapse and cosmology, {\fam \itfam \nineit Proc. Roy. Soc. (London)}, {\fam \bffam \ninebf A314}, 529--548, (1970). } 
{ \global \localindentsize =\parindent \global \setbox \numberbox =\hbox {1. } \global \advance \localindentsize by \wd \numberbox \par \vskip 2.4 pt\noindent Helmholtz H. L. F.\par \penalty 10000\global \hangindent \localindentsize \global \hangafter =1 1. Ueber Integrale der hydrodynamischen gleichungen welche den wir\-belbewegungen entspre\-chen, {\fam \itfam \nineit Jour. f\accent 127 ur die reine und ang. Math.}, {\fam \bffam \ninebf 55}, 25--55, (1858). } 
{ \global \localindentsize =\parindent \global \setbox \numberbox =\hbox {1. } \global \advance \localindentsize by \wd \numberbox \par \vskip 2.4 pt\noindent Hwa R. C. and Teplitz V. L.\par \penalty 10000\global \hangindent \localindentsize \global \hangafter =1 1. {\fam \itfam \nineit Homology and Feynman integrals}, W. A. Benjamin, (1966). } 
{ \global \localindentsize =\parindent \global \setbox \numberbox =\hbox {1. } \global \advance \localindentsize by \wd \numberbox \par \vskip 2.4 pt\noindent Jones V. F. R.\par \penalty 10000\global \hangindent \localindentsize \global \hangafter =1 1. A Polynomial invariant for knots via Von Neumann algebras, {\fam \itfam \nineit Bull. Amer. Math. Soc.}, {\fam \bffam \ninebf 12}, 103--111, (1985). } 
{ \global \localindentsize =\parindent \global \setbox \numberbox =\hbox {1. } \global \advance \localindentsize by \wd \numberbox \par \vskip 2.4 pt\noindent Kirchhoff G.\par \penalty 10000\global \hangindent \localindentsize \global \hangafter =1 1. Ueber die Aufl\accent 127 osung der Gleichungen, auf welche man bei der Untersuchung der linearen Vertheilung galvanisher Str\accent 127 ome gef\accent 127 urt wird, {\fam \itfam \nineit Ann. der Physik und Chemie}, {\fam \bffam \ninebf 72}, 497--508, (1847). } 
{ \global \localindentsize =\parindent \global \setbox \numberbox =\hbox {1. } \global \advance \localindentsize by \wd \numberbox \par \vskip 2.4 pt\noindent Kontsevich M.\par \penalty 10000\global \hangindent \localindentsize \global \hangafter =1 1. Enumeration of rational curves via torus actions, {\fam \itfam \nineit Progr. Math.}, {\fam \bffam \ninebf 129}, 335--368, (in {\fam \itfam \nineit The moduli space of curves (Texel Island 1994), Birkh\accent 127 auser, Boston, MA} 1995). } 
{ \global \localindentsize =\parindent \global \setbox \numberbox =\hbox {1. } \global \advance \localindentsize by \wd \numberbox \par \vskip 2.4 pt\noindent Kontsevich M. and Manin Y.\par \penalty 10000\global \hangindent \localindentsize \global \hangafter =1 1. Gromov-Witten classes, quantum cohomology and enumerative geometry, {\fam \itfam \nineit Commun. Math. Phys.}, {\fam \bffam \ninebf 164}, 525--562, (1994). } 
{ \global \localindentsize =\parindent \global \setbox \numberbox =\hbox {1. } \global \advance \localindentsize by \wd \numberbox \par \vskip 2.4 pt\noindent Laplace P. S.\par \penalty 10000\global \hangindent \localindentsize \global \hangafter =1 1. Proof of the theorem that the attractive force of a heavenly body could be so large that light could not flow out of it, {\fam \itfam \nineit Geographische Ephemeriden, verfasset von Einer Gesellschaft Gelehrten}, {\fam \bffam \ninebf I}, , (1799). } 
{ \global \localindentsize =\parindent \global \setbox \numberbox =\hbox {1. } \global \advance \localindentsize by \wd \numberbox \par \vskip 2.4 pt\noindent Lema\accent 94 \i tre G.\par \penalty 10000\global \hangindent \localindentsize \global \hangafter =1 1. L'univers en expansion, {\fam \itfam \nineit Ann. Soc. Sci. (Bruxelles)}, {\fam \bffam \ninebf A53}, 51--85, (1933). } 
{ \global \localindentsize =\parindent \global \setbox \numberbox =\hbox {1. } \global \advance \localindentsize by \wd \numberbox \par \vskip 2.4 pt\noindent Libgober A. S. and Wood J. W.\par \penalty 10000\global \hangindent \localindentsize \global \hangafter =1 1. Differentiable structures on complete intersections---I, {\fam \itfam \nineit Topology}, {\fam \bffam \ninebf 21}, 469--482, (1982). } 
{ \global \localindentsize =\parindent \global \setbox \numberbox =\hbox {1. } \global \advance \localindentsize by \wd \numberbox \par \vskip 2.4 pt\noindent Lieb E. H. and Temperley H. N. V.\par \penalty 10000\global \hangindent \localindentsize \global \hangafter =1 1. Relations between the percolation and colouring problem and other graph-theoretical problems associated with regular planar lattices: Some exact results for the percolation problem, {\fam \itfam \nineit Proc. Roy. Soc. Lond.}, {\fam \bffam \ninebf A322}, 251, (1971). } 
{ \global \localindentsize =\parindent \global \setbox \numberbox =\hbox {1. } \global \advance \localindentsize by \wd \numberbox \par \vskip 2.4 pt\noindent Listing J. B.\par \penalty 10000\global \hangindent \localindentsize \global \hangafter =1 1. Der Census r\accent 127 aumlicher Complexe oder Veallgemeinerung des Euler'schen Satzes von den Polyedern, {\fam \itfam \nineit Abhandlungen der k\accent 127 oniglichen Gesellschaften zu G\accent 127 ottingen}, {\fam \bffam \ninebf 10}, 97--180, (1861). } 
{ \global \localindentsize =\parindent \global \setbox \numberbox =\hbox {2. } \global \advance \localindentsize by \wd \numberbox \par \global \hangindent \localindentsize \global \hangafter =1 2. Vorstudien zur Topologie, {\fam \itfam \nineit G\accent 127 ottinger Studien}, {\fam \bffam \ninebf }, 811--875, (1847). } 
{ \global \localindentsize =\parindent \global \setbox \numberbox =\hbox {1. } \global \advance \localindentsize by \wd \numberbox \par \vskip 2.4 pt\noindent Maxwell J. C.\par \penalty 10000\global \hangindent \localindentsize \global \hangafter =1 1. {\fam \itfam \nineit A treatise on electricity and magnetism vol. I, (1873) ed. 3 (1904)}, Oxford University Press, (1904). } 
{ \global \localindentsize =\parindent \global \setbox \numberbox =\hbox {2. } \global \advance \localindentsize by \wd \numberbox \par \global \hangindent \localindentsize \global \hangafter =1 2. {\fam \itfam \nineit A treatise on electricity and magnetism vol II, (1873) ed. 3 (1904)}, Oxford University Press, (1904). } 
{ \global \localindentsize =\parindent \global \setbox \numberbox =\hbox {1. } \global \advance \localindentsize by \wd \numberbox \par \vskip 2.4 pt\noindent Michell J. (Rev.)\par \penalty 10000\global \hangindent \localindentsize \global \hangafter =1 1. On the means of discovering the distance, magnitude, etc., of the fixed stars, in consequence of the diminution of their light, in case such a diminution should be found to take place in any of them, and such other data should be procured from observations, as would be further necessary for that purpose, {\fam \itfam \nineit Phil. Trans. Roy. Soc. (London)}, {\fam \bffam \ninebf 74}, 35--57, (1784). } 
{ \global \localindentsize =\parindent \global \setbox \numberbox =\hbox {1. } \global \advance \localindentsize by \wd \numberbox \par \vskip 2.4 pt\noindent Mickelsson J.\par \penalty 10000\global \hangindent \localindentsize \global \hangafter =1 1. Geometry of spin and statistics in classical and quantum mechanics, {\fam \itfam \nineit Phys. Rev.}, {\fam \bffam \ninebf D3}, 1375--1378, (1984). } 
{ \global \localindentsize =\parindent \global \setbox \numberbox =\hbox {1. } \global \advance \localindentsize by \wd \numberbox \par \vskip 2.4 pt\noindent Milnor J.\par \penalty 10000\global \hangindent \localindentsize \global \hangafter =1 1. On manifolds homeomorphic to the 7-sphere, {\fam \itfam \nineit Ann. Math.}, {\fam \bffam \ninebf 64}, 399--405, (1956). } 
{ \global \localindentsize =\parindent \global \setbox \numberbox =\hbox {1. } \global \advance \localindentsize by \wd \numberbox \par \vskip 2.4 pt\noindent Montonen C. and Olive D.\par \penalty 10000\global \hangindent \localindentsize \global \hangafter =1 1. Magnetic monopoles as gauge particles, {\fam \itfam \nineit Phys. Lett.}, {\fam \bffam \ninebf B72}, 117--120, (1977). } 
{ \global \localindentsize =\parindent \global \setbox \numberbox =\hbox {1. } \global \advance \localindentsize by \wd \numberbox \par \vskip 2.4 pt\noindent Morandi G.\par \penalty 10000\global \hangindent \localindentsize \global \hangafter =1 1. {\fam \itfam \nineit Quantum Hall effect. Topological problems in condensed matter physics}, Bibliopolis, (1988). } 
{ \global \localindentsize =\parindent \global \setbox \numberbox =\hbox {1. } \global \advance \localindentsize by \wd \numberbox \par \vskip 2.4 pt\noindent Morrison D. R. and Plesser R. M.\par \penalty 10000\global \hangindent \localindentsize \global \hangafter =1 1. Summing the instantons: quantum cohomology and mirror symmetry in toric varieties, {\fam \itfam \nineit Nucl. Phys.}, {\fam \bffam \ninebf B440}, 279--354, (1995). } 
{ \global \localindentsize =\parindent \global \setbox \numberbox =\hbox {1. } \global \advance \localindentsize by \wd \numberbox \par \vskip 2.4 pt\noindent Morse M.\par \penalty 10000\global \hangindent \localindentsize \global \hangafter =1 1. {\fam \itfam \nineit Calculus of variations in the large}, Amer. Math. Soc. Colloq. Publ., (1934). } 
{ \global \localindentsize =\parindent \global \setbox \numberbox =\hbox {1. } \global \advance \localindentsize by \wd \numberbox \par \vskip 2.4 pt\noindent Nash C.\par \penalty 10000\global \hangindent \localindentsize \global \hangafter =1 1. {\fam \itfam \nineit Differential Topology and Quantum Field Theory}, Academic Press, (1991). } 
{ \global \localindentsize =\parindent \global \setbox \numberbox =\hbox {2. } \global \advance \localindentsize by \wd \numberbox \par \global \hangindent \localindentsize \global \hangafter =1 2. A comment on Witten's topological Lagrangian, {\fam \itfam \nineit Mod. Phy. Lett. A}, {\fam \bffam \ninebf 7}, 1953--1958, (1992). } 
{ \global \localindentsize =\parindent \global \setbox \numberbox =\hbox {1. } \global \advance \localindentsize by \wd \numberbox \par \vskip 2.4 pt\noindent Nielsen H. and Olesen P.\par \penalty 10000\global \hangindent \localindentsize \global \hangafter =1 1. Vortex line models for dual strings, {\fam \itfam \nineit Nucl. Phys.}, {\fam \bffam \ninebf B61}, 45--61, (1973). } 
{ \global \localindentsize =\parindent \global \setbox \numberbox =\hbox {1. } \global \advance \localindentsize by \wd \numberbox \par \vskip 2.4 pt\noindent Oppenheimer J. R. and Snyder H.\par \penalty 10000\global \hangindent \localindentsize \global \hangafter =1 1. On continued gravitational contraction, {\fam \itfam \nineit Phys. Rev}, {\fam \bffam \ninebf 56}, 455--459, (1939). } 
{ \global \localindentsize =\parindent \global \setbox \numberbox =\hbox {1. } \global \advance \localindentsize by \wd \numberbox \par \vskip 2.4 pt\noindent Osborn H.\par \penalty 10000\global \hangindent \localindentsize \global \hangafter =1 1. Topological charges for $N=4$ supersymmetric gauge theories and monopoles of spin $1$, {\fam \itfam \nineit Phys. Lett.}, {\fam \bffam \ninebf 83B}, 321--326, (1979). } 
{ \global \localindentsize =\parindent \global \setbox \numberbox =\hbox {1. } \global \advance \localindentsize by \wd \numberbox \par \vskip 2.4 pt\noindent Osterwalder K. and Schrader R.\par \penalty 10000\global \hangindent \localindentsize \global \hangafter =1 1. Axioms for Euclidean Green's functions, {\fam \itfam \nineit Commun. Math. Phys.}, {\fam \bffam \ninebf 31}, 83--112, (1973). } 
{ \global \localindentsize =\parindent \global \setbox \numberbox =\hbox {2. } \global \advance \localindentsize by \wd \numberbox \par \global \hangindent \localindentsize \global \hangafter =1 2. Axioms for Euclidean Green's functions 2, {\fam \itfam \nineit Commun. Math. Phys.}, {\fam \bffam \ninebf 42}, 281--305, (1975). } 
{ \global \localindentsize =\parindent \global \setbox \numberbox =\hbox {3. } \global \advance \localindentsize by \wd \numberbox \par \global \hangindent \localindentsize \global \hangafter =1 3. {\fam \itfam \nineit Springer Lecture notes in physics {\fam \bffam \ninebf 25}: Constructive quantum field theory}, Springer, (1973). } 
{ \global \localindentsize =\parindent \global \setbox \numberbox =\hbox {1. } \global \advance \localindentsize by \wd \numberbox \par \vskip 2.4 pt\noindent Penrose R.\par \penalty 10000\global \hangindent \localindentsize \global \hangafter =1 1. Gravitational collapse and space-time singularities, {\fam \itfam \nineit Phys. Rev. Lett.}, {\fam \bffam \ninebf 14}, 57--59, (1965). } 
{ \global \localindentsize =\parindent \global \setbox \numberbox =\hbox {1. } \global \advance \localindentsize by \wd \numberbox \par \vskip 2.4 pt\noindent Pham F.\par \penalty 10000\global \hangindent \localindentsize \global \hangafter =1 1. {\fam \itfam \nineit Introduction \accent 18 a l'\accent 19 etude topologique des singulariti\accent 19 es de Landau. M\accent 19 emorial des Sciences Math\-\accent 19 emat\-iques, Fasc. 164}, Gauthier-Villars, (1967). } 
{ \global \localindentsize =\parindent \global \setbox \numberbox =\hbox {1. } \global \advance \localindentsize by \wd \numberbox \par \vskip 2.4 pt\noindent Poincar\accent 19 e H.\par \penalty 10000\global \hangindent \localindentsize \global \hangafter =1 1. Sur le probl\accent 18 eme de trois corps et les \accent 19 equations de la dynamique, {\fam \itfam \nineit Acta Mathematica}, {\fam \bffam \ninebf 13}, 1--270, (1890). } 
{ \global \localindentsize =\parindent \global \setbox \numberbox =\hbox {2. } \global \advance \localindentsize by \wd \numberbox \par \global \hangindent \localindentsize \global \hangafter =1 2. Sur les courbes d\accent 19 efinies par une \accent 19 equation diff\accent 19 erentielle, {\fam \itfam \nineit C. Rend. Acad. Sc.}, {\fam \bffam \ninebf 90}, 673--675, (1880). } 
{ \global \localindentsize =\parindent \global \setbox \numberbox =\hbox {3. } \global \advance \localindentsize by \wd \numberbox \par \global \hangindent \localindentsize \global \hangafter =1 3. M\accent 19 emoire sur les courbes d\accent 19 efinies par une \accent 19 equation diff\accent 19 erentielle, {\fam \itfam \nineit J. de Math.}, {\fam \bffam \ninebf 7}, 375--422, (1881). } 
{ \global \localindentsize =\parindent \global \setbox \numberbox =\hbox {4. } \global \advance \localindentsize by \wd \numberbox \par \global \hangindent \localindentsize \global \hangafter =1 4. M\accent 19 emoire sur les courbes d\accent 19 efinies par une \accent 19 equation diff\accent 19 erentielle, {\fam \itfam \nineit J. de Math.}, {\fam \bffam \ninebf 8}, 251--296, (1882). } 
{ \global \localindentsize =\parindent \global \setbox \numberbox =\hbox {5. } \global \advance \localindentsize by \wd \numberbox \par \global \hangindent \localindentsize \global \hangafter =1 5. Sur les courbes d\accent 19 efinies par les \accent 19 equations diff\accent 19 erentielle, {\fam \itfam \nineit J. de Math.}, {\fam \bffam \ninebf 1}, 167--244, (1885). } 
{ \global \localindentsize =\parindent \global \setbox \numberbox =\hbox {6. } \global \advance \localindentsize by \wd \numberbox \par \global \hangindent \localindentsize \global \hangafter =1 6. Sur les courbes d\accent 19 efinies par les \accent 19 equations diff\accent 19 erentielle, {\fam \itfam \nineit J. de Math.}, {\fam \bffam \ninebf 2}, 151--217, (1886). } 
{ \global \localindentsize =\parindent \global \setbox \numberbox =\hbox {7. } \global \advance \localindentsize by \wd \numberbox \par \global \hangindent \localindentsize \global \hangafter =1 7. Analyse des travaux scientifiques de Henri Poincar\accent 19 e, faite par lui m\accent 94 eme, {\fam \itfam \nineit Acta Math.}, {\fam \bffam \ninebf 38}, 1--135, (1921). } 
{ \global \localindentsize =\parindent \global \setbox \numberbox =\hbox {8. } \global \advance \localindentsize by \wd \numberbox \par \global \hangindent \localindentsize \global \hangafter =1 8. Analysis situs, {\fam \itfam \nineit J. \accent 19 Ec. Polyt.}, {\fam \bffam \ninebf 1}, 1--121, (1895). } 
{ \global \localindentsize =\parindent \global \setbox \numberbox =\hbox {9. } \global \advance \localindentsize by \wd \numberbox \par \global \hangindent \localindentsize \global \hangafter =1 9. Sur l'Analysis situs, {\fam \itfam \nineit C. R. Acad. Sc.}, {\fam \bffam \ninebf 115}, 633--636, (1892). } 
{ \global \localindentsize =\parindent \global \setbox \numberbox =\hbox {10. } \global \advance \localindentsize by \wd \numberbox \par \global \hangindent \localindentsize \global \hangafter =1 10. Sur les nombres de Betti, {\fam \itfam \nineit C. R. Acad. Sc.}, {\fam \bffam \ninebf 128}, 629--630, (1899). } 
{ \global \localindentsize =\parindent \global \setbox \numberbox =\hbox {11. } \global \advance \localindentsize by \wd \numberbox \par \global \hangindent \localindentsize \global \hangafter =1 11. Compl\accent 19 ement \accent 18 a l'Analysis situs, {\fam \itfam \nineit Rend. Circ. Matem. Palermo}, {\fam \bffam \ninebf 13}, 285--343, (1899). } 
{ \global \localindentsize =\parindent \global \setbox \numberbox =\hbox {12. } \global \advance \localindentsize by \wd \numberbox \par \global \hangindent \localindentsize \global \hangafter =1 12. Second compl\accent 19 ement \accent 18 a l'Analysis situs, {\fam \itfam \nineit Proc. Lond. math. Soc.}, {\fam \bffam \ninebf 32}, 277--308, (1900). } 
{ \global \localindentsize =\parindent \global \setbox \numberbox =\hbox {13. } \global \advance \localindentsize by \wd \numberbox \par \global \hangindent \localindentsize \global \hangafter =1 13. Sur l'Analysis situs, {\fam \itfam \nineit C. R. Acad. Sc.}, {\fam \bffam \ninebf 133}, 707--709, (1901). } 
{ \global \localindentsize =\parindent \global \setbox \numberbox =\hbox {14. } \global \advance \localindentsize by \wd \numberbox \par \global \hangindent \localindentsize \global \hangafter =1 14. Sur certaines surfaces alg\accent 19 ebriques; troisi\accent 18 eme compl\accent 19 ement \accent 18 a l'Analysis situs, {\fam \itfam \nineit Bull. Soc. Math. Fr.}, {\fam \bffam \ninebf 30}, 49--70, (1902). } 
{ \global \localindentsize =\parindent \global \setbox \numberbox =\hbox {15. } \global \advance \localindentsize by \wd \numberbox \par \global \hangindent \localindentsize \global \hangafter =1 15. Sur la connexion des surfaces alg\accent 19 ebriques, {\fam \itfam \nineit C. R. Acad. Sc.}, {\fam \bffam \ninebf 133}, 969--973, (1901). } 
{ \global \localindentsize =\parindent \global \setbox \numberbox =\hbox {16. } \global \advance \localindentsize by \wd \numberbox \par \global \hangindent \localindentsize \global \hangafter =1 16. Sur les cycles des surfaces alg\accent 19 ebriques; quatri\accent 18 eme compl\accent 19 ement \accent 18 a l'Analysis situs, {\fam \itfam \nineit J. Math. pures et appl.}, {\fam \bffam \ninebf 8}, 169--214, (1902). } 
{ \global \localindentsize =\parindent \global \setbox \numberbox =\hbox {17. } \global \advance \localindentsize by \wd \numberbox \par \global \hangindent \localindentsize \global \hangafter =1 17. Cinqui\accent 18 eme compl\accent 19 ement \accent 18 a l'Analysis situs, {\fam \itfam \nineit Rend. Circ. Matem. Palermo}, {\fam \bffam \ninebf 18}, 45--110, (1904). } 
{ \global \localindentsize =\parindent \global \setbox \numberbox =\hbox {18. } \global \advance \localindentsize by \wd \numberbox \par \global \hangindent \localindentsize \global \hangafter =1 18. Sur un th\accent 19 eor\accent 18 eme de g\accent 19 eom\accent 19 etrie, {\fam \itfam \nineit Rend. Circ. Matem. Palermo}, {\fam \bffam \ninebf 33}, 375--407, (1912). } 
{ \global \localindentsize =\parindent \global \setbox \numberbox =\hbox {1. } \global \advance \localindentsize by \wd \numberbox \par \vskip 2.4 pt\noindent Polyakov A. M.\par \penalty 10000\global \hangindent \localindentsize \global \hangafter =1 1. Spectrum of particles in the quantum field theory, {\fam \itfam \nineit JETP Lett.}, {\fam \bffam \ninebf 20}, 194--195, (1974). } 
{ \global \localindentsize =\parindent \global \setbox \numberbox =\hbox {1. } \global \advance \localindentsize by \wd \numberbox \par \vskip 2.4 pt\noindent Pont J.-C.\par \penalty 10000\global \hangindent \localindentsize \global \hangafter =1 1. {\fam \itfam \nineit La topologie alg\accent 19 ebrique des origines \accent 18 a Poincar\accent 19 e}, Presses Universitaires de France, (1974). } 
{ \global \localindentsize =\parindent \global \setbox \numberbox =\hbox {1. } \global \advance \localindentsize by \wd \numberbox \par \vskip 2.4 pt\noindent Prasad M. K. and Sommerfield C. M.\par \penalty 10000\global \hangindent \localindentsize \global \hangafter =1 1. Exact classical solution for the 't Hooft monopole and the Julia-Zee dyon, {\fam \itfam \nineit Phys. Rev. Lett.}, {\fam \bffam \ninebf 35}, 760--762, (1975). } 
{ \global \localindentsize =\parindent \global \setbox \numberbox =\hbox {1. } \global \advance \localindentsize by \wd \numberbox \par \vskip 2.4 pt\noindent Riemann B.\par \penalty 10000\global \hangindent \localindentsize \global \hangafter =1 1. Theorie der Abel'schen Functionen, {\fam \itfam \nineit Jour. f\accent 127 ur die reine und ang. Math.}, {\fam \bffam \ninebf 54}, , (1857). } 
{ \global \localindentsize =\parindent \global \setbox \numberbox =\hbox {1. } \global \advance \localindentsize by \wd \numberbox \par \vskip 2.4 pt\noindent Rohlin V. A.\par \penalty 10000\global \hangindent \localindentsize \global \hangafter =1 1. New results in the theory of 4 dimensional manifolds, {\fam \itfam \nineit Dok. Akad. Nauk. U. S. S. R.}, {\fam \bffam \ninebf 84}, 221--224, (1952). } 
{ \global \localindentsize =\parindent \global \setbox \numberbox =\hbox {1. } \global \advance \localindentsize by \wd \numberbox \par \vskip 2.4 pt\noindent Saari D. G. and Xia Z. (eds)\par \penalty 10000\global \hangindent \localindentsize \global \hangafter =1 1. {\fam \itfam \nineit Hamiltonian mechanics and celestial mechanics}, Amer. Math. Soc., (1996). } 
{ \global \localindentsize =\parindent \global \setbox \numberbox =\hbox {1. } \global \advance \localindentsize by \wd \numberbox \par \vskip 2.4 pt\noindent Schwarzschild K.\par \penalty 10000\global \hangindent \localindentsize \global \hangafter =1 1. \accent 127 Uber das Gravitationsfeld eines Masses nach der Einsteinschen Theorie, {\fam \itfam \nineit Sitzungberichte K\accent 127 oniglich Preuss. Akad. Wiss., Physik-Math. Kl.}, {\fam \bffam \ninebf }, 189--196, (1916). } 
{ \global \localindentsize =\parindent \global \setbox \numberbox =\hbox {2. } \global \advance \localindentsize by \wd \numberbox \par \global \hangindent \localindentsize \global \hangafter =1 2. \accent 127 Uber das Gravitationsfeld einer Kugal aus inkompressibler Fl\accent 127 ussigkeit nach der Einsteinschen Theorie, {\fam \itfam \nineit Sitzungberichte K\accent 127 oniglich Preuss. Akad. Wiss., Physik-Math. Kl.}, {\fam \bffam \ninebf }, 424--434, (1916). } 
{ \global \localindentsize =\parindent \global \setbox \numberbox =\hbox {1. } \global \advance \localindentsize by \wd \numberbox \par \vskip 2.4 pt\noindent Schwinger J.\par \penalty 10000\global \hangindent \localindentsize \global \hangafter =1 1. Sources and magnetic charges, {\fam \itfam \nineit Phys. Rev.}, {\fam \bffam \ninebf 173}, 1536--1544, (1968). } 
{ \global \localindentsize =\parindent \global \setbox \numberbox =\hbox {1. } \global \advance \localindentsize by \wd \numberbox \par \vskip 2.4 pt\noindent Segal G.\par \penalty 10000\global \hangindent \localindentsize \global \hangafter =1 1. {\fam \itfam \nineit Two dimensional conformal field theories and modular functors}, I. A. M. P. Congress, Swansea, 1988, {edited by: Davies I., Simon B. and Truman A.}, Institute of Physics, (1989). } 
{ \global \localindentsize =\parindent \global \setbox \numberbox =\hbox {1. } \global \advance \localindentsize by \wd \numberbox \par \vskip 2.4 pt\noindent Seiberg N. and Witten E.\par \penalty 10000\global \hangindent \localindentsize \global \hangafter =1 1. Electric-magnetic duality, monopole condensation, and confinement in $N=2$ supersymmetric Yang--Mills theory, {\fam \itfam \nineit Nucl. Phys.}, {\fam \bffam \ninebf B426}, 19--52, (1994; Erratum-ibid., {\fam \bffam \ninebf B430}, 485--486, 1994). } 
{ \global \localindentsize =\parindent \global \setbox \numberbox =\hbox {2. } \global \advance \localindentsize by \wd \numberbox \par \global \hangindent \localindentsize \global \hangafter =1 2. Monopoles, duality and chiral symmetry breaking in $N=2$ supersymmetric QCD, {\fam \itfam \nineit Nucl. Phys.}, {\fam \bffam \ninebf B431}, 484--550, (1994). } 
{ \global \localindentsize =\parindent \global \setbox \numberbox =\hbox {1. } \global \advance \localindentsize by \wd \numberbox \par \vskip 2.4 pt\noindent Simon B.\par \penalty 10000\global \hangindent \localindentsize \global \hangafter =1 1. Holonomy, the quantum adiabatic theorem and Berry's phase, {\fam \itfam \nineit Phys. Rev. Lett.}, {\fam \bffam \ninebf 51}, 2167--2170, (1983). } 
{ \global \localindentsize =\parindent \global \setbox \numberbox =\hbox {1. } \global \advance \localindentsize by \wd \numberbox \par \vskip 2.4 pt\noindent Smale S.\par \penalty 10000\global \hangindent \localindentsize \global \hangafter =1 1. Generalised Poincar\accent 19 e's conjecture in dimensions greater than four, {\fam \itfam \nineit Ann. Math.}, {\fam \bffam \ninebf 74}, 391--406, (1961). } 
{ \global \localindentsize =\parindent \global \setbox \numberbox =\hbox {2. } \global \advance \localindentsize by \wd \numberbox \par \global \hangindent \localindentsize \global \hangafter =1 2. Topology and mechanics I, {\fam \itfam \nineit Inventiones Math.}, {\fam \bffam \ninebf 10}, 305--331, (1970). } 
{ \global \localindentsize =\parindent \global \setbox \numberbox =\hbox {3. } \global \advance \localindentsize by \wd \numberbox \par \global \hangindent \localindentsize \global \hangafter =1 3. Topology and mechanics II, {\fam \itfam \nineit Inventiones Math.}, {\fam \bffam \ninebf 11}, 45-64, (1970). } 
{ \global \localindentsize =\parindent \global \setbox \numberbox =\hbox {1. } \global \advance \localindentsize by \wd \numberbox \par \vskip 2.4 pt\noindent Streater R. F.\par \penalty 10000\global \hangindent \localindentsize \global \hangafter =1 1. Outline of axiomatic relativistic quantum field theory, {\fam \itfam \nineit Rep. Prog. Phys.}, {\fam \bffam \ninebf 38}, 771--846, (1975). } 
{ \global \localindentsize =\parindent \global \setbox \numberbox =\hbox {1. } \global \advance \localindentsize by \wd \numberbox \par \vskip 2.4 pt\noindent Strominger A. and Vafa C.\par \penalty 10000\global \hangindent \localindentsize \global \hangafter =1 1. Microscopic origin of the Bekenstein--Hawking entropy, {\fam \itfam \nineit Phys. Lett.}, {\fam \bffam \ninebf B379}, 99--104, (1996). } 
{ \global \localindentsize =\parindent \global \setbox \numberbox =\hbox {1. } \global \advance \localindentsize by \wd \numberbox \par \vskip 2.4 pt\noindent Sundman K. F.\par \penalty 10000\global \hangindent \localindentsize \global \hangafter =1 1. Nouvelles recherche sur le probl\accent 18 eme des trois corps, {\fam \itfam \nineit Acta Soc. Sci. Fenn.}, {\fam \bffam \ninebf 35}, 1--27, (1909). } 
{ \global \localindentsize =\parindent \global \setbox \numberbox =\hbox {2. } \global \advance \localindentsize by \wd \numberbox \par \global \hangindent \localindentsize \global \hangafter =1 2. Recherches sur le probl\accent 18 eme des trois corps, {\fam \itfam \nineit Acta Soc. Sci. Fenn.}, {\fam \bffam \ninebf 34}, 1--43, (1907). } 
{ \global \localindentsize =\parindent \global \setbox \numberbox =\hbox {3. } \global \advance \localindentsize by \wd \numberbox \par \global \hangindent \localindentsize \global \hangafter =1 3. Memoire sur le probl\accent 18 eme des trois corps, {\fam \itfam \nineit Acta Mathematica}, {\fam \bffam \ninebf 36}, 105--179, (1912). } 
{ \global \localindentsize =\parindent \global \setbox \numberbox =\hbox {1. } \global \advance \localindentsize by \wd \numberbox \par \vskip 2.4 pt\noindent Tait P. G.\par \penalty 10000\global \hangindent \localindentsize \global \hangafter =1 1. {\fam \itfam \nineit Scientific papers (On knots I, II, III) 273--437}, Cambridge University Press, (1898). } 
{ \global \localindentsize =\parindent \global \setbox \numberbox =\hbox {1. } \global \advance \localindentsize by \wd \numberbox \par \vskip 2.4 pt\noindent Taubes C. H.\par \penalty 10000\global \hangindent \localindentsize \global \hangafter =1 1. A framework for Morse theory for the Yang-Mills functional, {\fam \itfam \nineit Inventiones Math.}, {\fam \bffam \ninebf 94}, 327--402, (1988). } 
{ \global \localindentsize =\parindent \global \setbox \numberbox =\hbox {2. } \global \advance \localindentsize by \wd \numberbox \par \global \hangindent \localindentsize \global \hangafter =1 2. Min-Max theory for the Yang-Mills-Higgs Equations, {\fam \itfam \nineit Commun. Math. Phys.}, {\fam \bffam \ninebf 97}, 473--540, (1985). } 
{ \global \localindentsize =\parindent \global \setbox \numberbox =\hbox {1. } \global \advance \localindentsize by \wd \numberbox \par \vskip 2.4 pt\noindent Thom R.\par \penalty 10000\global \hangindent \localindentsize \global \hangafter =1 1. Topological models in biology, {\fam \itfam \nineit Topology}, {\fam \bffam \ninebf 8}, 313--335, (1969). } 
{ \global \localindentsize =\parindent \global \setbox \numberbox =\hbox {2. } \global \advance \localindentsize by \wd \numberbox \par \global \hangindent \localindentsize \global \hangafter =1 2. {\fam \itfam \nineit Stabilit\accent 19 e structurelle et morphogen\accent 18 ese}, Benjamin, (1972). } 
{ \global \localindentsize =\parindent \global \setbox \numberbox =\hbox {3. } \global \advance \localindentsize by \wd \numberbox \par \global \hangindent \localindentsize \global \hangafter =1 3. {\fam \itfam \nineit Mod\accent 18 eles math\accent 19 ematiques de la morphogen\accent 18 ese}, Acad. Naz. Lincei, Pisa, (1971). } 
{ \global \localindentsize =\parindent \global \setbox \numberbox =\hbox {1. } \global \advance \localindentsize by \wd \numberbox \par \vskip 2.4 pt\noindent Thomson W. H. (Lord Kelvin)\par \penalty 10000\global \hangindent \localindentsize \global \hangafter =1 1. On vortex atoms, {\fam \itfam \nineit Proc. Roy. Soc. Edin.}, {\fam \bffam \ninebf 34}, 15--24, (1867). } 
{ \global \localindentsize =\parindent \global \setbox \numberbox =\hbox {2. } \global \advance \localindentsize by \wd \numberbox \par \global \hangindent \localindentsize \global \hangafter =1 2. On vortex motion, {\fam \itfam \nineit Trans. Roy. Soc. Edin.}, {\fam \bffam \ninebf 25}, 217--260, (1869). } 
{ \global \localindentsize =\parindent \global \setbox \numberbox =\hbox {3. } \global \advance \localindentsize by \wd \numberbox \par \global \hangindent \localindentsize \global \hangafter =1 3. Vortex statics, {\fam \itfam \nineit Proc. Roy. Soc. Edin.}, {\fam \bffam \ninebf 1875--76 session}, , (1875). } 
{ \global \localindentsize =\parindent \global \setbox \numberbox =\hbox {4. } \global \advance \localindentsize by \wd \numberbox \par \global \hangindent \localindentsize \global \hangafter =1 4. {\fam \itfam \nineit Mathematical and physical papers volume IV}, Cambridge University Press, (1910). } 
{ \global \localindentsize =\parindent \global \setbox \numberbox =\hbox {5. } \global \advance \localindentsize by \wd \numberbox \par \global \hangindent \localindentsize \global \hangafter =1 5. On the duties of ether for electricity and magnetism, {\fam \itfam \nineit Phil. Mag.}, {\fam \bffam \ninebf 50}, 305--307, (1900). } 
{ \global \localindentsize =\parindent \global \setbox \numberbox =\hbox {1. } \global \advance \localindentsize by \wd \numberbox \par \vskip 2.4 pt\noindent 't Hooft G.\par \penalty 10000\global \hangindent \localindentsize \global \hangafter =1 1. Renormalization of massless Yang--Mills fields, {\fam \itfam \nineit Nucl. Phys.}, {\fam \bffam \ninebf B33}, 173--199, (1971). } 
{ \global \localindentsize =\parindent \global \setbox \numberbox =\hbox {2. } \global \advance \localindentsize by \wd \numberbox \par \global \hangindent \localindentsize \global \hangafter =1 2. Magnetic monopoles in unified gauge theories, {\fam \itfam \nineit Nucl. Phys.}, {\fam \bffam \ninebf B79}, 276--284, (1974). } 
{ \global \localindentsize =\parindent \global \setbox \numberbox =\hbox {3. } \global \advance \localindentsize by \wd \numberbox \par \global \hangindent \localindentsize \global \hangafter =1 3. Computation of the quantum effects due to a four dimensional pseudoparticle, {\fam \itfam \nineit Phys. Rev.}, {\fam \bffam \ninebf D14}, 3432--3450, (1976). } 
{ \global \localindentsize =\parindent \global \setbox \numberbox =\hbox {1. } \global \advance \localindentsize by \wd \numberbox \par \vskip 2.4 pt\noindent 't Hooft G. and Veltman M.\par \penalty 10000\global \hangindent \localindentsize \global \hangafter =1 1. Regularization and renormalization of gauge fields, {\fam \itfam \nineit Nucl. Phys.}, {\fam \bffam \ninebf B44}, 189--213, (1972). } 
{ \global \localindentsize =\parindent \global \setbox \numberbox =\hbox {1. } \global \advance \localindentsize by \wd \numberbox \par \vskip 2.4 pt\noindent Thorne K. S.\par \penalty 10000\global \hangindent \localindentsize \global \hangafter =1 1. {\fam \itfam \nineit Black holes and time warps: Einstein's outrageous legacy}, W. W. Norton, (1994). } 
{ \global \localindentsize =\parindent \global \setbox \numberbox =\hbox {1. } \global \advance \localindentsize by \wd \numberbox \par \vskip 2.4 pt\noindent Tscheuschner R. D.\par \penalty 10000\global \hangindent \localindentsize \global \hangafter =1 1. Topological spin-statistics relation in quantum field theory, {\fam \itfam \nineit Int. Jour. Theor. Phys.}, {\fam \bffam \ninebf 28}, 1269--1310, (1989). } 
{ \global \localindentsize =\parindent \global \setbox \numberbox =\hbox {1. } \global \advance \localindentsize by \wd \numberbox \par \vskip 2.4 pt\noindent Ward R. S.\par \penalty 10000\global \hangindent \localindentsize \global \hangafter =1 1. On self-dual gauge fields, {\fam \itfam \nineit Phys. Lett.}, {\fam \bffam \ninebf 61A}, 81--82, (1977). } 
{ \global \localindentsize =\parindent \global \setbox \numberbox =\hbox {1. } \global \advance \localindentsize by \wd \numberbox \par \vskip 2.4 pt\noindent Wilczek F. and Zee A.\par \penalty 10000\global \hangindent \localindentsize \global \hangafter =1 1. Linking numbers, spin and statistics of solitons, {\fam \itfam \nineit Phys. Rev. Lett.}, {\fam \bffam \ninebf 51}, 2250--2252, (1983). } 
{ \global \localindentsize =\parindent \global \setbox \numberbox =\hbox {1. } \global \advance \localindentsize by \wd \numberbox \par \vskip 2.4 pt\noindent Witten E.\par \penalty 10000\global \hangindent \localindentsize \global \hangafter =1 1. Dyons of charge $e\theta /2\pi $, {\fam \itfam \nineit Phys. Lett.}, {\fam \bffam \ninebf 86B}, 283--287, (1979). } 
{ \global \localindentsize =\parindent \global \setbox \numberbox =\hbox {2. } \global \advance \localindentsize by \wd \numberbox \par \global \hangindent \localindentsize \global \hangafter =1 2. {\fam \itfam \nineit Some geometrical applications of quantum field theory}, I. A. M. P. Congress, Swansea, 1988, {edited by: Davies I., Simon B. and Truman A.}, Institute of Physics, (1989).} 
{ \global \localindentsize =\parindent \global \setbox \numberbox =\hbox {3. } \global \advance \localindentsize by \wd \numberbox \par \global \hangindent \localindentsize \global \hangafter =1 3. Quantum field theory and the Jones polynomial, {\fam \itfam \nineit Commun. Math. Phys.}, {\fam \bffam \ninebf 121}, 351--400, (1989). } 
{ \global \localindentsize =\parindent \global \setbox \numberbox =\hbox {4. } \global \advance \localindentsize by \wd \numberbox \par \global \hangindent \localindentsize \global \hangafter =1 4. Topological quantum field theory, {\fam \itfam \nineit Commun. Math. Phys.}, {\fam \bffam \ninebf 117}, 353--386, (1988). } 
{ \global \localindentsize =\parindent \global \setbox \numberbox =\hbox {5. } \global \advance \localindentsize by \wd \numberbox \par \global \hangindent \localindentsize \global \hangafter =1 5. Monopoles and four-manifolds, {\fam \itfam \nineit Math. Res. Lett.}, {\fam \bffam \ninebf 1}, 769--796, (1994). } 
{ \global \localindentsize =\parindent \global \setbox \numberbox =\hbox {6. } \global \advance \localindentsize by \wd \numberbox \par \global \hangindent \localindentsize \global \hangafter =1 6. Supersymmetry and Morse theory, {\fam \itfam \nineit J. Diff. Geom.}, {\fam \bffam \ninebf 17}, 661--692, (1982). } 
{ \global \localindentsize =\parindent \global \setbox \numberbox =\hbox {1. } \global \advance \localindentsize by \wd \numberbox \par \vskip 2.4 pt\noindent Yau S.-T. (ed.)\par \penalty 10000\global \hangindent \localindentsize \global \hangafter =1 1. {\fam \itfam \nineit Essays on Mirror Manifolds}, International Press, Hong Kong, (1992). } 
{ \global \localindentsize =\parindent \global \setbox \numberbox =\hbox {1. } \global \advance \localindentsize by \wd \numberbox \par \vskip 2.4 pt\noindent Zwanziger D.\par \penalty 10000\global \hangindent \localindentsize \global \hangafter =1 1. Exactly soluble nonrelativistic model of particles with both electric and magnetic charge, {\fam \itfam \nineit Phys. Rev.}, {\fam \bffam \ninebf 176}, 1480--1488, (1968). } 
{ \global \localindentsize =\parindent \global \setbox \numberbox =\hbox {2. } \global \advance \localindentsize by \wd \numberbox \par \global \hangindent \localindentsize \global \hangafter =1 2. Exactly soluble nonrelativistic model of particles with both electric and magnetic charge, {\fam \itfam \nineit Phys. Rev.}, {\fam \bffam \ninebf 176}, 1480--1488, (1968). } 

\bye
\cite[807,1]  
\cite[933,1]  
\cite[745,1]  
\cite[746,2]  
\cite[727,1]  
\cite[728,2]  
\cite[108,3]  
\cite[485,4]  
\cite[87,1]   
\cite[137,1]  
\cite[69,1]   
\cite[33,1]   
\cite[967,1]  
\cite[748,1]  
\cite[822,1]  
\cite[214,1]  
\cite[711,1]  
\cite[174,1]  
\cite[810,1]  
\cite[970,1]  
\cite[763,1]  
\cite[875,1]  
\cite[938,2]  
\cite[32,1]   
\cite[169,1]  
\cite[143,2]  
\cite[808,1]  
\cite[940,1]  
\cite[976,1]  
\cite[946,1]  
\cite[885,1]  
\cite[895,2]  
\cite[924,1]  
\cite[948,1]  
\cite[952,1]  
\cite[953,2]  
\cite[38,1]   
\cite[62,1]   
\cite[111,2]  
\cite[500,3]  
\cite[930,4]  
\cite[931,1]  
\cite[506,1]  
\cite[954,1]  
\cite[956,2]  
\cite[823,1]  
\cite[968,1]  
\cite[157,1]  
\cite[176,2]  
\cite[102,1]  
\cite[88,1]   
\cite[227,1]  
\cite[905,1]  
\cite[603,1]  
\cite[960,1]  
\cite[572,1]  
\cite[945,1]  
\cite[74,1]   
\cite[75,2]   
\cite[757,1]  
\cite[978,2]  
\cite[977,1]  
\cite[890,1]  
\cite[889,1]  
\cite[887,1]  
\cite[829,1]  
\cite[906,1]  
\cite[90,1]   
\cite[881,1]  
\cite[961,1]  
\cite[756,1]  
\cite[891,1]  
\cite[894,1]  
\cite[971,1] 
\cite[215,1]  
\cite[882,1]  
\cite[883,2]  
\cite[879,1]  
\cite[880,2]  
\cite[892,1]  
\cite[969,1]  
\cite[249,1]  
\cite[760,1]  
\cite[951,1]  
\cite[964,1]  
\cite[580,1]  
\cite[626,1]  
\cite[649,2]  
\cite[910,1]  
\cite[888,1]  
\cite[767,1]  
\cite[926,1]  
\cite[927,2]  
\cite[928,3]  
\cite[884,1]  
\cite[904,1]  
\cite[837,1]  
\cite[860,2]  
\cite[861,3]  
\cite[862,4]  
\cite[863,5]  
\cite[864,6]  
\cite[869,7]  
\cite[839,8]  
\cite[838,9]  
\cite[840,10] 
\cite[841,11] 
\cite[842,12] 
\cite[843,13] 
\cite[844,14] 
\cite[845,15] 
\cite[846,16] 
\cite[847,17] 
\cite[848,18] 
\cite[898,1]  
\cite[876,1]  
\cite[903,1]  
\cite[828,1]  
\cite[259,1]  
\cite[941,1]  
\cite[870,1]  
\cite[871,2]  
\cite[899,1]  
\cite[598,1]  
\cite[936,1]  
\cite[934,2]  
\cite[809,1]  
\cite[581,1]  
\cite[872,2]  
\cite[873,3]  
\cite[925,1]  
\cite[949,1]  
\cite[962,1]  
\cite[965,2]  
\cite[963,3]  
\cite[831,1]  
\cite[265,1]  
\cite[24,2]   
\cite[742,1]  
\cite[743,2]  
\cite[744,3]  
\cite[733,1]  
\cite[824,2]  
\cite[825,3]  
\cite[826,4]  
\cite[830,5]  
\cite[915,1]  
\cite[897,2]  
\cite[109,3]  
\cite[916,1]  
\cite[893,1]  
\cite[551,1]  
\cite[929,1]  
\cite[966,1]  
\cite[896,1]  
\cite[606,2]  
\cite[416,3]  
\cite[388,4]  
\cite[805,5]  
\cite[70,6]   
\cite[798,1]  
\cite[900,1]  
\cite[900,2]  
\bye